\documentclass[structabstract]{aa}  
\usepackage{graphicx}
\usepackage{txfonts}
\usepackage{url}
\usepackage{subfig, color}
\usepackage{natbib,twoopt}
\usepackage[breaklinks=true]{hyperref} 
\bibpunct{(}{)}{;}{a}{}{,} 
\newcommandtwoopt{\citeads}[3][][]{\href{http://adsabs.harvard.edu/abs/#3}%
{\citealp[#1][#2]{#3}}} 
\newcommandtwoopt{\citepads}[3][][]{\href{http://adsabs.harvard.edu/abs/#3}%
{\citep[#1][#2]{#3}}} 
\newcommandtwoopt{\citetads}[3][][]{\href{http://adsabs.harvard.edu/abs/#3}%
{\citet[#1][#2]{#3}}}
\newcommandtwoopt{\citeyearads}[3][][]%
{\href{http://adsabs.harvard.edu/abs/#3}{\citeyear[#1][#2]{#3}}}

\usepackage[normalem]{ulem}

\begin{document}
   \title{UPMASK: unsupervised photometric membership assignment in stellar clusters}
   \author{A. Krone-Martins
             \inst{1}
          \and
          A. Moitinho\inst{1}
          }

   \institute{SIM - Faculdade de Ci\^encias da Universidade de Lisboa, Ed. C8, Campo Grande, 1749-016 Lisboa, Portugal\\
              \email{algol@sim.ul.pt}
             }

   \date{Received , ; accepted}

 
  \abstract
   {}
  {We develop a method for membership assignment in stellar clusters using only photometry and positions. The method is aimed to be unsupervised, data driven, model free, and to rely on as few assumptions as possible.}
  {
The approached followed in this work for membership assessment is based on an iterative process, principal component analysis, clustering algorithm, and kernel density estimations. The method, UPMASK, is able to take into account arbitrary error models.  An implementation in R was tested on simulated clusters that covered a broad range of ages, masses, distances, reddenings, and also on real data of cluster fields.} 
 {
Running UPMASK on simulations showed that the algorithm effectively separates cluster and field populations.
The overall spatial structure and distribution of cluster member stars in the colour-magnitude diagram were recovered under a broad variety of conditions.
For a set of 360 simulations, the resulting true positive rates (a measurement of purity) and member recovery rates (a measurement of completeness) at the 90\% membership probability level reached high values for a range of open cluster ages ($10^{7.1}-10^{9.5}$ yr), initial masses ($0.5-10\times10^3$M$_{\sun}$) and heliocentric distances ($0.5-4.0$ kpc). UPMASK was also tested on real data from the fields of open cluster Haffner~16 and of the closely projected clusters Haffner~10 and Czernik~29. These tests showed that even for moderate variable extinction and cluster superposition, the method yielded useful cluster membership probabilities and provided some insight into their stellar contents. 
The UPMASK implementation is available at the CRAN archive. 
}
{}

   \keywords{Methods: data analysis -- Methods: statistical -- open clusters and associations: general -- open clusters and associations: individual: Haffner 10, Haffner 16, Czernik 29}

   \maketitle


\section{Introduction}

Astronomical research using observations of star clusters often meets a fundamental challenge: being able to recognise and make use of the signature of cluster members in data sets that are heavily contaminated by field stars. The signature can be of different kinds. It can be an apparent over-density of stars in a stellar field image or map, a cluster sequence in photometric diagrams, clustering of stars with common kinematics in proper motion, or radial velocity space, or similar parallaxes, among others.

Ideally, given the constrained spatial origin of stars in a cluster, membership would be determined through a three-dimensional selection of stars at similar distances, at least for young clusters not too dynamically evolved or for the concentrated members of older clusters.
In practice, accurate parallactic distances are not known except for the closest stars \citepads{1997ESASP1200.....P} and indirect distance indicators  (spectroscopic, photometric, among others) will not be accurate enough to distinguish members based on distances alone, except in very favourable cases such as high latitude fields with little field contamination.

After distances, proper motions are considered to be the best member discriminator. The canonical Vasilevskis-Sanders method \citepads{1958AJ.....63..387V, 1971A&A....14..226S} and variations have been widely used in the literature for determinations of cluster memberships and kinematics \citepads{2004AN....325..740K,2006A&A...446..949D,2010A&A...516A...3K}. These methods assume bivariate Gaussian distributions for cluster and field motions. It has been argued that assumptions on the distribution might not be adequate in some situations, which has led other authors to develop non-parametric approaches \citepads[e.g.][ and references within]{1990A&A...235...94C, 2004A&A...426..819B,2006A&A...447..915J}. However, independently of the details of each method, precise proper motions are required, which are currently constrained to relatively short heliocentric distances, since  relatively bright stars ($V\sim 15-16$) and measurable angular motions are required. Moreover, in any case these methods will not be able to distinguish cluster members in directions dominated by the solar motion.

Spectroscopic radial velocities can also be used, alone or combined with proper motions, for assessing cluster memberships.  
Combined with photometry, spectroscopy provides a powerful means for selecting cluster members since it allows one to precisely pinpoint each spectral type in photometric diagrams, determine individual reddenings and use reddening as a member discriminator. But spectroscopy is a telescope-time-intensive approach, which limits its use in characterising large numbers of clusters. However, recent and planned large spectroscopic surveys of the Milky Way such as the SDSS SEGUE and APOGEE \citepads{2009AJ....137.4377Y, 2007AAS...21113208M}, RAVE \citepads{2006AJ....132.1645S}, Gaia-ESO \citepads{2012Msngr.147...25G}, and LAMOST-LEGUE \citepads{2012RAA....12..735D} will allow  membership determinations for some hundreds of clusters. Still, as in the case of proper motions, stringent limits are imposed by the required brightness of stars and solar motion.

Multi-band imaging (and the resulting photometric catalogues) is by far the most wide-spread technique for studying star clusters. Imaging allows obtaining data for large numbers of stars over wide areas and reaches depths not accessible to other techniques, or at least not accessible to other techniques without a large investment of observing time. 
However, the rich numbers of stars brought by imaging surveys also bring the problem of strong contamination by field stars.

The extent to which cluster members need to be identified is driven by the scientific question being addressed. In some cases, a simple perception of the cluster presence can be enough. These are the cases of determining cluster centres and radii from star counts and classic isochrone eye-fitting of prominent cluster sequences in photometric diagrams. Cluster luminosity and mass functions can also be determined without knowledge of the specific cluster members through statistical comparisons with control fields \citepads[e.g. ][ and many others]{1997AJ....113.1359M}.
However, under heavy field contamination visual techniques break down since the eye no longer unambiguously identifies the cluster signature. Determination of cluster reddenings, distances, and ages by classical methods becomes almost a guess. Similarly with simple statistical approaches: the cluster over-density (in physical or magnitude space) becomes marginal when compared with the fluctuations in the number of field stars, yielding uncertain, or even useless, radii and luminosity and mass functions.

Additionally, recognising the sequence is not enough in many studies and knowledge of individual memberships are required. Examples range from hunting brown dwarfs and other research on specific stellar types, to separating members from clusters superimposed along their lines of sight, just to mention a few.  This is also the case of determining cluster parameters through automated fitting of zero-age main-sequence (ZAMS) or isochrone models to photometric sequences. It has long been recognised that objective and precise model-fitting procedures (as opposed to eye fitting) yielding repeatable results and rigorously assessed uncertainties should be employed. However, only recently have such methods began to appear more frequently  in the literature. The trigger has been the profusion of cluster photometric data and the need to compare results from different studies in a systematic way, together with common availability of the necessary computing power.  Examples of these approaches are the $\tau^2$ method by \citetads{2006MNRAS.373.1251N} and the Bayesian inversion method by \citetads{2006ApJ...645.1436V}, which require samples of stars only composed of cluster members or with very little contamination. 
\citetads{2010A&A...516A...2M} adopted a cross-entropy method for finding the best isochrone fit, but without requiring a pure sample of cluster stars. Instead, they lessened the effect of field contamination in the fits by adopting a filtering and weighting scheme based on  the distribution of stars in coordinate space and multi-dimensional magnitude space -- this method was subsequently refined in \citetads{2012A&A...539A.125D}.

Another problem closely related to photometric cluster membership determinations is the determination of individual stellar parameters from photometry, including reddening and stellar classification.  Reddening can provide a criterion for membership determinations. For clusters unaffected by variable reddening, member stars appear reddened by the same amount, which will in general be different for background and foreground stars. This is specially applicable to the majority of open clusters, which are located at low Galactic latitudes where integrated extinction along the line of sight varies significantly with distance. Individual stellar classification together with photometric measurements allow a rough distance estimation and thus cleaning the sample by selecting stars within a range of distances.
These stellar parameter determinations of cluster members  are in turn linked to the problem of determining cluster parameters (reddening, distance, age, metallicity) by fitting isochrones to photometric cluster sequences. If this can be done, the values for all the member stars become determined, assuming they share these properties. Both problems have a noticeable aspect in common: the solutions are model dependent, relying on the adoption of reference lines (isochrones, ZAMS) and reddening law. They may also be sensitive to mismatches between the  passbands used in the observations and the ones assumed in the models.  

The motivation of this paper is the identification of star cluster members, buried in contaminated fields, in a purely data-driven way without relying on models. The paper focuses on the special case of membership determinations using exclusively multi-band photometry and projected on-sky positions. However, the method developed here - UPMASK - can be immediately applied to other types of data, including spectrophotometry (as produced by the forthcoming Gaia mission), radial velocities, and proper motions. In principle, and as discussed in more detail in Sect.~\ref{sect:remarks}, the method can be perfected in a way such that it accommodates missing data and other measurement errors distributions. This will be addressed in future work.

The remainder of this paper is organised as follows. Sect.~\ref{sect:method} describes the UPMASK method. Sects.~\ref{sect:simdatamain} and~\ref{sect:realdatamain} report on the results of UPMASK using simulated and real data, respectively. Limitations and possible extensions of the method are discussed in Sect.~\ref{sect:remarks}, followed by the conclusions of this work in Sect.~\ref{sect:conclusions}.\footnote{This article addresses \emph{star clusters} as well as \emph{clusters of data} (as identified by clustering algorithms). For economy and readability, the word \emph{cluster} will be used to describe either whenever the context is unambiguous.}

\section{The UPMASK method\label{sect:method}}
\subsection{Conceptual approach\label{sect:concept}}

Owing to their common origin and spatial confinement, members of stellar clusters share common properties. Statistically, this means that in most parameter spaces star cluster members are expected to be clustered together. The clustering region, however, can assume arbitrary formats such as tubular regions, or even be disconnected.
Field stars can also be clustered in certain parameter spaces but are not expected to cluster in most (e.g. the space defined by reddening, distance, and age).

In this paper, we define a stellar cluster as a spatial over-density of stars with a common origin.\footnote{Different schools have different definitions for star clusters. Another popular and more restrictive definition requires clusters to be gravitationally bound.} Conversely, field stars are expected to be spatially scattered as they do not share a single origin. According to this definition, random subsets of stellar cluster members are spatially concentrated. This is not necessarily the case with the measured photometric indexes. For example, arbitrary subsets of cluster members are not necessarily concentrated around a common colour index. 
In this view, the positional space has a favoured place in the UPMASK method.

These are the main assumptions on which the UPMASK method relies: cluster members will be clustered in most spaces, including positional space. Field stars, even if clustered in some spaces, are not expected to cluster in positional space.

Because of their position in the Galaxy, star clusters are affected by tidal interactions that can reshape their spatial density profiles, deviating from those of self-gravitating stellar systems, such as the King profile. Also, although the morphologies of star cluster colour-magnitude diagrams (CMDs)  are fairly well described by current stellar evolution models, not all evolutionary stages, mass ranges, and even colour-magnitude combinations are equally well described \citepads[e.g.][]{2006A&A...453..101L, 2010ARA&A..48..581S}. In particular, there is still much to be improved for younger clusters with pre-main sequence stars, especially for the less massive stars. Finally, one should not exclude the possibility that the data could show the unexpected, such as clusters with more than one epoch of star formation, which could be particularly relevant for globular clusters, merged clusters, or clusters projected on the same line of sight.

Therefore, another central point in the design of the method is that it should be data-driven. Apart from the spatial uniformity of field members, UPMASK does not assume a priori parametrisations, such as isochrones or King profiles, of any probability distributions involved.

\subsection{Dimensional transformation \label{sect:dimred}}

Although UPMASK is a data-driven approach, it is not the data, but the information they encode that drives the membership assignment. However, several observable spaces, and particularly multi-magnitude and multi-colour spaces, exhibit redundant information. This is manifested through correlations between the observables.

Ideally, membership assessment would be performed in spaces of physically uncorrelated variables. In these spaces, the distances between the observables representing these variables are maximised. By eliminating correlations between observables, and thus redundancy, these spaces have the smallest number of parameters encoding (most of) the information content of the original observables.

A common approach for building such spaces is provided by the principal component analysis (PCA). 
This is a linear transformation of the original data onto an orthogonal coordinate system where projections of the data on the axes are ordered by variance: the first principal component describes most of the variance in the data, and so on. By finding an  orthogonal coordinate system, PCA transforms linearly correlated variables to uncorrelated variables.

The examples in this paper use UBVRI photometric data-sets. In addition to using the original $U$, $B$, $V$, $R$ and $I$ magnitudes, PCA was also performed using the $U-B$, $B-V$, $V-I$, $R-I$ colours and the (almost) reddening-free index $Q$\footnote{$Q=(U-B) - 0.72\times(B-V).$} as additional observables. The reason is that PCA is designed for linear correlations between the observables, but the interdependencies between the UBVRI bands are non-linear (principal components are expected to vary between different spectral types) and some help had to be given by providing additional observables with a higher capability of separating the intervening physical factors.

The resulting principal components were found to have most of the variance in the first four components, with the fourth providing a minor contribution. We interpret this as an indication that these components contain information on the main four physical quantities that intuitively are expected to contribute to the variance in UBVRI photometric data: distance, reddening, temperature, and, to a lesser degree, metallicity or surface gravity. In the end, the justification for using linear PCA is the good performance of UMASK reported in Sects.~\ref{sect:simdatamain} and ~\ref{sect:realdatamain}. In this sense, linear PCA is to be taken as a useful heuristic with a physical motivation.

Although the exact number varies from data-set to data-set, the first three principal components are responsible for 99\% of the variance in the data, with the fourth component contributing to almost all of the remaining variance. For instance, in the Haffner 16  data-set analysed in Section \ref{haf16} the contributions from the four most important PCs are 58.8\%, 31.3\%, 9.0\%, and 0.9\%.

We note that since the method is data-driven, it is also expected to work on other kinds of parameter spaces. Naturally, its success in segregating populations depends on the discriminating capability of physical quantities encoded by the chosen observables.

\subsection{Implementation of the method \label{sect:implem}}

The concept described above was implemented in the R language \citepads{R}. The implementation relies on two iterative processes. The first process, called the \emph{UPMASK kernel}, implements the core of the concept. The second process, which we call the \emph{outer loop}, adds the capability of taking measurement errors into account, as well as of relaxing the dependency on the initial conditions chosen for the adopted clustering algorithms.

\subsubsection{Main iterative process: UPMASK kernel\label{sect:mainiter}}

The UPMASK kernel is the core of the UPMASK method. It is an iterative loop composed of three main steps.

First, a PCA is performed on the selected observables, positions excluded. Since the observables will in general have different variances and span different ranges, they are first scaled to unit variance. This scaling is applied to avoid that observables with the highest variance dominate the principal components. The PCA result will allow  selecting the most significant principal components. As discussed in Sect.~\ref{sect:dimred}, physical motivations and information content of broadband UBVRI data have led us to consider four principal components. 

Then, a clustering analysis is conducted on the selected principal components. Although the method concept is independent of the adopted data-clustering algorithm, a choice has to be made. In this paper we adopted one of the simplest algorithms, $k$-means with a random initialisation, but others can be used as well. The number of clusters, $k$, is not directly set by the user. Instead, the user sets the number of stars per cluster. The number of clusters will then be the total number of stars under analysis, divided by the number of stars per cluster, rounded up to the next integer. The tests performed in this work had best results with values between 10 and 25 stars per clustering. Applications of the method that adopted more complex algorithms might improve the results or completely eliminate the need of user input. 

Finally, the stars in each of the clusters output by the previous step are tested for clustering in positional space (details in the next subsection). If the data in a cluster are also clustered in positional space, it is retained for the next iteration. Otherwise, its stars are catalogued as field stars and are not considered in subsequent iterations. At this point, some stars classified as field stars may be members of stellar clusters. This will be addressed by the re-sampling introduced in the outer loop described in Sect.~\ref{sect:outiter}.

\begin{figure}
   \centering
   \includegraphics[width=0.6\columnwidth]{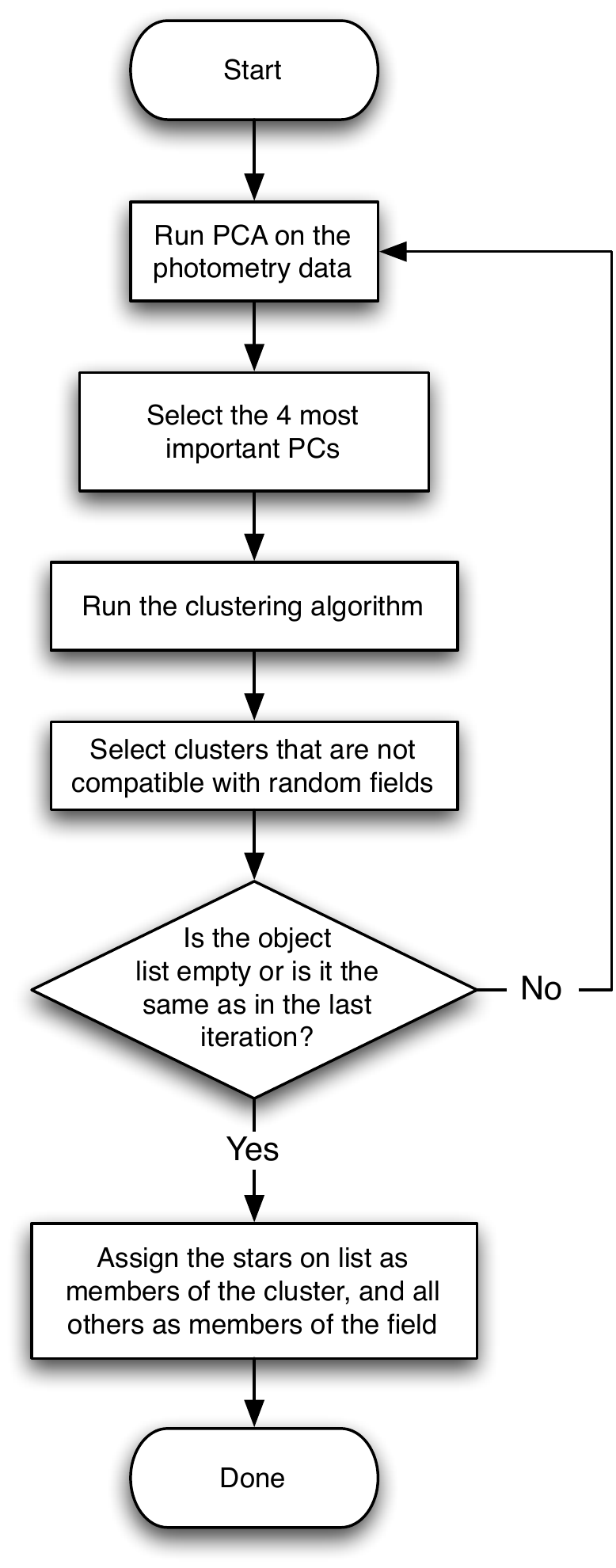}
      \caption{Flowchart of the UPMASK Kernel.}
         \label{Fig:upmask-kernel}
   \end{figure}

These steps are repeated until no more stars are added to the list of field stars. A graphical representation of the steps composing the UPMASK kernel can be found in Fig.~\ref{Fig:upmask-kernel}.

\subsubsection{Detection of spatial clustering}
To assess whether a certain distribution of stars is clustered in positional space, the UPMASK method employs kernel density estimations. Density parameters determined for real data and randomly generated uniform fields are compared. If the values obtained for the real data and for the random fields are compatible, the stars are classified as field members. The detailed process is as follows:

First, a two-dimensional normal kernel density estimation is performed on the positions of the stars within a $k$-means cluster, adopting the bandwidth recommended by \citet{MASS} for such kernels. The resulting density estimation function $\Phi(x,y)$ is then sampled, resulting in a $\mathbf{\Phi}$ set. From this set, a distance between its maximum value (corresponding to the maximum spatial density) and its mean value in units of standard deviations is computed: 

\begin{equation}
D({\mathbf{\Phi}})=\frac{max(\mathbf{\Phi}) - \langle\mathbf{\Phi}\rangle}{\sigma_\mathbf{\Phi}}, 
\end{equation}

where $\sigma_\mathbf{\Phi}$ is the standard deviation of the set $\mathbf{\Phi}$.

Then, using the same number of stars and region size as in the real data, a large set of realisations of random fields with uniform position distribution is generated. For each of these random realisations, a two-dimensional normal kernel density estimation $\Psi_i(x,y)$ is computed and also sampled, resulting in a $\mathbf{\Psi}_i$ set. Afterwards, a set $\mathbf{D}_\mathbf{\Psi}$ is constructed grouping the parameters $D_i(\mathbf{\Psi}_i)$ computed from each random realisation. Finally, the data is considered as not compatible with a random realisation of a uniform field if

\[
D(\mathbf{\Phi}) \ge \langle\mathbf{D}_\mathbf{\Psi}\rangle + T\times\sigma_{\mathbf{D}_\mathbf{\Psi}},
\]
\noindent where $\langle\mathbf{D}_\mathbf{\Psi}\rangle$ is the mean of the $\mathbf{D}_\mathbf{\Psi}$ set, $\sigma_{\mathbf{D}_\mathbf{\Psi}}$ is its standard deviation, and T is the threshold level above $\sigma$.

The number of random realisations performed is 2000. This value was empirically chosen. For larger numbers the computed parameters vary by less than $0.5\%$. As this is a time-consuming step, the current implementation of UPMASK uses a look-up table. In this strategy the parameters are computed only the first time they are needed in the analysis (per number of stars and region area), and further requests of these values are retrieved from the look-up table.

Finally, we adopted the thresholding level $T=1$, but this threshold level can be tuned for applications that require higher purity in the final members.

\subsubsection{Outer loop\label{sect:outiter}}

Although the iterative UPMASK kernel described in Sect.~\ref{sect:mainiter} already performs a membership assignment, it does not implement the whole UPMASK concept presented in Sect.~\ref{sect:concept}. The result is a binary classification, with stars listed as either belonging to the field  or to a stellar cluster. As mentioned in Sect.~\ref{sect:mainiter}, it is expected that the field catalogue will include cluster stars. Conversely, the catalogue  of star cluster members is expected to include field stars.

A first goal of the outer loop is to break this binary classification by assigning membership probabilities. A second goal is to take the effects of observational errors into account in the membership assignment process. It works as follows:

First, for each star in the data set, each original observable (in our case the magnitudes) is replaced by a random draw from a Gaussian distribution with the mean equal to the observable's value and $\sigma$ equal to the measurement error. For observables with manifestly non-Gaussian error distributions, other error models can be easily plugged into the method.

Then, the UPMASK kernel described in section \ref{sect:mainiter} is run on the newly generated data-set. If the kernel requires initialisation of variables, it is done before processing the data-set. In the current implementation, the $k$-means algorithm is initialised from random means. Upon termination, the kernel will have classified the stars in this set as cluster members or field stars.

\begin{figure}
   \centering
   \includegraphics[width=0.6\columnwidth]{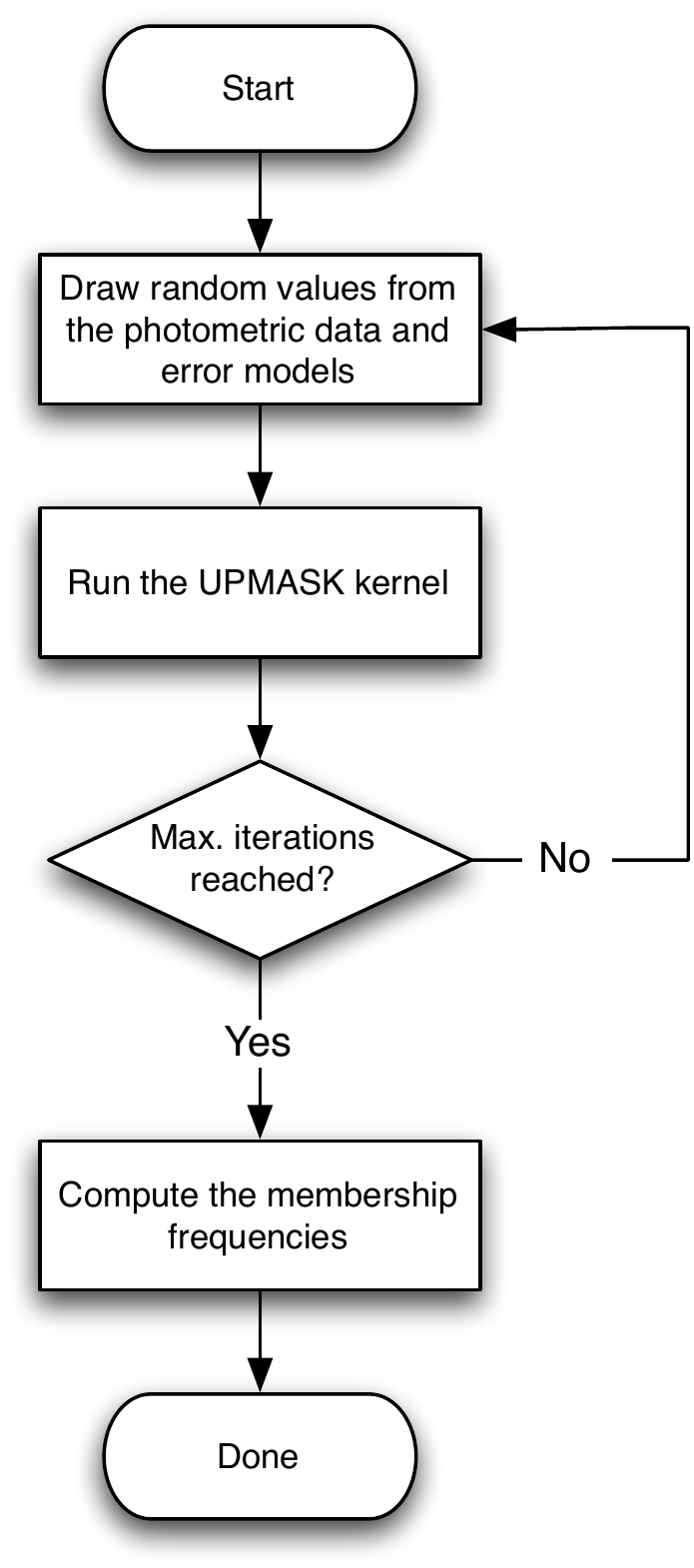}
      \caption{Flowchart of the Outer Loop.}
         \label{Fig:outeriteration}
   \end{figure}

The two steps described above are repeated until a user-defined maximum number of iterations is reached. Then, the results from all the runs are taken, and for each star the fraction of runs is computed for which it was assigned as cluster member. This fraction is an indicator of how certain the method is in assigning a star as a member of a stellar cluster. In a frequentist sense it is identified with a \emph{membership probability}. Since outer iterations run the UPMASK kernel many times, the $k$-means clustering algorithm is also randomly initialised many times. This reduces the dependency of the final results on the choice of the clustering algorithm's initialisation conditions. A flowchart representing the outer iterations of UPMASK is shown in Fig.~\ref{Fig:outeriteration}.

\section{Application to simulated data\label{sect:simdatamain}}

UPMASK was validated by running the method on synthetic data sets. In this section we describe the characteristics of these data sets and present an assessment of the method's classification performance, caveats, and limitations.

\subsection{Simulated data\label{sect:simdata} }    

Synthetic catalogues of UBVRI photometry of cluster fields were generated to simulate observations of clusters spanning a range of initial mass, age, and distance  (see Table~\ref{table:simparam}) along a line-of-sight in the Milky Way.
The field of view was set at $12{\arcmin}\times 12{\arcmin}$, the simulated CCD frames were defined to be $2048\times 2048$ pixels and the limiting magnitude at $U,B,V,R,I\sim 19.5$, $20.5$, $21.5$, $22$, $22$ mag to match typical open cluster photometric tables found in the literature.

The underlying Galactic field was created with the TRILEGAL simulator\footnote{\url{http://stev.oapd.inaf.it/cgi-bin/trilegal}} \citepads{2005A&A...436..895G} for the direction $(l, b) = (120^{\circ},0^{\circ})$ with an applied extinction of 1mag/kpc. The photometric system of the simulations was set to UBVRIJHK \citepads{1990PASP..102.1181B, 2006AJ....131.1184M}. The other TRILEGAL parameters (except for the field of view and magnitude limit) were left at their defaults. Since TRILEGAL does not output individual stellar coordinates, these were randomly generated with a uniform distribution across the simulated CCD frame.

The open cluster photometry and CCD positions were created using the MASSCLEAN\footnote{\url{http://www.physics.uc.edu/~bogdan/massclean.html}} package \citepads{2009AJ....138.1724P}. For each simulated cluster distance, the adopted visual absorption, $A_V$, was the value determined in TRILEGAL as corresponding to the same distance. This was done to ensure that the location of the cluster sequence was consistent with the field in the colour space.

MASSCLEAN generates error-free photometry, and thus, magnitude dependent noise was added to the simulated photometry. An exponential error model was used, with errors starting at 0.01 mag for the brightest sources, reaching about 0.06 mag at 1.5 mag before the limiting magnitude, and then steeply rising to asymptotically reach infinity at the limiting magnitude.

\begin{table}
\caption{Parameter range covered by the simulations.}            
\label{table:simparam}     
\centering                 
\begin{tabular}{cc}       
\hline\hline                
Parameter & Values \\  
\hline                        
   Age [$10^n$yr] &  7.1, 7.7, 8.0, 8.3, 8.7, 8.8, 9.0, 9.2, 9.5 \\     
   Initial mass [$10^3$M$_{\sun}$] & 0.5, 1, 2.5, 5, 10 \\
   Distance [kpc] & 0.5, 1.0, 1.5, 2.0, 2.5, 3.0, 3.5, 4.0  \\
\hline                                  
\end{tabular}
\end{table}

\subsection{Results on simulated data}

\subsubsection{Global analysis \label{sect:ga}} 

One central aspect in the analysis of the simulations is the purity level of the sample of identified cluster members. Since UPMASK outputs individual membership probabilities for each star, the purity analysis must be performed on the subset of the data whose membership probability is above a certain cut-off. 

As a measure of purity, we define the true positive rate (TPR) as the ratio between the number of real cluster members in the high-probability subset and its total size. A method capable of creating the high-probability subset that consists purely of cluster members would attain a TPR of 1 (or $100\%$) independently of the degree of contamination in the global data-set.

\begin{figure}
   \centering
   \includegraphics[width=0.8\columnwidth]{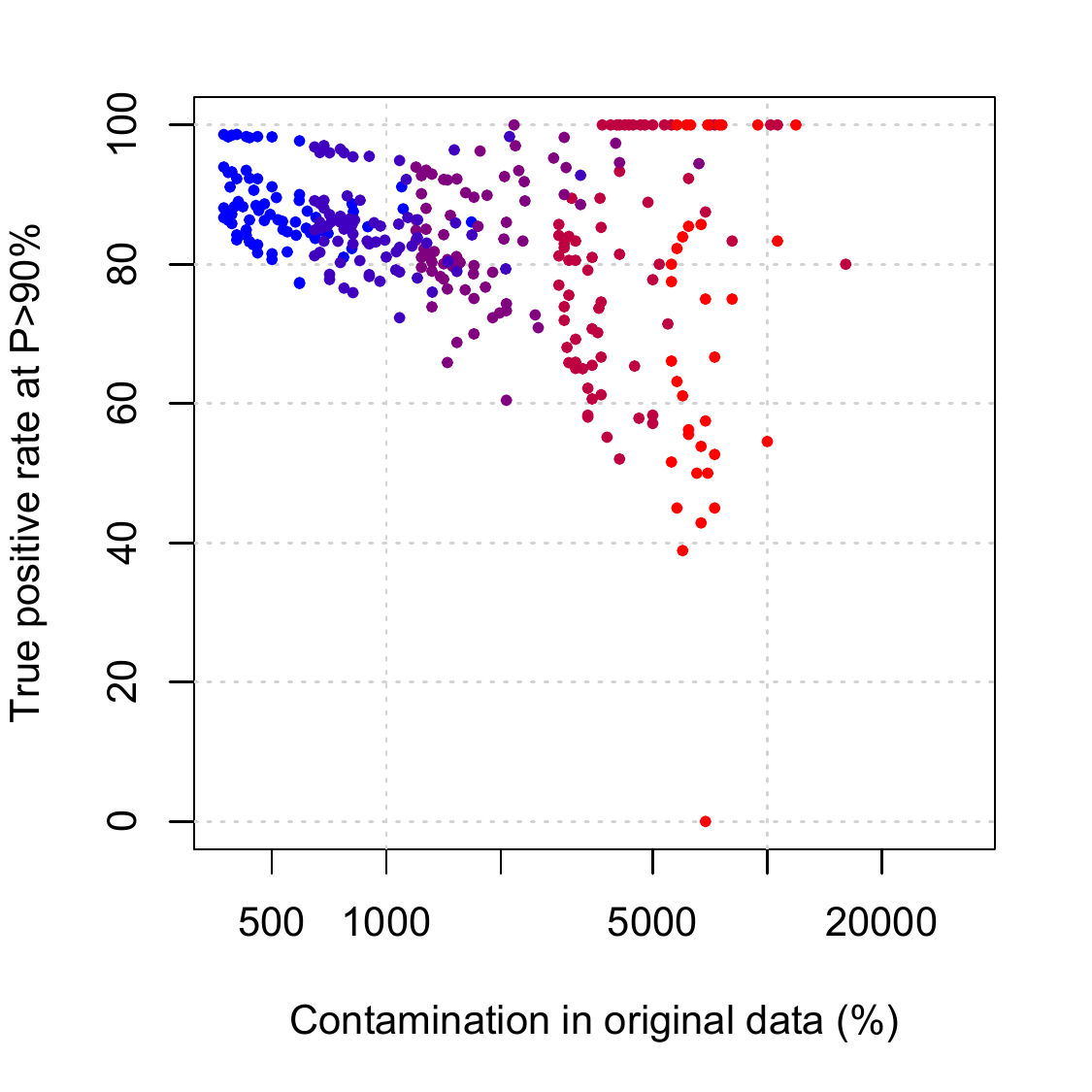}
      \caption{The true positive rate of the subsample with membership probabilities higher than 90$\%$ (TPR$_{90}$) against the field contamination in the original data-set. Each point corresponds to one of the simulations in the grid defined in Tab.~\ref{table:simparam}. The point colours indicate the cluster masses: the bluer, the more massive, the redder, the less massive.}
         \label{Fig:TPR90}
   \end{figure}

In this work, we chose the cut-off level at a membership probability of $90\%$ and represent the corresponding TPR as TPR$_{90}$. Fig.~\ref{Fig:TPR90} depicts the TPR$_{90}$ for all the simulated clusters described in Tab.~\ref{table:simparam} against the contamination rate of the corresponding original data-sets. The contamination rate is defined as the ratio between the total number of stars in a field and the number of simulated cluster stars.

As seen in Fig.~\ref{Fig:TPR90}, even under heavy contamination the resulting TPR$_{90}$ is in general fairly high. As expected, the TPR is depends on the contamination rate of the data-set. For a fixed Galactic direction and limiting magnitude, the contamination will increase with the cluster's age, distance, and reddening and decrease with increasing initial mass. Naturally, one does not expect to obtain a simple relation between the TPR and the contamination rate.

These results, obtained over a grid of 360 fields, spanning a wide range of conditions, show that more than $63\%$ of the simulated clusters have a TPR$_{90}$ $\ge 80\%$. This is a remarkable performance for an unsupervised method that is only based on very weak assumptions on the objects under analysis. 

Fig.~\ref{Fig:TPR90} also shows that the TPR covers a broad range of values depending on the contamination rate. Given that the Galactic background is the same for all simulations, the contamination rate depends on the initial mass of the cluster. In this figure, the masses follow the contamination, with the lowest masses in the rightmost part of the plot and the most massive clusters in the left part. We note that even for the lowest initial masses (500 M$_{\sun}$ and 1000 M$_{\sun}$) the TPR$_{90}$ reaches high values.

By considering only membership probabilities above 90\%, it is expected that a number of cluster stars will not have been included in our list of members. In other words, the list has a high purity, but is not complete. Depending on the research goals, one can increase completeness by relaxing the cut-off to lower membership probability values such as $50\%$. The trade-off will be a lower purity, or a lower TPR.

\begin{figure}
   \centering
   \includegraphics[width=1\columnwidth]{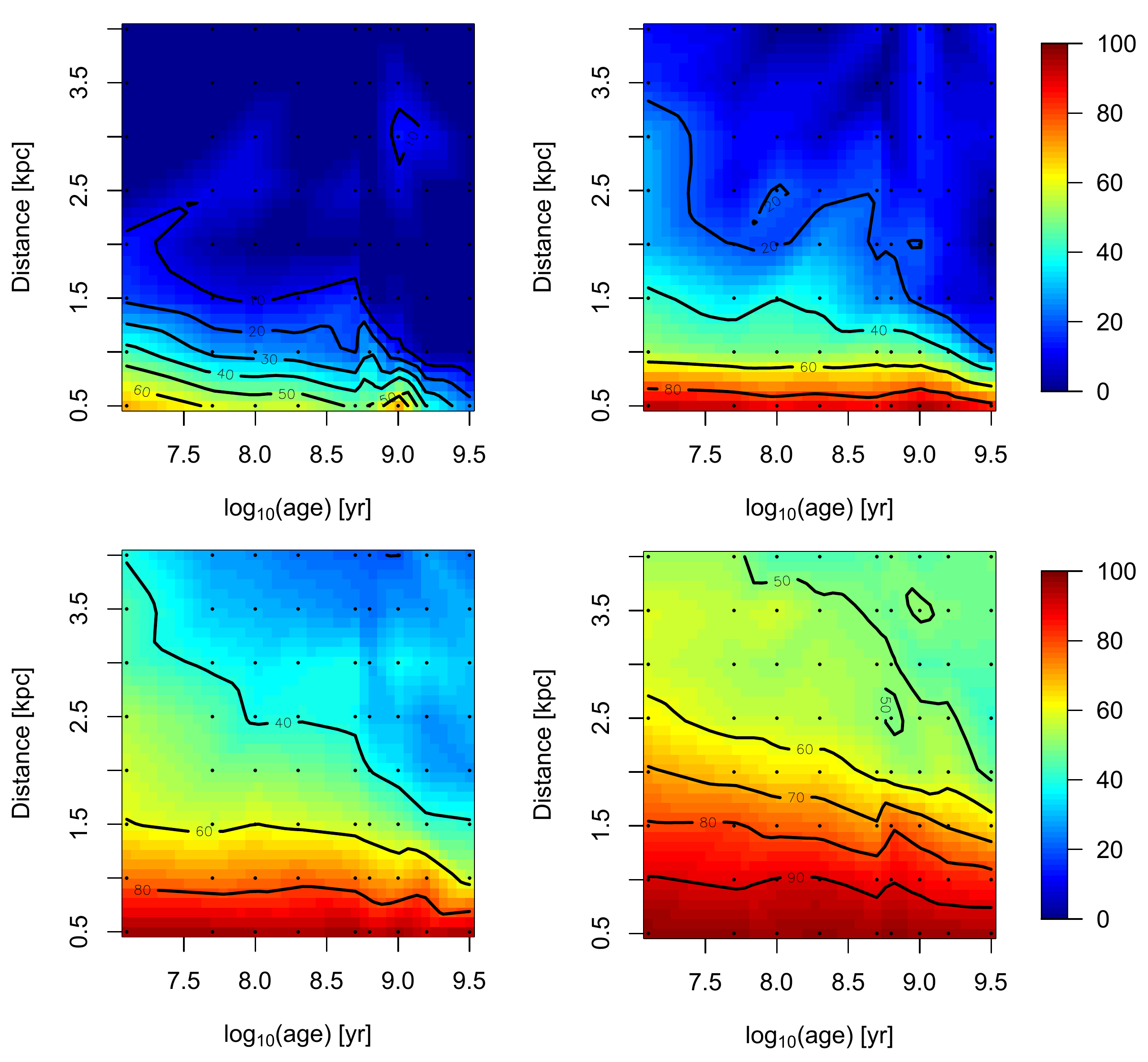}
      \caption{Dependency of member recovery rate (colour scale) for membership probability above $90\%$ on cluster age and distance for four initial masses. Upper left: 500 M$_{\sun}$.  Upper right: 1000 M$_{\sun}$. Lower left: 2500 M$_{\sun}$. Lower right: 5000 M$_{\sun}$. The $10^4$ M$_{\sun}$ simulations have an MRR$_{90}\gg80$ across most of the considered age-distance plane and are not represented in the figure. The dots mark the regions of the plane sampled by the simulations.}
         \label{Fig:MRR90}
   \end{figure}

To measure the completeness of the sample of recovered members, we defined the member recovery rate (MRR) as the ratio between the number of members in the sample and the total number of members. The MRR will be 1 (or $100\%$) when all the members are classified as such and 0 when no members are recovered. Considering a cut-off at $90\%$ membership probability, we obtained the MRR$_{90}$ represented in Fig.~\ref{Fig:MRR90} for clusters over the simulated grid.

As expected, the MRR is higher for high-mass, and thus richer, clusters (Fig.~\ref{Fig:MRR90} lower right) than for lower mass objects (Fig. \ref{Fig:MRR90} upper left) in the entire age-distance plane. However, Fig.~\ref{Fig:MRR90} also shows that even for the poorer clusters  ($500$M$_{\sun}$ initial mass), UPMASK has recovered with high purity a fraction of the members significant enough to enable subsequent  analyses such as isochrone fitting, up to distances of more than 1 kpc for most ages (photometric sequences of more distant and older cluster are more embedded in the field distribution and are therefore intrinsically harder to separate). For the high-mass clusters, UPMASK can still yield good recovery rates (MRR$_{90} \sim 50\%$) at distances of 3.5 kpc.

\subsubsection{Case studies\label{sect:casestudies}}

In this section we examine more closely the simulation results. Given the large number of simulated fields (360), we have chosen two representative cases, that are however not the most favorable ones.

 \begin{figure}
   \centering
  \begin{center}
  \subfloat[]{
    \includegraphics[width=0.5\linewidth, trim=0cm 0.5cm 0cm 0.5cm, clip=true]{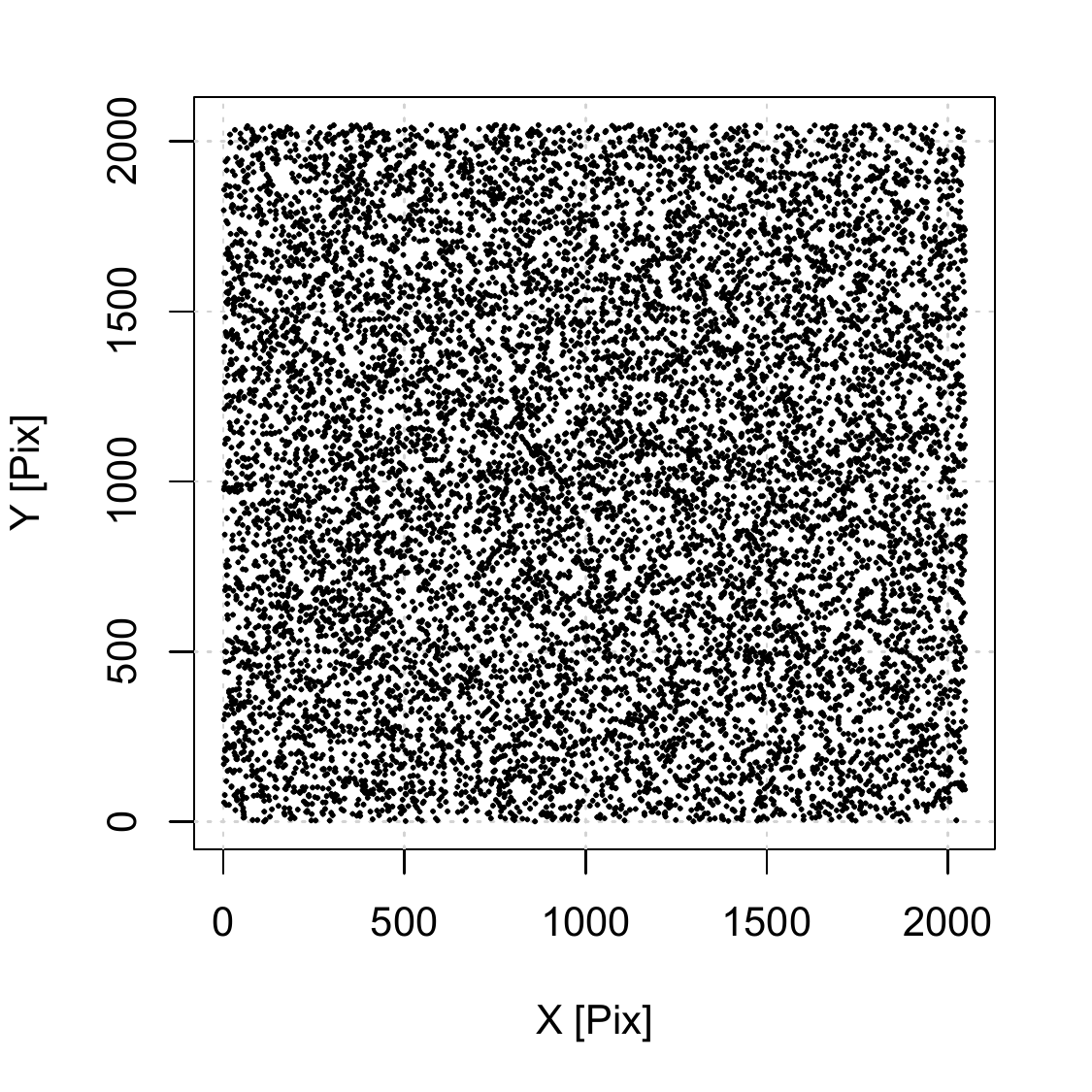}
  }
  \subfloat[]{
    \includegraphics[width=0.5 \linewidth, trim=0cm 0.5cm 0cm 0.5cm, clip=true]{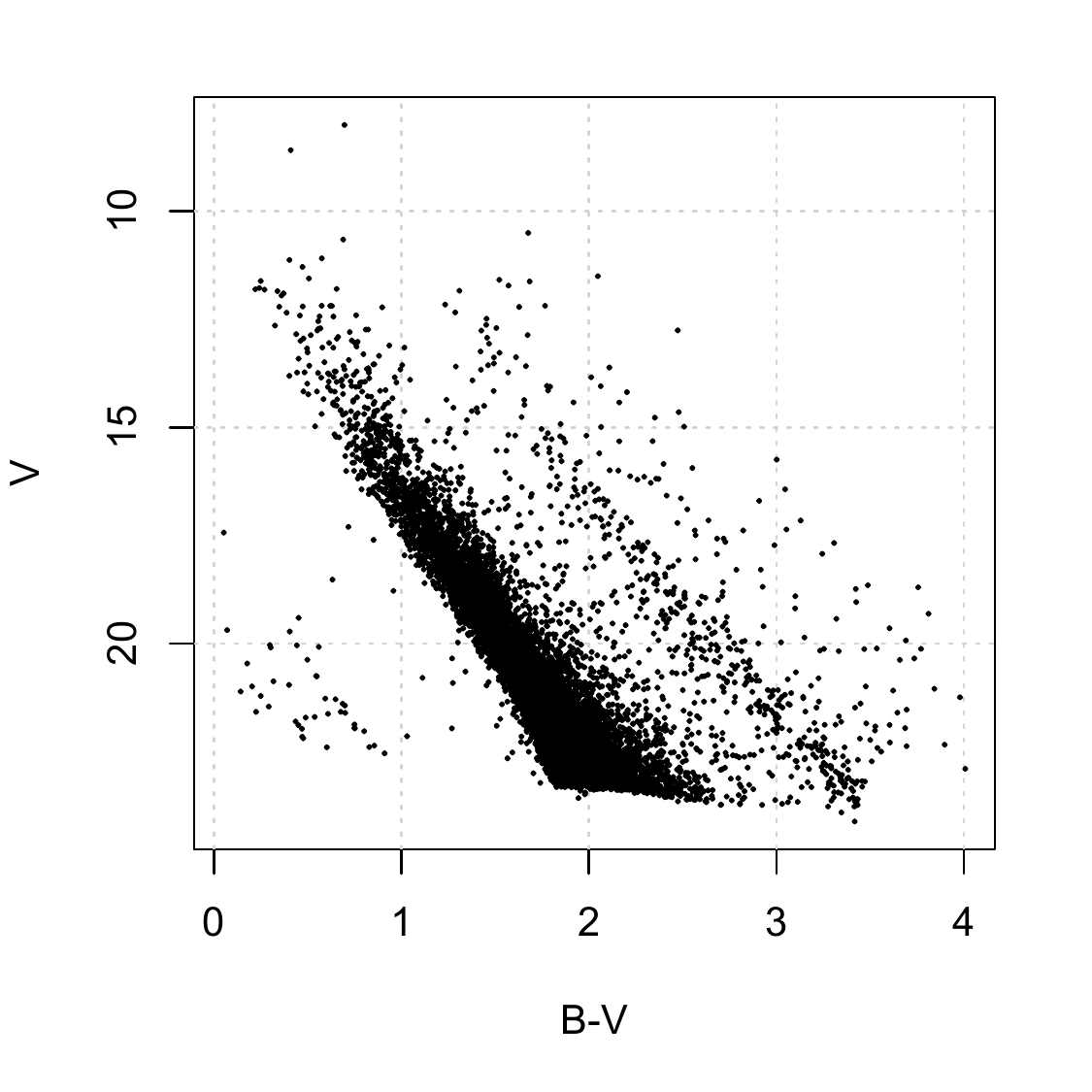}
  }\\
  \subfloat[]{
    \includegraphics[width=0.5 \linewidth, trim=0cm 0.5cm 0cm 0.5cm, clip=true]{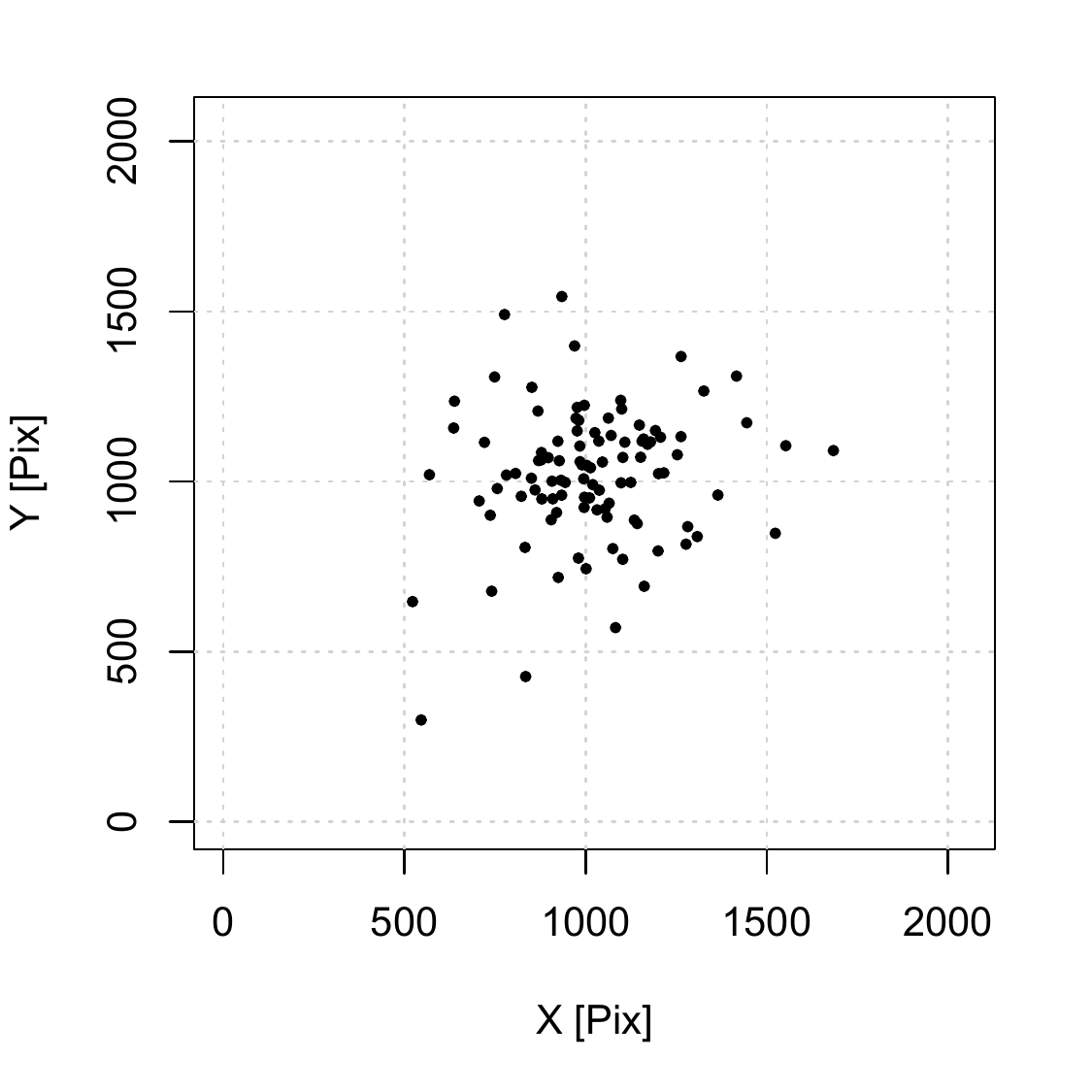}
  }
  \subfloat[]{
    \includegraphics[width=0.5 \linewidth, trim=0cm 0.5cm 0cm 0.5cm, clip=true]{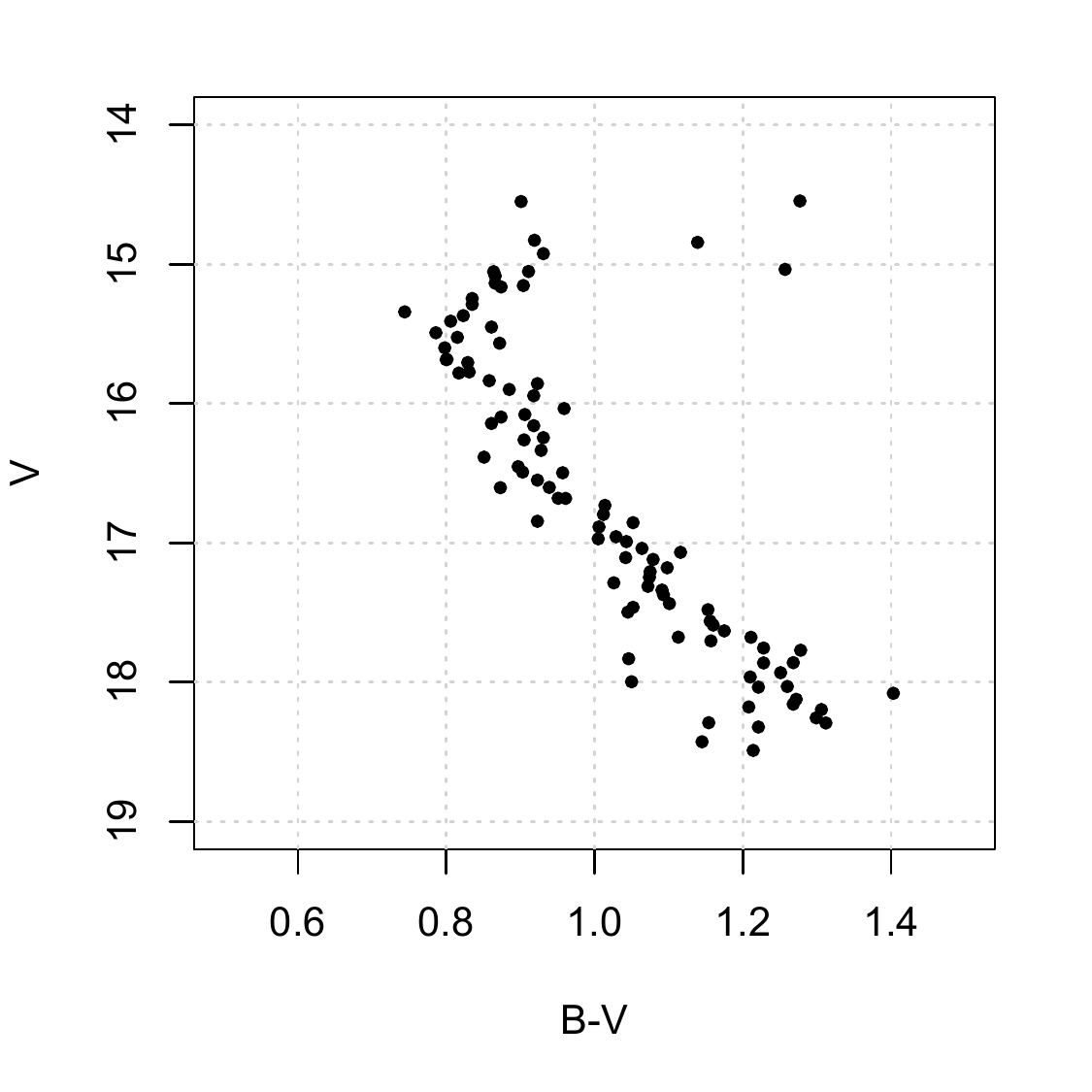}
  }\\
  \subfloat[]{
    \includegraphics[width=0.5 \linewidth, trim=0cm 0.5cm 0cm 0.5cm, clip=true]{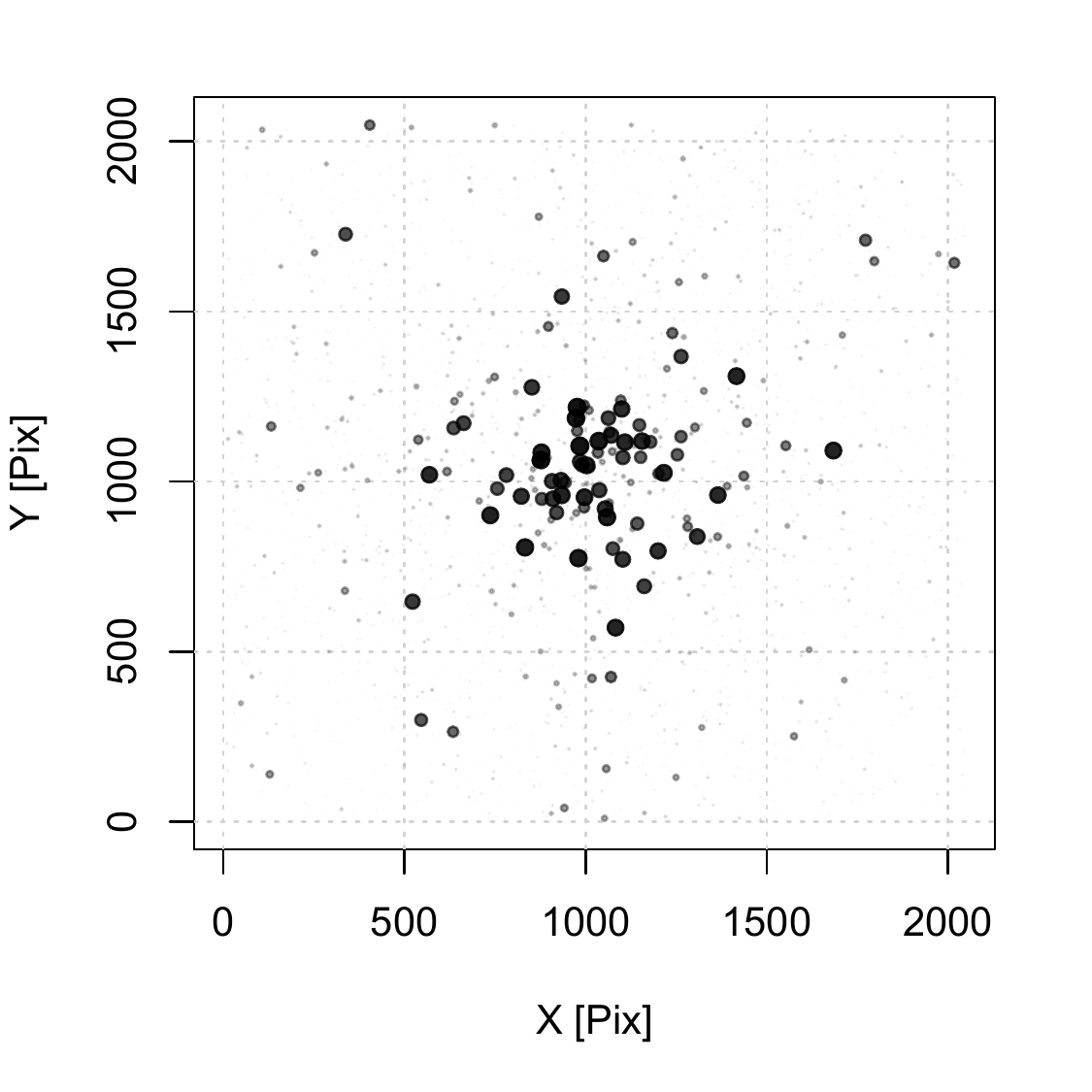}
  }
  \subfloat[]{
    \includegraphics[width=0.5 \linewidth, trim=0cm 0.5cm 0cm 0.5cm, clip=true]{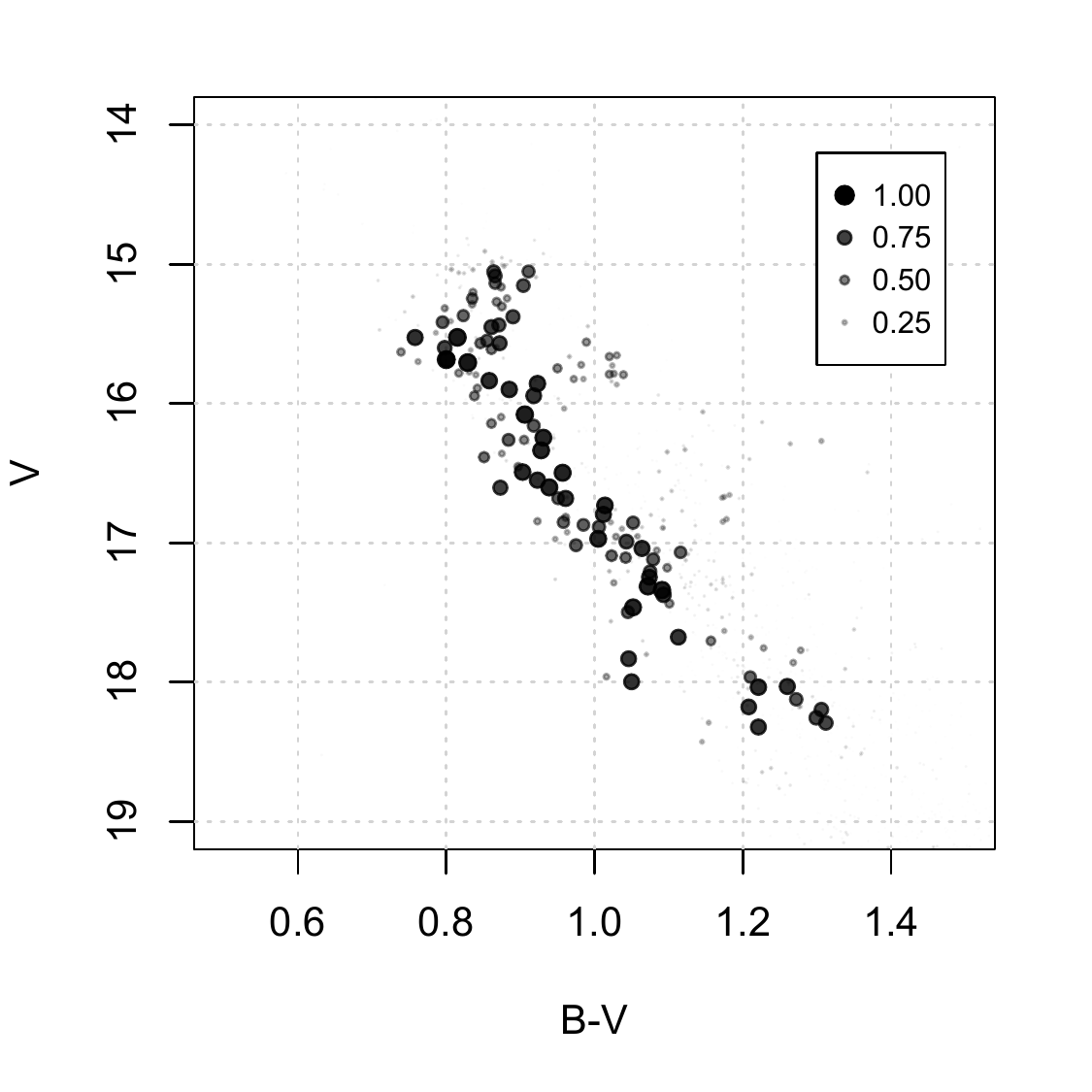}
 }
  \end{center}
      \caption{Results of UPMASK on a simulated cluster with $10^3$ M$_{\sun}$ and $10^{9.5}$ yr at 1.5kpc. In (a) and (b) we plot the distribution of field+cluster stars in positional space and in V vs. (B-V) space. In (c) and (d) we show the distributions of cluster members only.  In (e) and (f) we plot the distributions of the cluster members identified by UPMASK. Point size and transparency encode membership probability.}
         \label{Fig:ExampleSimulatedClusters1}
   \end{figure}

 \begin{figure}
   \centering
  \begin{center}
  \subfloat[]{
    \includegraphics[width=0.5\linewidth, trim=0cm 0.5cm 0cm 0.5cm, clip=true]{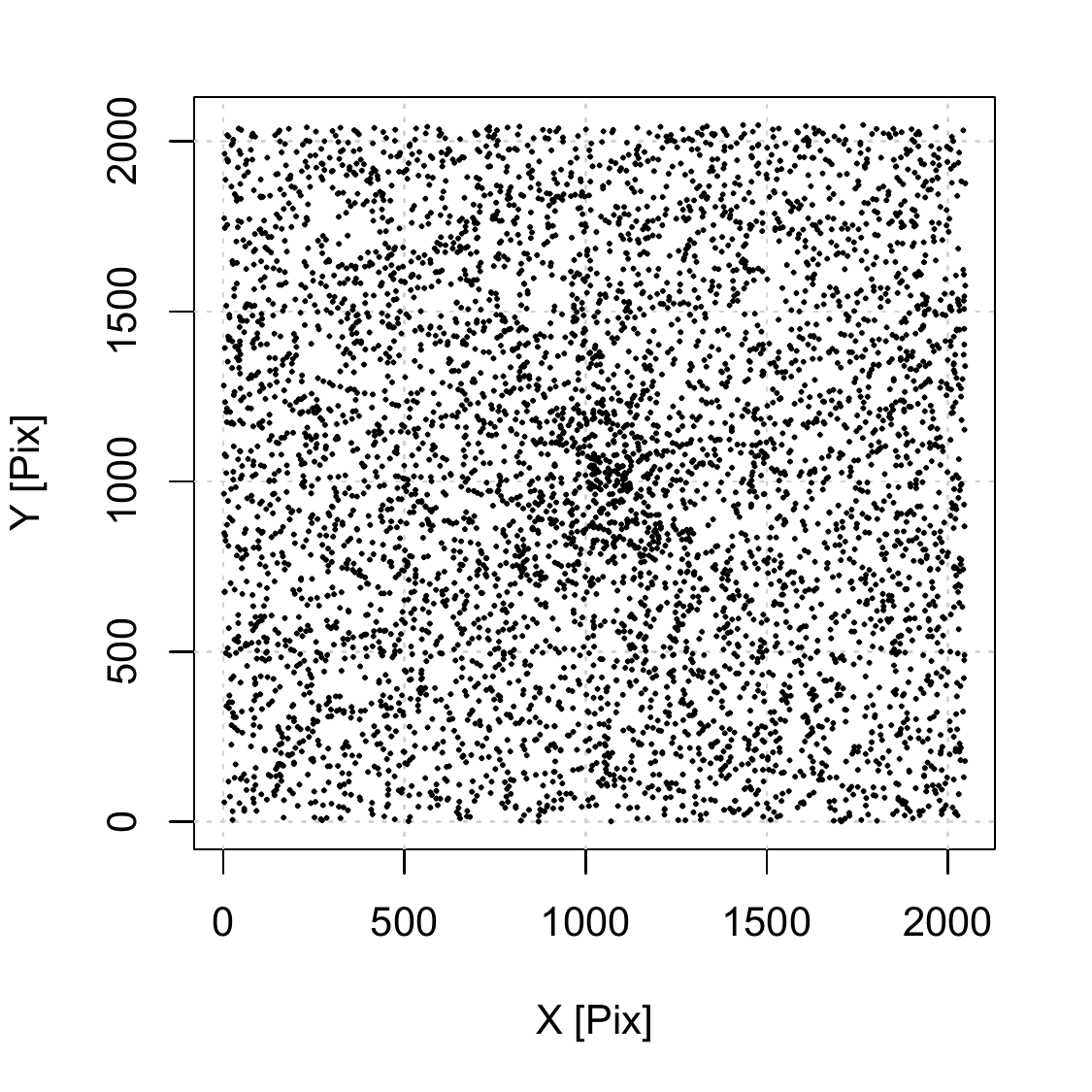}
  }
  \subfloat[]{
    \includegraphics[width=0.5 \linewidth, trim=0cm 0.5cm 0cm 0.5cm, clip=true]{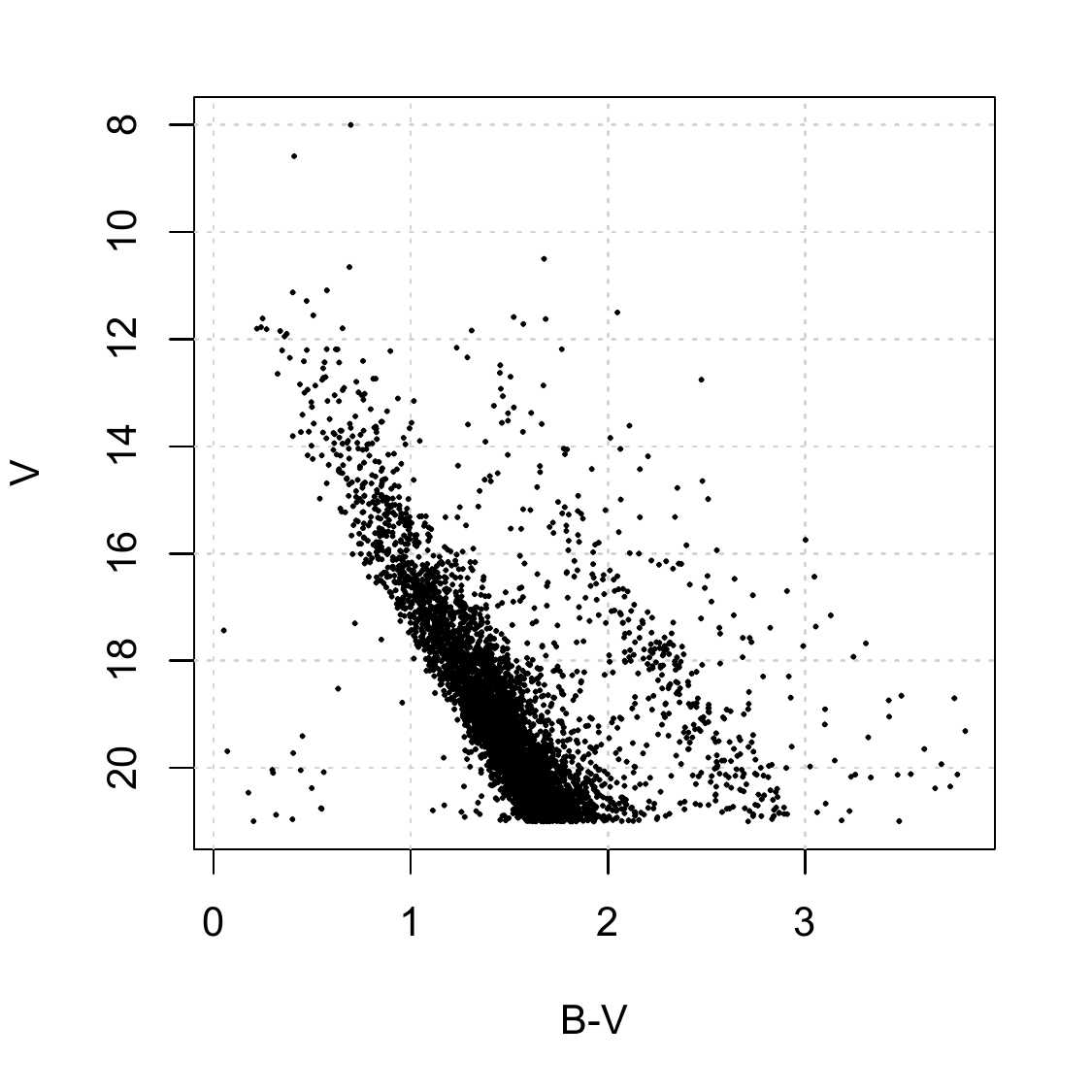}
  }\\
  \subfloat[]{
    \includegraphics[width=0.5 \linewidth, trim=0cm 0.5cm 0cm 0.5cm, clip=true]{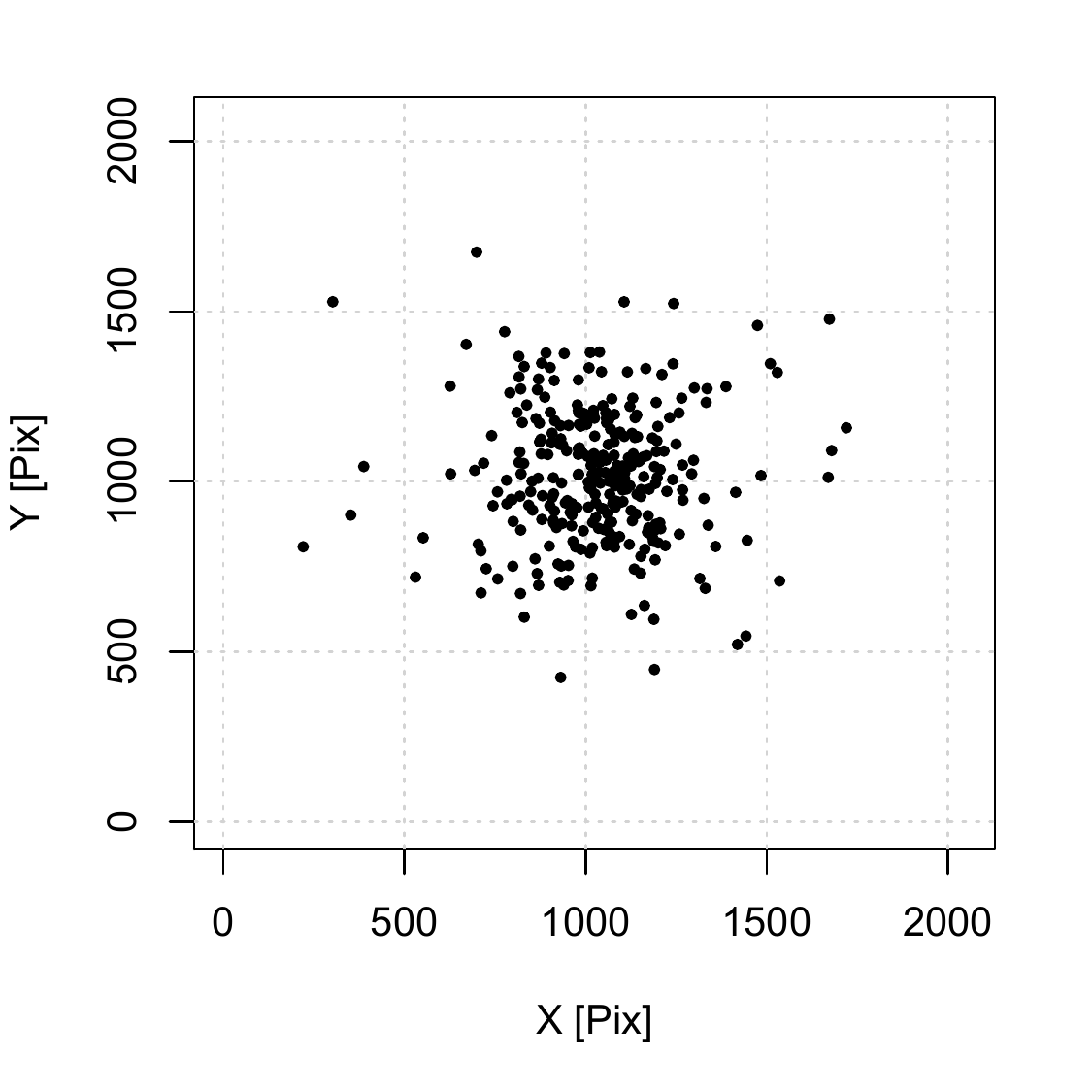}
  }
  \subfloat[]{
    \includegraphics[width=0.5 \linewidth, trim=0cm 0.5cm 0cm 0.5cm, clip=true]{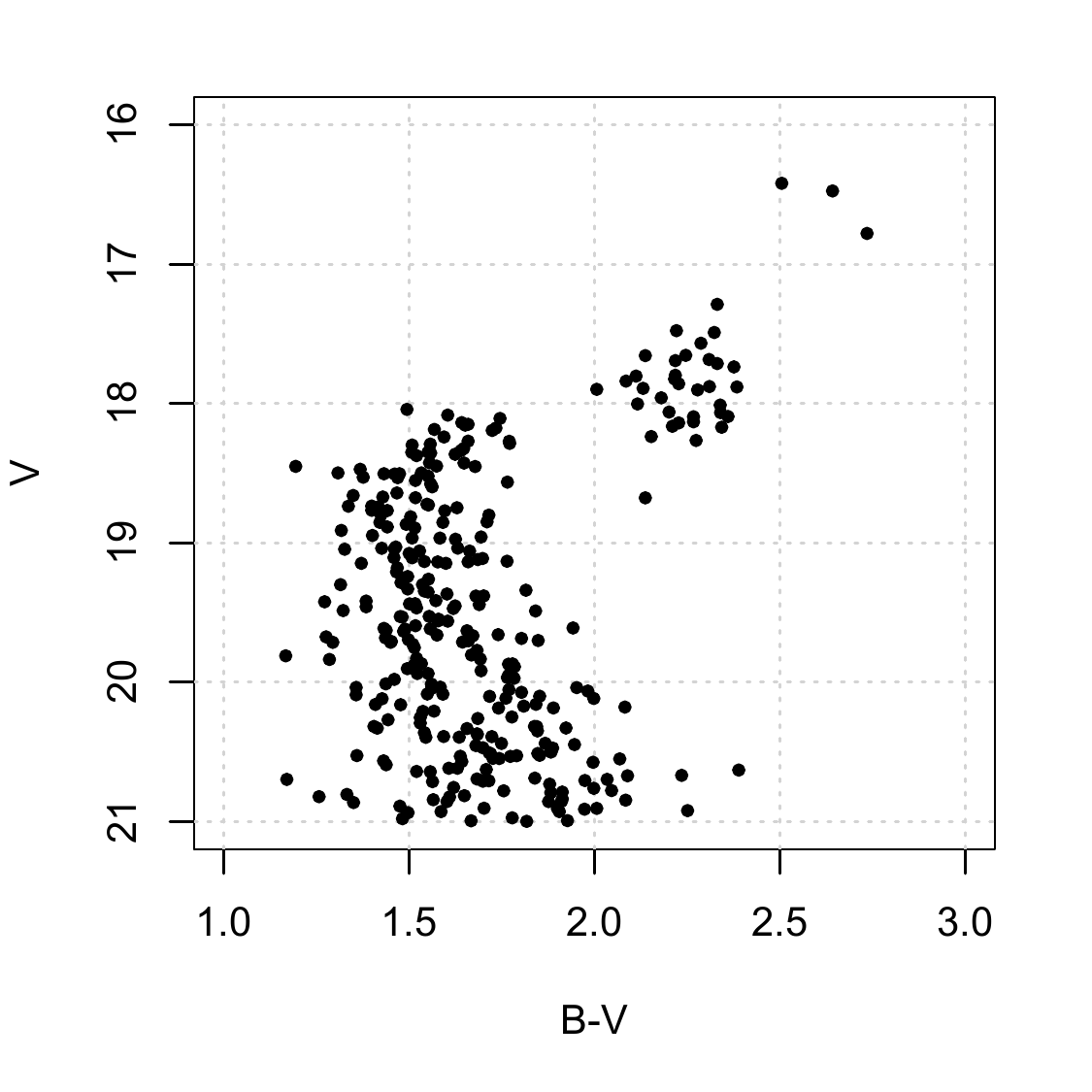}
  }\\
  \subfloat[]{
    \includegraphics[width=0.5 \linewidth, trim=0cm 0.5cm 0cm 0.5cm, clip=true]{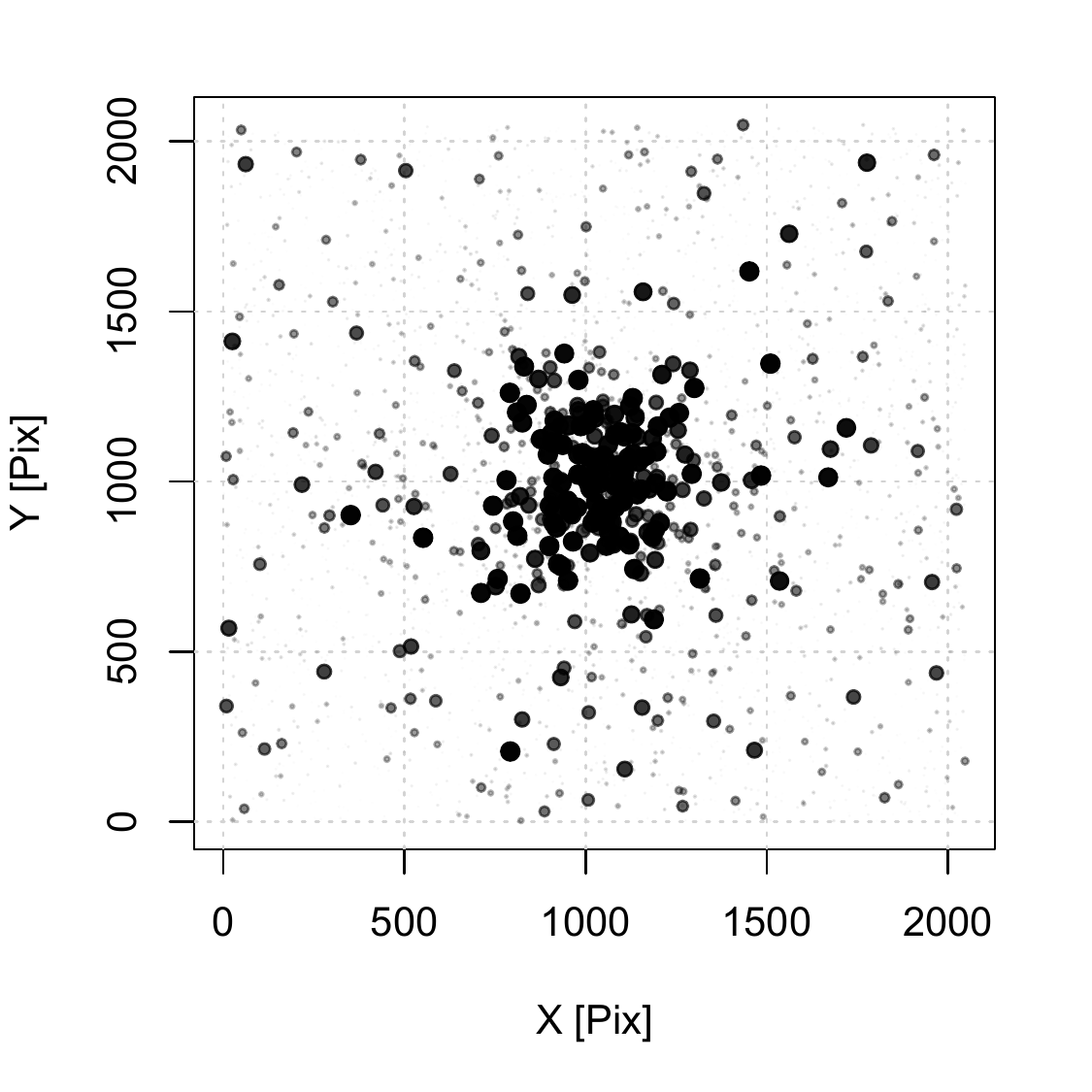}
  }
  \subfloat[]{
    \includegraphics[width=0.5 \linewidth, trim=0cm 0.5cm 0cm 0.5cm, clip=true]{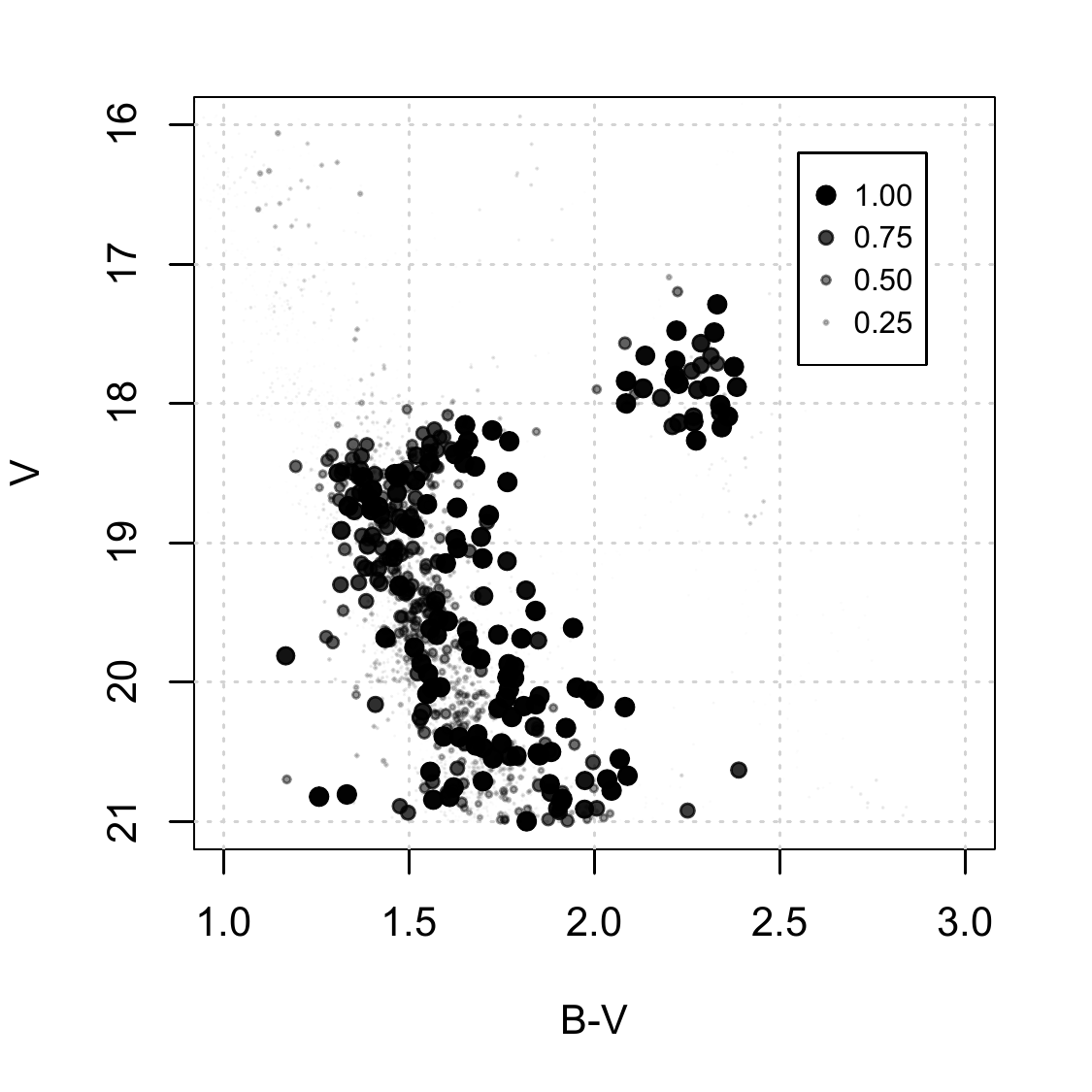}
 }
  \end{center}
      \caption{As Fig. \ref{Fig:ExampleSimulatedClusters1}, but for a simulated cluster with $5\times10^3$ M$_{\sun}$ and $1$ Gyr at $4.0$ kpc.}
         \label{Fig:ExampleSimulatedClusters2}
   \end{figure}
      

The first case is a cluster with an initial mass of $10^3$ M$_{\sun}$, has an age of $10^{9.5}$ yr ($\sim3.16$ Gyr) and is located at 1.5 kpc. Fig.~\ref{Fig:ExampleSimulatedClusters1} illustrates the results. Panels (a) and (b) show the distribution of the initial cluster+field data in positional and V vs. (B-V) spaces. In these plots, one cannot even guess at the existence of a cluster. This particular example is heavily contaminated by field stars: while there are 12249 field stars, there are only 101 cluster stars, which are represented in panels (c) and (d). The resulting membership assignment by UPMASK is represented in panels (e) and (f).

Although from the global analysis presented in section \ref{sect:ga} one could expect this cluster to present poorer results, its members were well recovered by the UPMASK method, mostly with high ($\gg 0.5$) membership probabilities. 

For some regions in the CMD minor voids are opened (ex. between B-V = 1.1 and 1.2 mag). Gaps in the CMD of star clusters can be real \citepads[e.g.][]{2000ApJ...544L..65D}, but this is not the case of the present simulation. Here, the gaps are a consequence of not requiring continuity between the different $k$-means clusters classified as members/non-members. Nonetheless, they can be filled by a non-parametric post-processing phase that would determine whether stars lying between high-probability clusters could belong to the star cluster. An alternative, but parametric, approach could consist of first fitting an isochrone to the identified members and then using an isochrone proximity criterion for selecting stars in the gapped regions.

In any case, the comparison of the simulated cluster with the recovered members shown in Fig.~\ref{Fig:ExampleSimulatedClusters1} shows that the general features of the distribution of stars in the original cluster were captured.


The second case represents a more massive cluster with {\bf $5\times10^{3}$} M$_{\sun}$ and an age of $1$ Gyr that is located at $4.0$ kpc. The UPMASK results and the initial data are represented in Fig. \ref{Fig:ExampleSimulatedClusters2}.

Because this cluster is more massive than that from the first example, it shows up as a region with higher density in the position space (Fig. \ref{Fig:ExampleSimulatedClusters2} a). It is, nonetheless, hardly distinguishable in the CMD (Fig. \ref{Fig:ExampleSimulatedClusters2} b), except for a hint of a concentration of red-clump stars.

Even in a highly contaminated data-set, the method was able to separate the field and cluster populations. It was also able to recover the cluster's red-clump stars, as can be seen by comparing the simulated members and the stars assigned as members (Fig.~\ref{Fig:ExampleSimulatedClusters2} d and f). Although, as discussed in the previous example, gaps may appear in the recovered sample of cluster members, one of the strengths of UPMASK is precisely allowing for such gaps. In the present case, it has allowed us to identify  the main-sequence and the red-clump stars, which are disconnected. Finally, the overall morphological structure of the cluster in positional space was also well recovered (Figs. \ref{Fig:ExampleSimulatedClusters2} c and e).

 \begin{figure}
   \centering
  \begin{center}
  \subfloat[]{
    \includegraphics[width=0.5\linewidth, trim=0cm 0.5cm 0cm 0.5cm, clip=true]{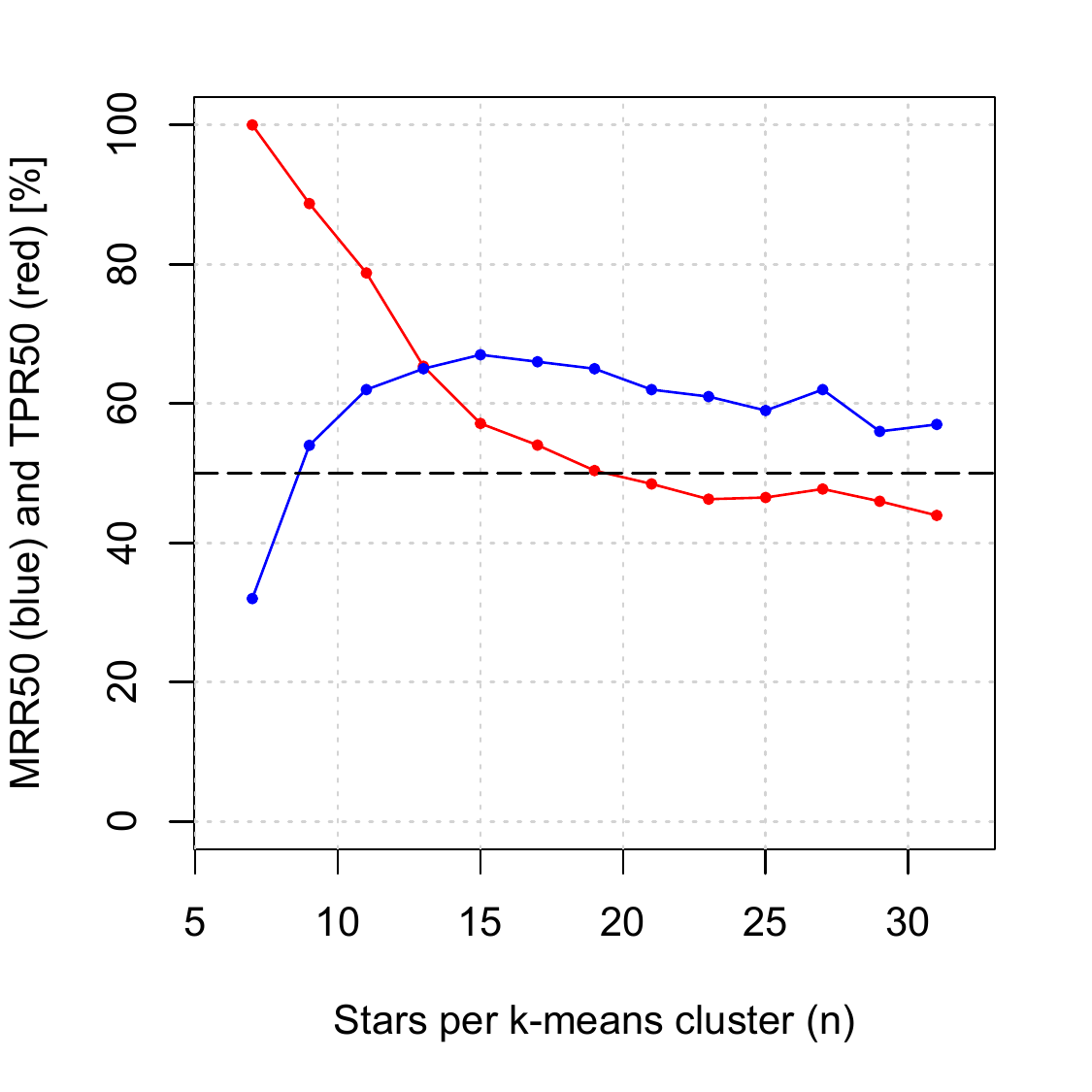}
  }
  \subfloat[]{
    \includegraphics[width=0.5 \linewidth, trim=0cm 0.5cm 0cm 0.5cm, clip=true]{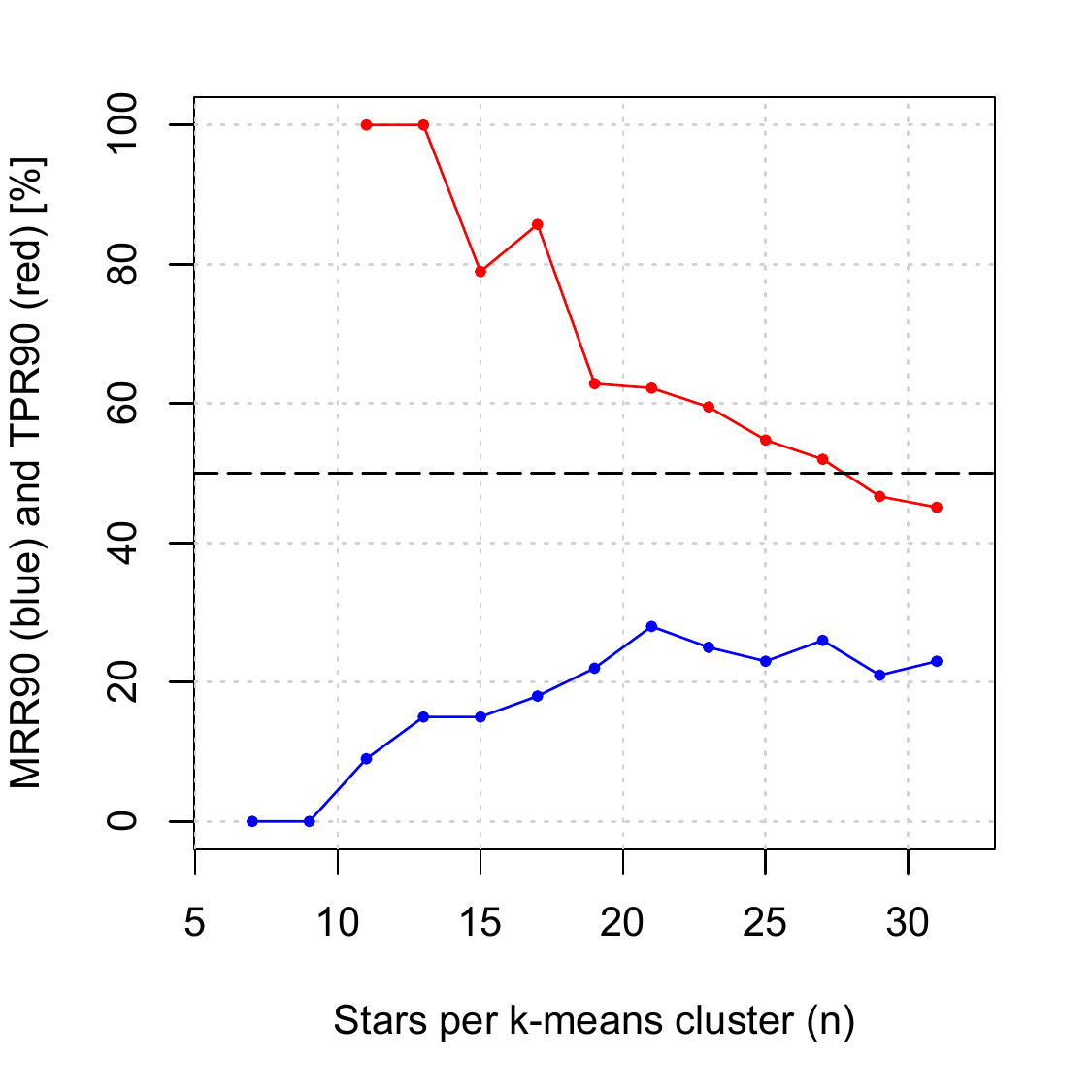}
  }\\
  \subfloat[]{
      \includegraphics[width=0.5\linewidth, trim=0cm 0.5cm 0cm 0.5cm, clip=true]{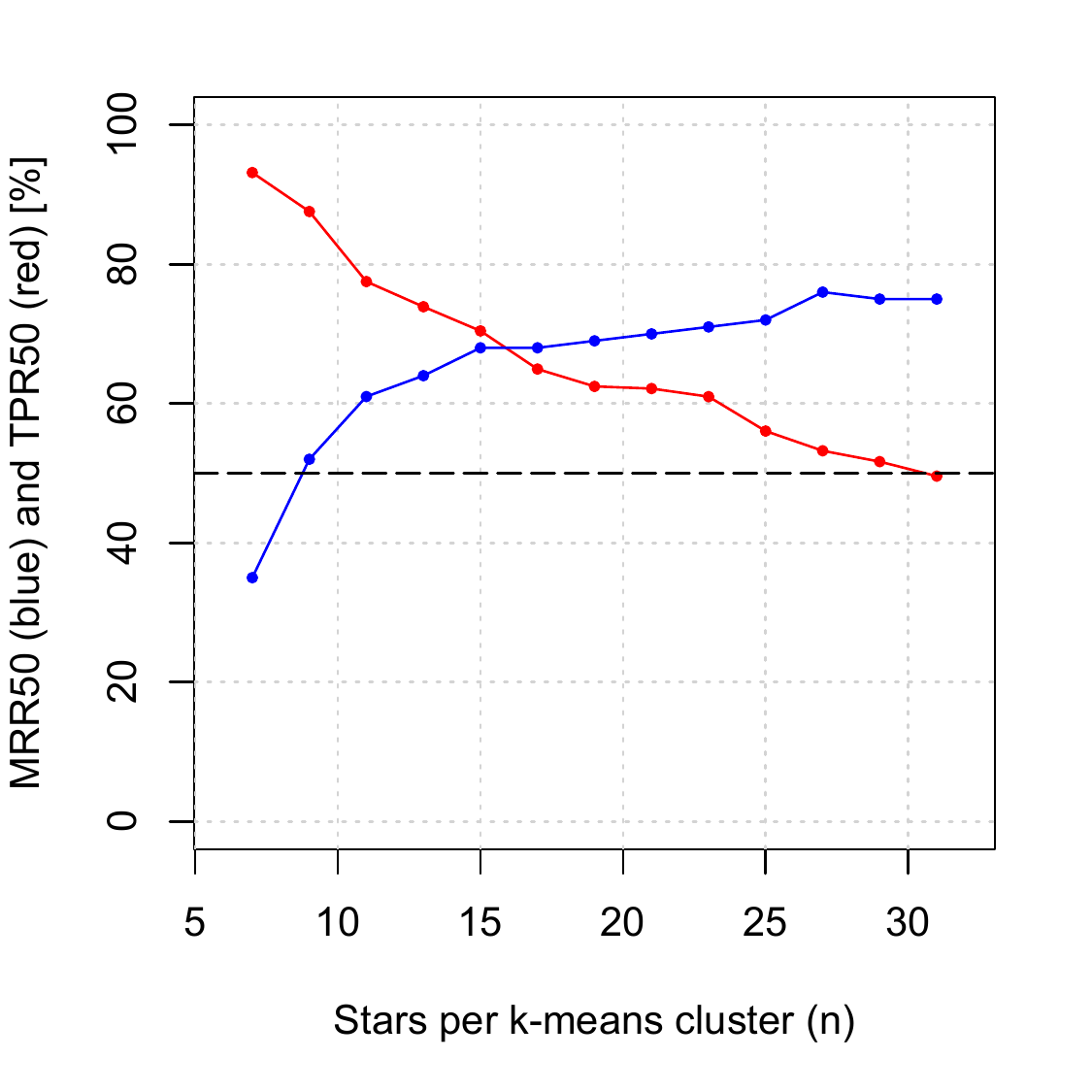}
  }
  \subfloat[]{
    \includegraphics[width=0.5 \linewidth, trim=0cm 0.5cm 0cm 0.5cm, clip=true]{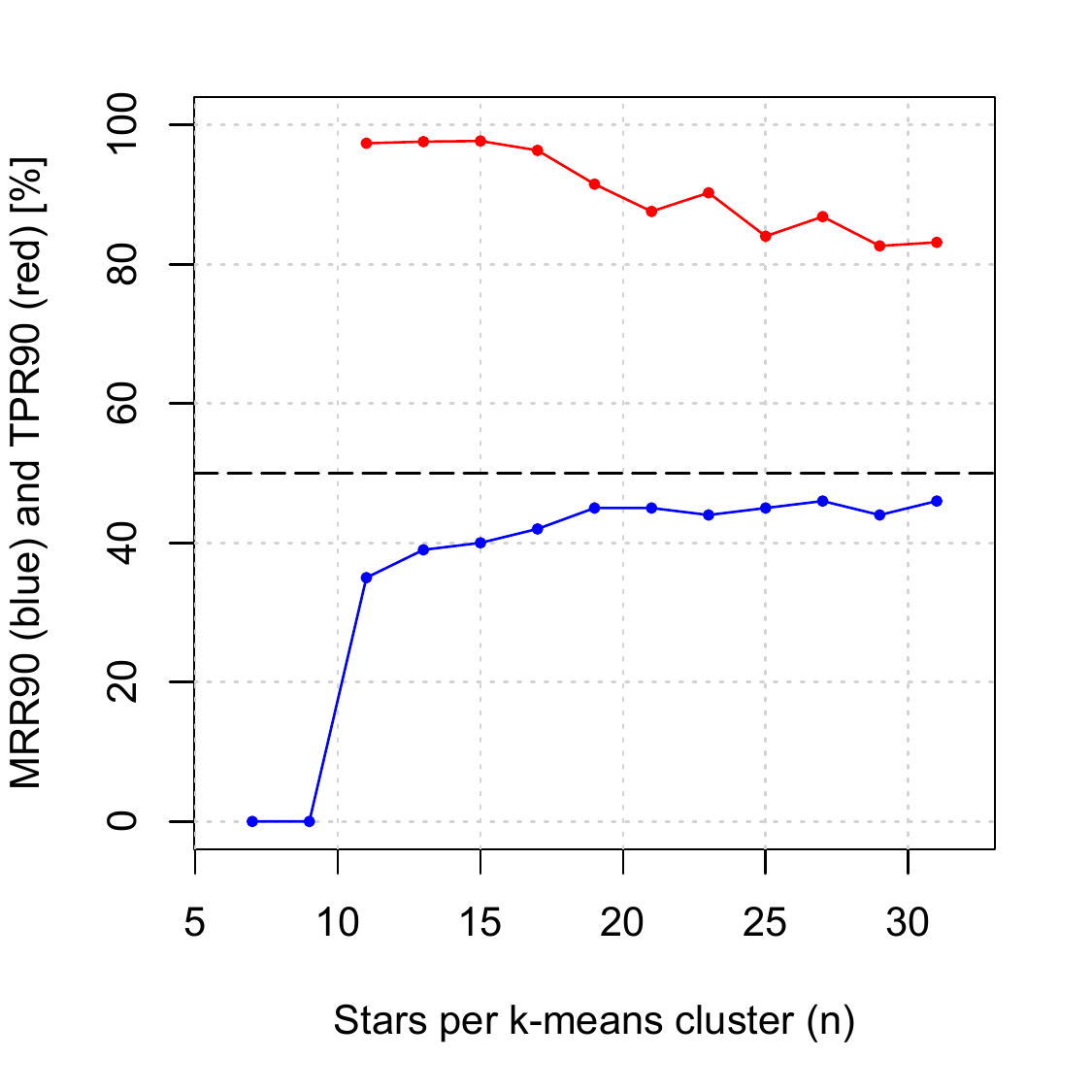}
  }
  \end{center}
   \caption{Dependency of the completeness and purity indicators at 50\% and 90\% membership probability levels, with the adopted number of stars per $k$-means cluster. In the top panels we show the cluster from Fig. \ref{Fig:ExampleSimulatedClusters1}. In the bottom panels we show the cluster from Fig. \ref{Fig:ExampleSimulatedClusters2}}
   \label{Fig:ExampleSimulatedClusterskDep}
 \end{figure}

\paragraph{The effect of the number of stars per cluster when using the $k$-means algorithm}
\mbox{ }\\
The UPMASK method does not rely on the choice of a specific algorithm for the clustering step in the  UPMASK kernel (see Sect.~\ref{sect:mainiter}). Nonetheless, a choice must be made, and the current implementation has used the $k$-means algorithm. The question then arises of how to specify the number of $k$-means clusters, or equivalently, as we have done, the number, $n$, of stars per cluster. In Sect.~\ref{sect:mainiter} we mentioned that best results were achieved with $n$ between 10 and 25. Here we analyse of the effects of varying $n$ for both simulated clusters studied in this section.
 
Fig.~\ref{Fig:ExampleSimulatedClusterskDep} exemplifies the dependency of purity and completeness at the 50\% and 90\% membership levels for different values of $n$. The TPR$_{90}$ decreases continuously from $n>13$ and the MMR$_{90}$ reaches values around $\sim25\%$ for $20>n>30$. Meanwhile, the MMR$_{50}$ quickly reaches $\sim70$\% for $n=15$, and then slowly decreases. These figures indicate that adopting high values of $n$ may result in more contaminated assignments at a fixed membership probability level. A noticeable feature however, is that for a low value $n=7 - 10$ not only TPR$_{50}$ is very high, but MMR$_{50}$ is also high (within its range). This is an indication that for less densely populated clusters the adoption of lower values of $n$ is recommended, as one could naively expect.

Although fine-tuning of $n$ can always be performed for studies requiring very high purity, the analysis in this section indicates that values in the range of $\sim10$ to $\sim25$ can be adopted in most applications. According to the cases studied here,  a value of $n=15$ stars could be taken as an optimal default value.

 
\paragraph{Effect of the field size}
\mbox{ }\\
It is known that for proper-motion-based membership studies, unavoidable pre-selection of areas in proper motion space and/or in coordinate space affects the derived membership probabilities \citep[e.g. ][]{2010A&A...510A..78S}.
Here we investigate how the the relative areas of the field and cluster under analysis might influence the output of UPMASK.

\begin{figure}
   \centering
  \begin{center}
  \subfloat[]{
    \includegraphics[width=0.5\linewidth, trim=0cm 0.5cm 0cm 0.5cm, clip=true]{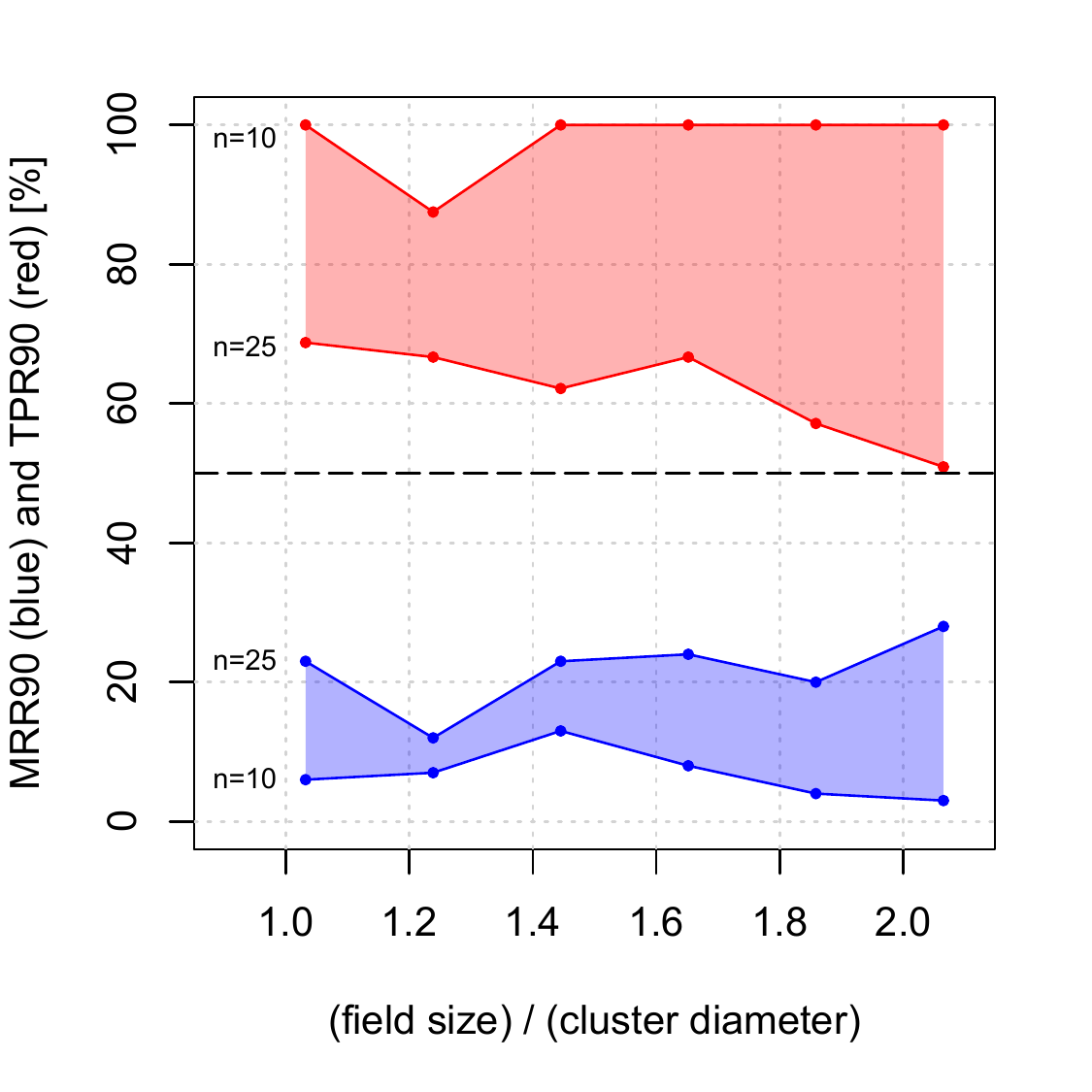}
  }
  \subfloat[]{
    \includegraphics[width=0.5 \linewidth, trim=0cm 0.5cm 0cm 0.5cm, clip=true]{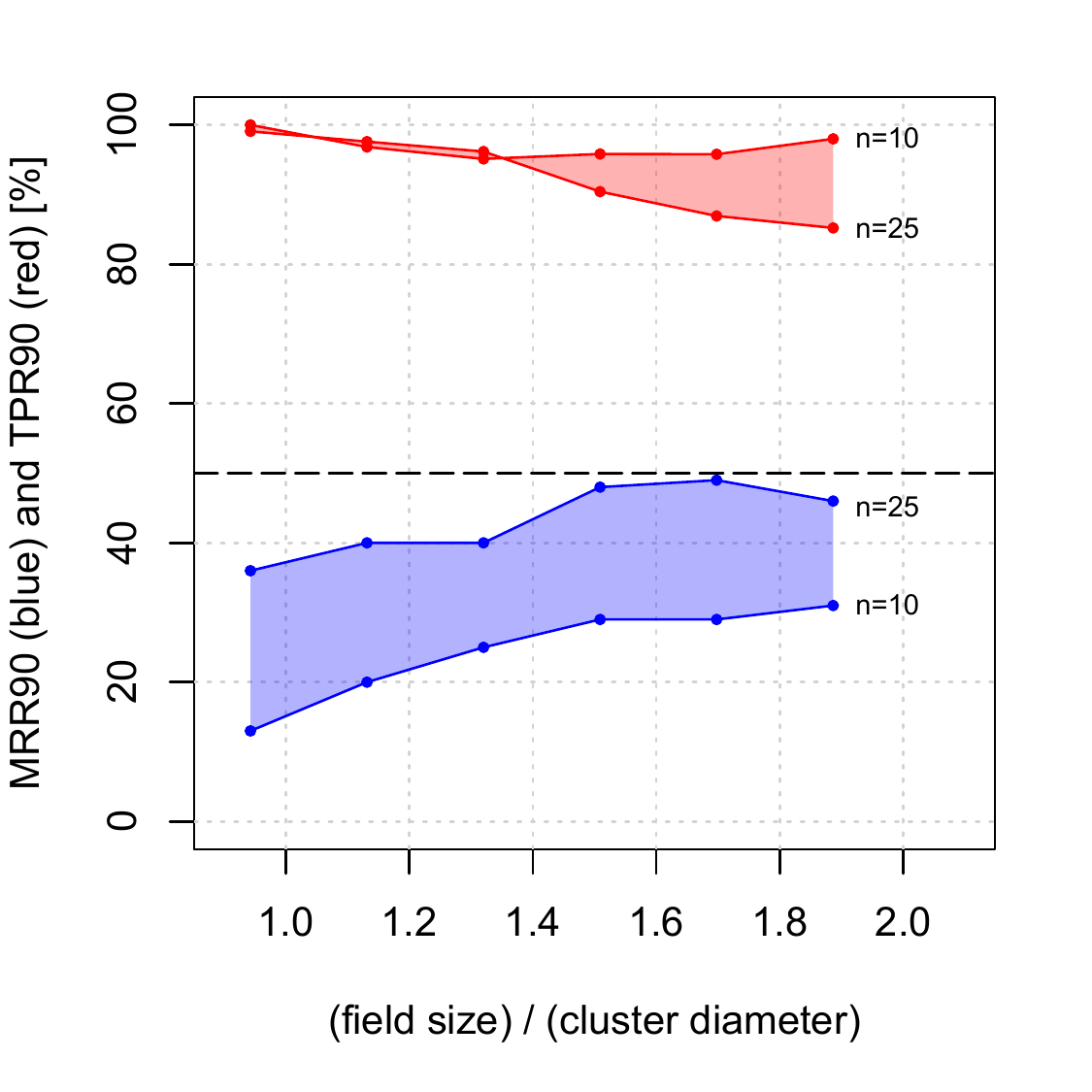}
  }
  \end{center}
   \caption{Dependency of the completeness and purity indicators at the 90\% membership probability level with the field of view. In the left panel we show the cluster from Fig. \ref{Fig:ExampleSimulatedClusters1}. In the right panel we show the cluster from Fig. \ref{Fig:ExampleSimulatedClusters2}. Field size refers to the linear dimension of the simulated square field of view.}
   \label{Fig:ExampleSimulatedClustersFieldSizeDep}
 \end{figure}

The dependency of the true positive and the membership recovery rates as a function of the ratio between the size of the observed field of view and the cluster diameter is shown in Fig.~\ref{Fig:ExampleSimulatedClustersFieldSizeDep}. Here, the cluster radius is considered as four times the King's profile core radius, since we found that this is roughly the radius at which star counts become indistinguishable from those of the field. The current implementation of UPMASK does not perform the cluster/field disentangling on fields much smaller than the cluster radius. We are investigating variations of the spatial density estimation step in the UPMASK kernel for adapting to smaller fields.

In Fig.~\ref{Fig:ExampleSimulatedClustersFieldSizeDep} we present the results for the lower mass, nearby cluster in the left plot and the more massive, farther removed object in the right plot. In both cases, the TRP$_{90}$ (in red) and MMR$_{90}$ (in blue) are shown for two extreme values of the number $n$ of stars per $k$-means cluster. The completeness, represented by MMR$_{90}$, remains fairly constant at the 10-20\% level for the lighter cluster over a wide range of field/cluster size ratios. For the same cluster, we observe that the purity indicator, TRP$_{90}$, shows a stronger variation: while a lower value of $n$ resulted in highly pure sets, even reaching 100\% for most size ratios, the results for a higher $n$ were poorer and degraded as the field of view (FoV) grew. This test confirms that lower values of the number of stars per $k$-means cluster should be adopted if the best solutions are required for less massive objects, even though an acceptable result is obtained with TRP$_{90}$ above 50\% for the worst case that was analysed. For the farther removed, but more massive cluster the results for purity are very stable at high levels above $\sim 85\%$. The completeness, on the other hand, favours somewhat more extended fields, which can increase the MMR$_{90}$ up to $\sim 10\%$. Strictly speaking, from the point of view of the method concept, this increase is unexpected because there is no immediate reason for such an increase beyond the cluster limits. Our current explanation is that it could be due to the recipe for the spatial clustering test. This will be analysed in future work.

Finally, as previously mentioned and according to Fig.~\ref{Fig:MRR90}, the simulated clusters analysed here do not occupy particularly favourable regions of the mass-distance space. Nevertheless, the tests showed that even in these cases, useful purity and completeness results are obtained. These examples show that the UPMASK method is able to separate cluster/field populations and moreover recovers the overall morphological structure and features of the distribution of member stars on the CMD, as readily seen in Figs.~\ref{Fig:ExampleSimulatedClusters1} and~\ref{Fig:ExampleSimulatedClusters2}.

\subsubsection{Fields with lower counts}

Since UPMASK is based on statistical procedures, it is conceivable that for fields with low stellar counts, small number statistics could affect the results. To investigate this, we analysed simulations of a low-mass cluster (500$M_{\sun}$ initial mass, Myr, 81 stars) at three different Galactic coordinates: 
$(l, b)_{gal} = \{ (120^\circ, 0^\circ), (180^\circ, 15^\circ), (180^\circ, 25^\circ) \}$, corresponding to 4956, 1975, and 916 field stars respectively.

Following the analysis in Sect.~\ref{sect:casestudies}, for lower counts it is advisable to adopt lower values for the number $n$ of stars in each $k$-means cluster. Adopting $n=7$, the TPR$_{90}$ obtained for the three coordinates are $\sim97.5\%$, $\sim94.4\%$, and $\sim100.0\%$, respectively. Although high purity values were obtained, the completeness in this last case (the least populated field) was low, with only 12\% of the cluster stars recovered with a membership probability $>90\%$. At the $50\%$ membership probability level, however, we obtained a purity of $90.5\%$ and a completeness of $47\%$. In the other cases, the completeness at $90\%$ membership probability was $48\%$ and $42\%$, respectively.

Despite the evoked reasons for adopting a low $n$, we also tested  $n=25$. The TRP$_{90}$  obtained were  $\sim66.1$\%, $\sim73.8\%$, and $\sim85.5\%$. As expected, these are significantly lower than the purity figures obtained for $n=7$. Conversely, the completeness is significantly higher, reaching $99\%$, $73\%$, and $58\%$, respectively.

These tests show that the choice of the number of stars per $k$-means cluster does influence the best results in low count situations. Nevertheless, the default value of $n=15$ would produce acceptable results in the analysed simulations.

\section{Application to real data\label{sect:realdatamain}}

Following the validation of UPMASK with simulated data presented above, we tested the method on real data-sets. Here we present results using UBVRI photometry of two cluster fields taken from \citetads{2001A&A...370..436M}, which have the same field of view and limiting magnitude as those in the simulations. The first field is centred on the open cluster \object{Haffner~16}. The second field adds complexity by including two open clusters, \object{Haffner~10} and \object{Czernik~29}. Only stars with measurements in all five UBVRI bands were considered.

\subsection{Haffner 16}
\label{haf16}
\begin{figure}
   \centering
  \begin{center}
  \subfloat[]{
    \includegraphics[width=0.5\linewidth, trim=0cm 0.5cm 0cm 0.5cm, clip=true]{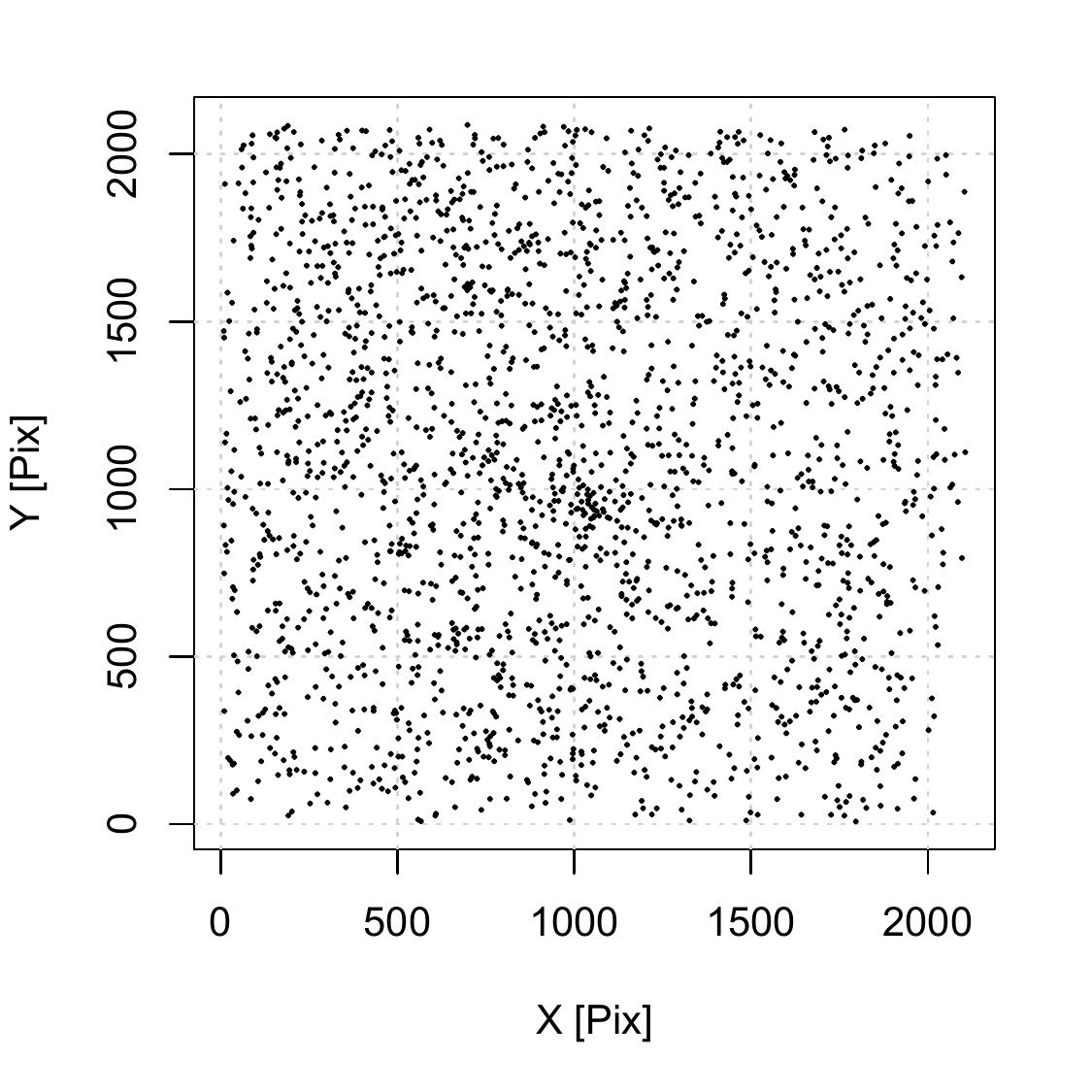}
  }
  \subfloat[]{
    \includegraphics[width=0.5 \linewidth, trim=0cm 0.5cm 0cm 0.5cm, clip=true]{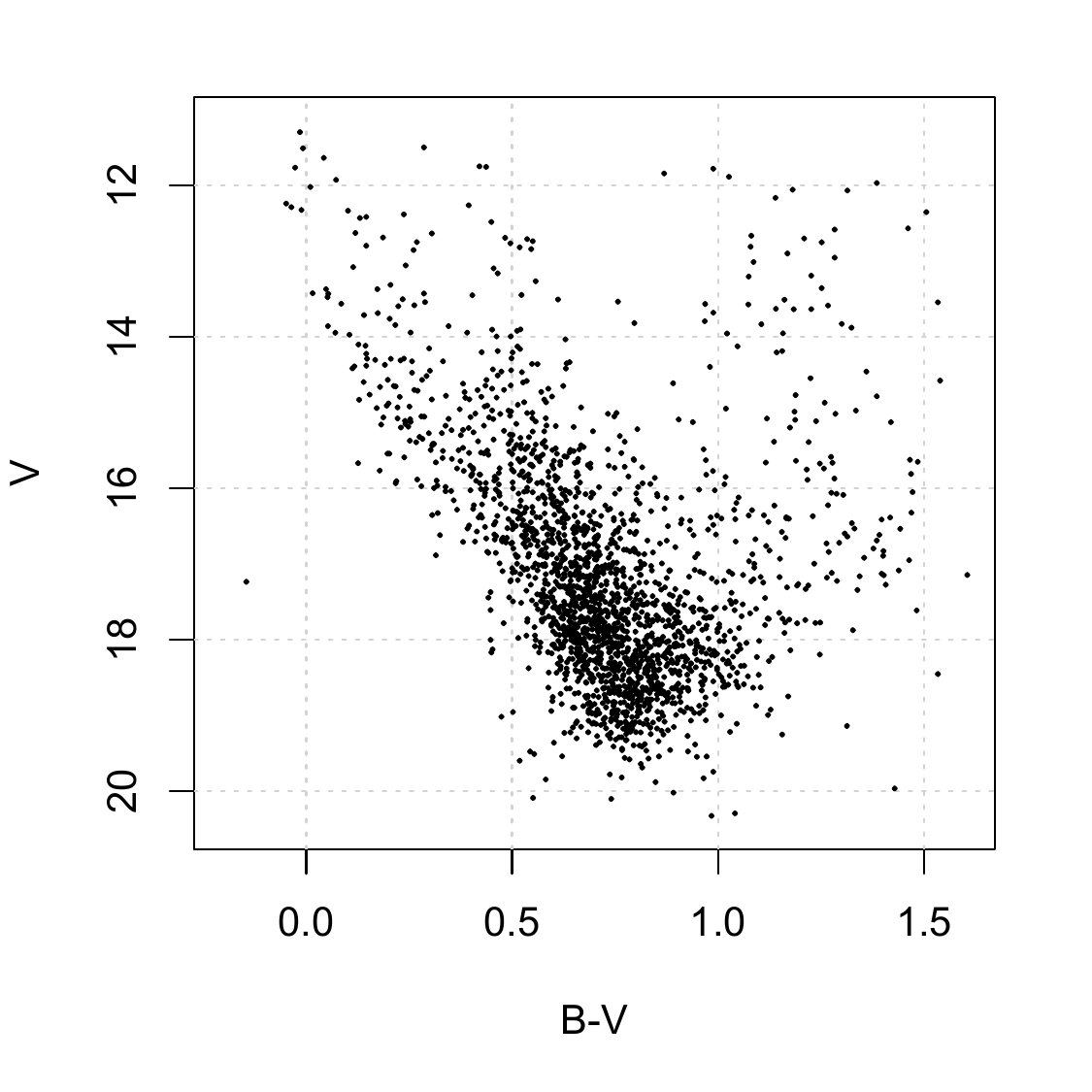}
  }\\
  \subfloat[]{
    \includegraphics[width=0.5 \linewidth, trim=0cm 0.5cm 0cm 0.5cm, clip=true]{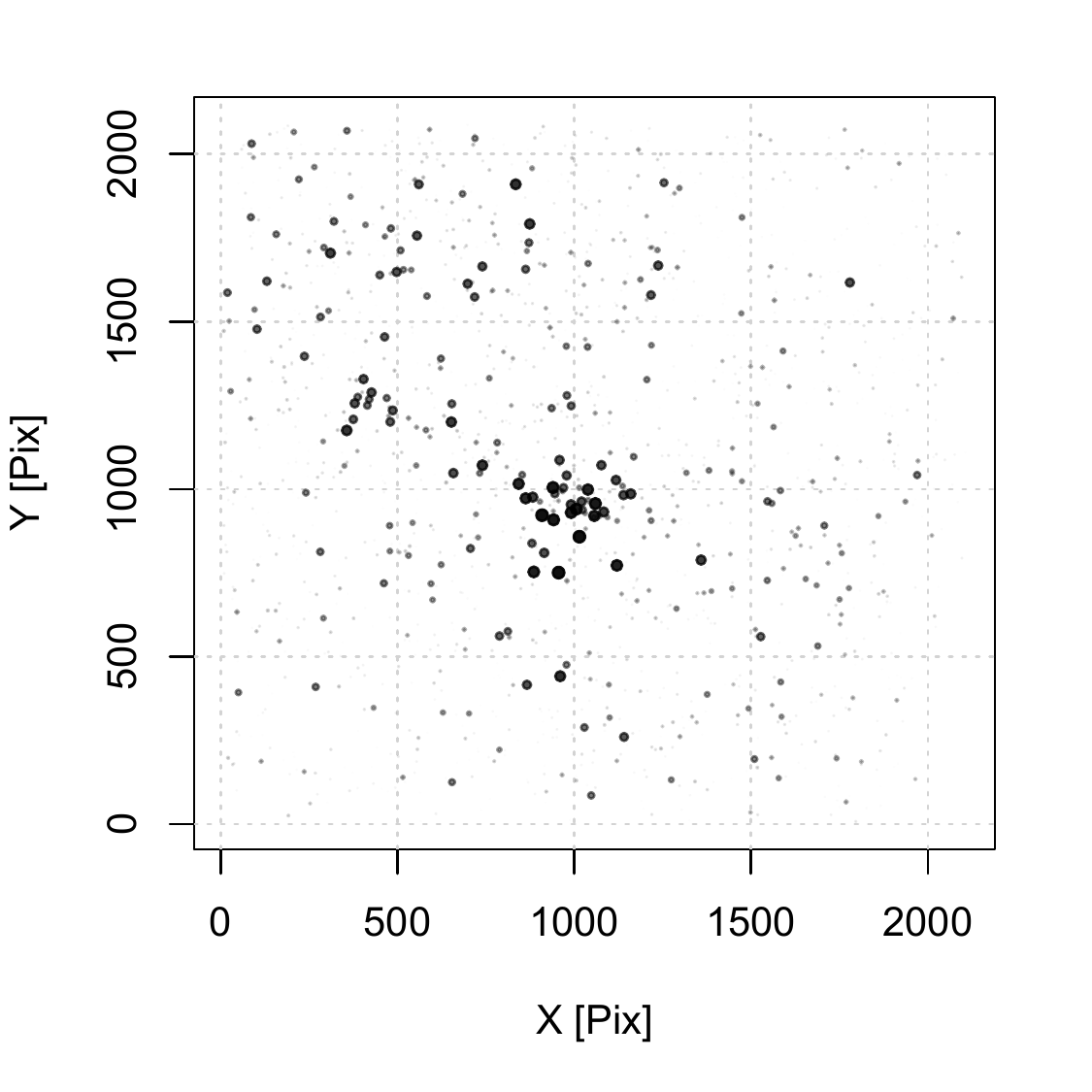}
  }
  \subfloat[]{
    \includegraphics[width=0.5 \linewidth, trim=0cm 0.5cm 0cm 0.5cm, clip=true]{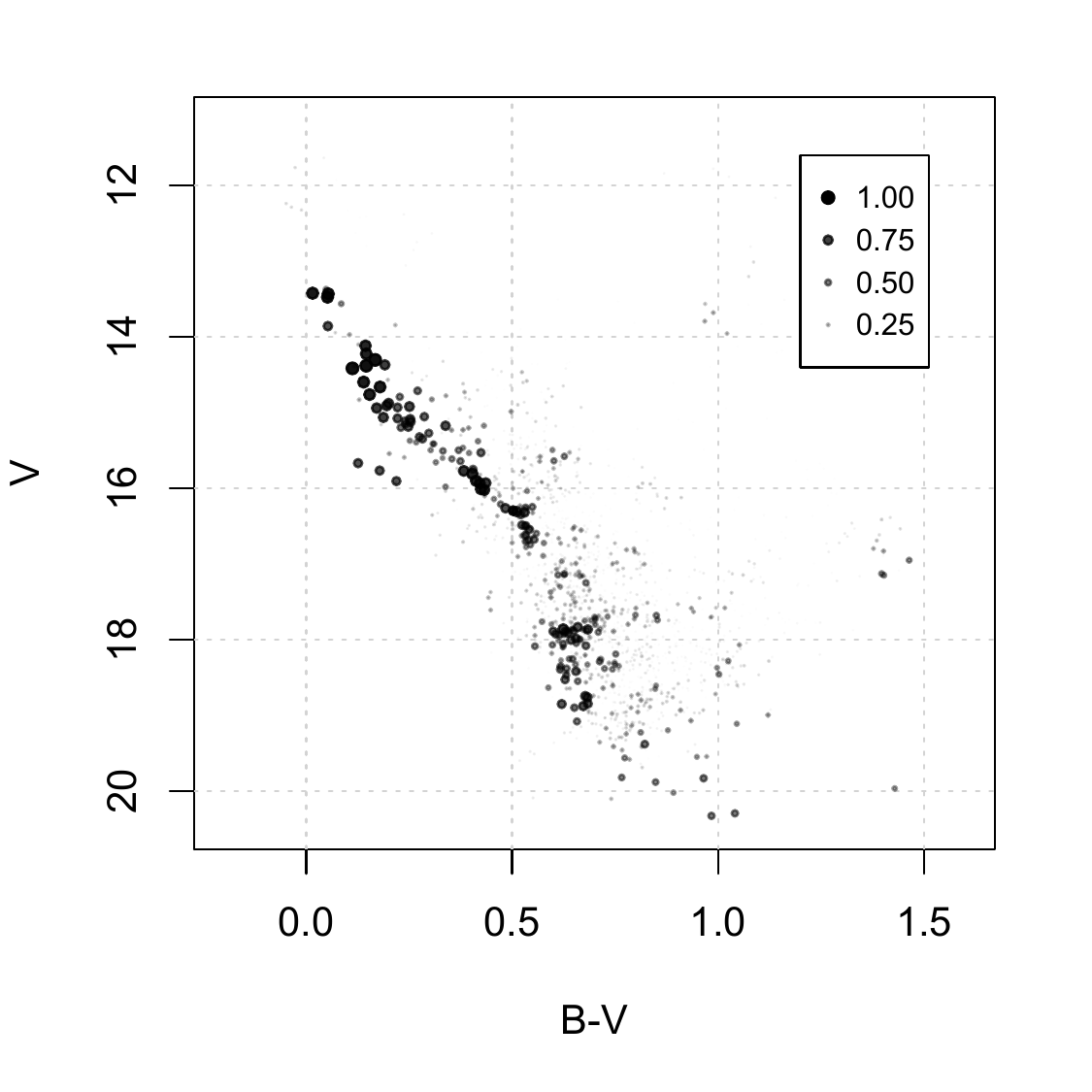}
 }\\
  \subfloat[]{
    \includegraphics[width=0.5 \linewidth, trim=0cm 0.5cm 0cm 0.5cm, clip=true]{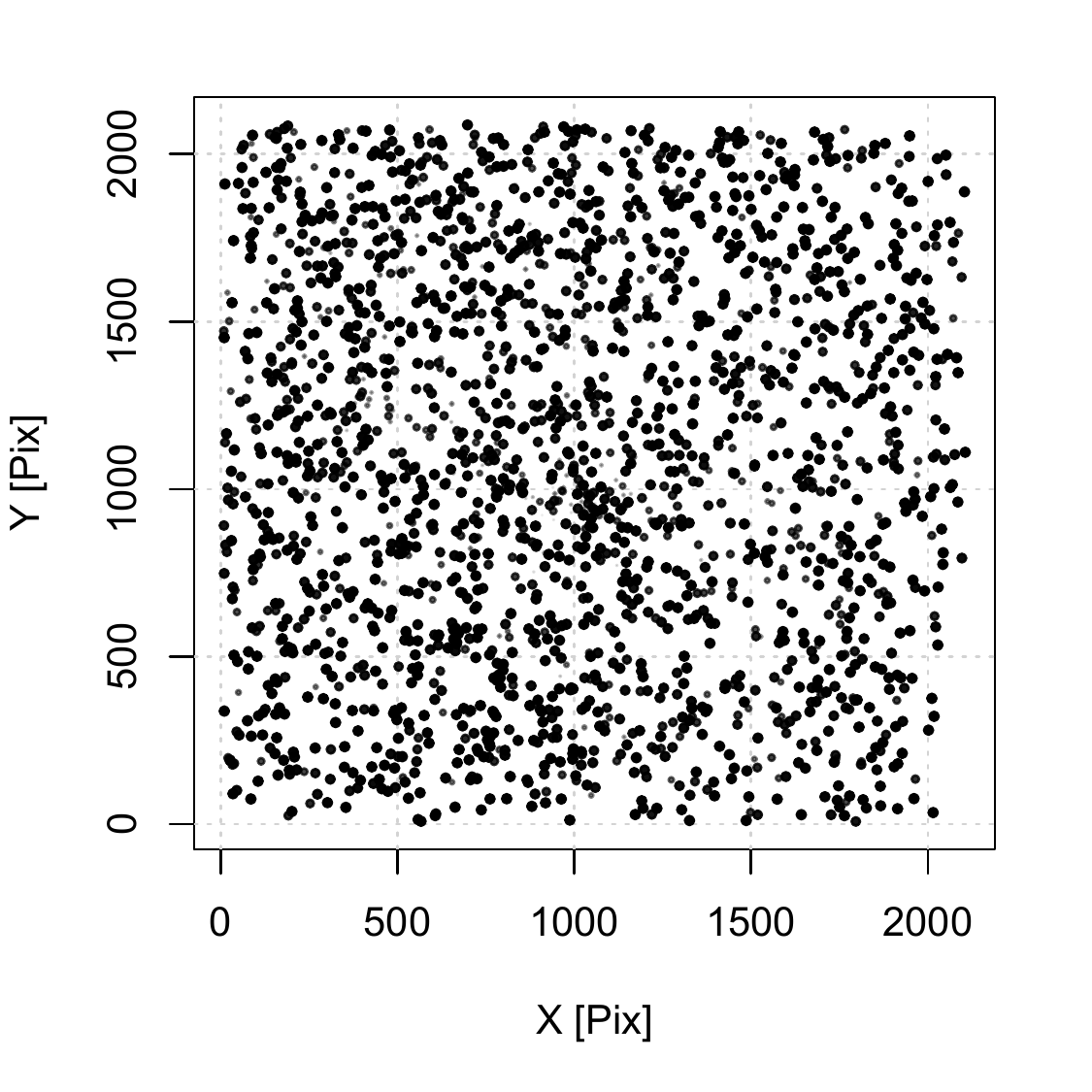}
  }
  \subfloat[]{
    \includegraphics[width=0.5 \linewidth, trim=0cm 0.5cm 0cm 0.5cm, clip=true]{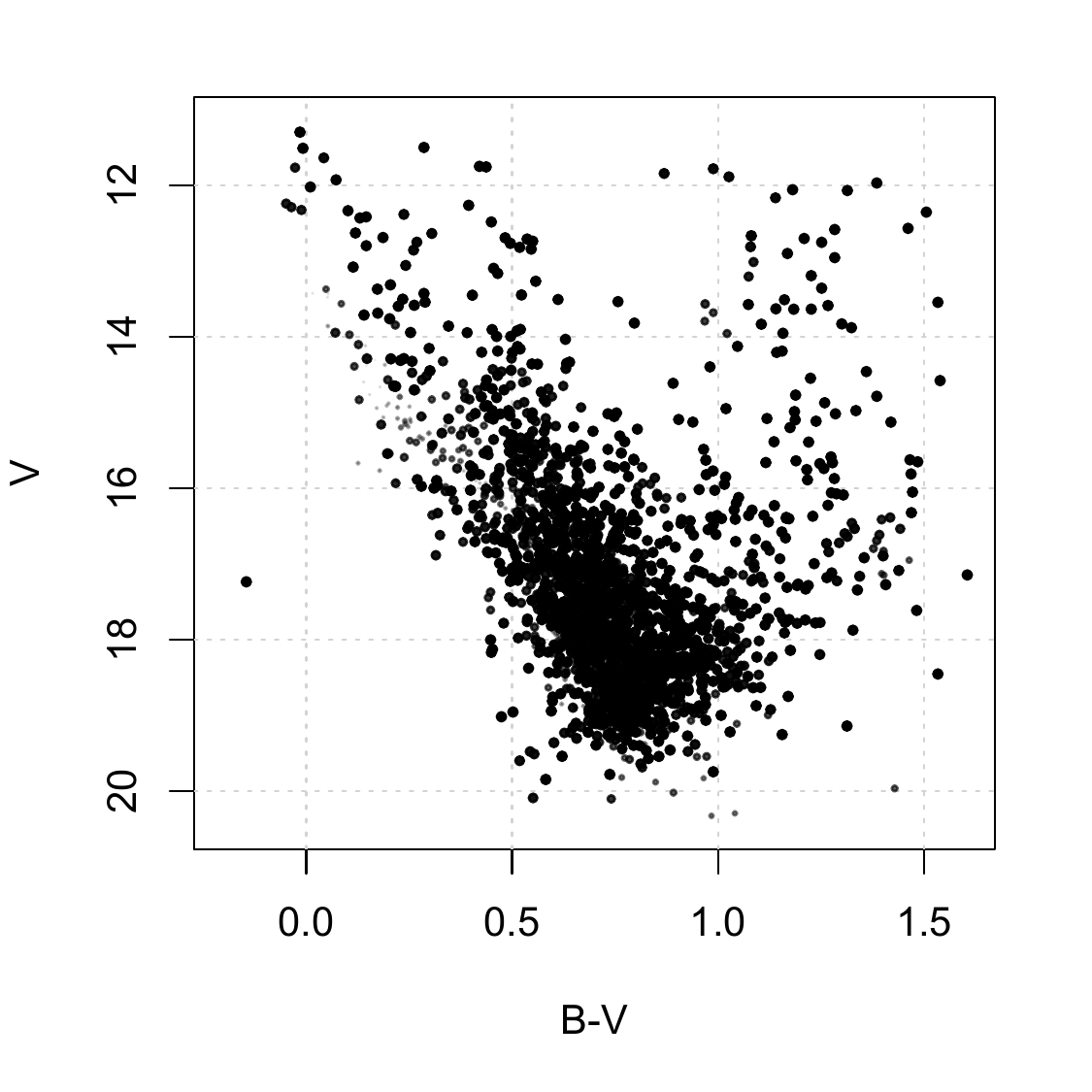}
 }
  \end{center}
      \caption{Results of the UPMASK run on the UBVRI data of Haffner~16. \textbf{(a):} Distribution of all observed objects in positional space (X-Y CCD coordinates). \textbf{(b):}  Distribution of all observed objects in the V vs. (B-V)  CMD. \textbf{(c)}  and \textbf{(d):}  Cluster membership probabilities are encoded as point sizes and transparencies in the CMD and positional spaces (less probable = smaller and more transparent). \textbf{(e)} and \textbf{(f):} the same encoding, but for field membership probabilities.}
         \label{Fig:Haffner16}
   \end{figure}

The original data for Haffner~16 and the results from the UPMASK run are presented in Fig.~\ref{Fig:Haffner16}. For positional data, we used the CCD pixel coordinates.  On-sky coordinates (right ascension and declination, $\alpha$ and $\delta$) could have been used instead. This is expected to have no effect because UPMASK sets no particular requirement on the shape of the spatial clustering.
 
Fig.~\ref{Fig:Haffner16}d shows that for $V < 17$ mag the cluster main sequence emerges in the CMD from the high-probability members alone. Immediately below, we find a gap extending almost to $V \sim 18$ mag, after which there is a new group of stars assigned with high membership probabilities. At these fainter magnitudes the U photometry becomes incomplete and has larger errors, which could partially explain the scatter in the faint group. Although it is beyond the scope of this paper to embark on a detailed interpretation of the observations, we note that the faint group appears as a blue bump in the overall distribution of stars.  
This bump corresponds mostly to the stars with high membership probability seen off the cluster centre in the upper left of Fig.~\ref{Fig:Haffner16} (c). Together with the red clump at $\textrm{(B-V)}\sim 1.4$ mag, this might indicate the presence of a distant and older cluster in the field. Another possible scenario is that we are seeing the signature of the metal-poor thick disc reported in various fields in this region of the Galaxy \citepads{2007AJ....133.1058C, 2008MNRAS.385.1597C, 2010ApJ...718..683C}.

Returning to the upper main sequence, we find some dispersion in the high-probability members. This need not be due to residual field contamination. The thickness of the main sequence is well compatible with being due to unresolved binary members. In fact, these stars are also clustered in positional space.

Another noticeable of feature UPMASK can be seen in Figs.~\ref{Fig:Haffner16} (e) and (f). Here we see that the method avoided opening holes in the CMD or positional spaces for the field distribution. Holes in the distribution of the remaining field stars are a well-known artefact in parametric methods such as those based on the Vasilevskis-Sanders approach.  However, comparing Fig. \ref{Fig:Haffner16} (e) and (a), one can notice a residual concentration of stars in the centre of the positional space in (e). Some of these stars are very likely cluster members, but have been assigned low membership probabilities and thus still appear as a concentration in the field. 

  \begin{figure}
   \centering
  \begin{center}
  \subfloat[]{
    \includegraphics[width=0.5\linewidth, trim=0cm 0.5cm 0cm 0.5cm, clip=true]{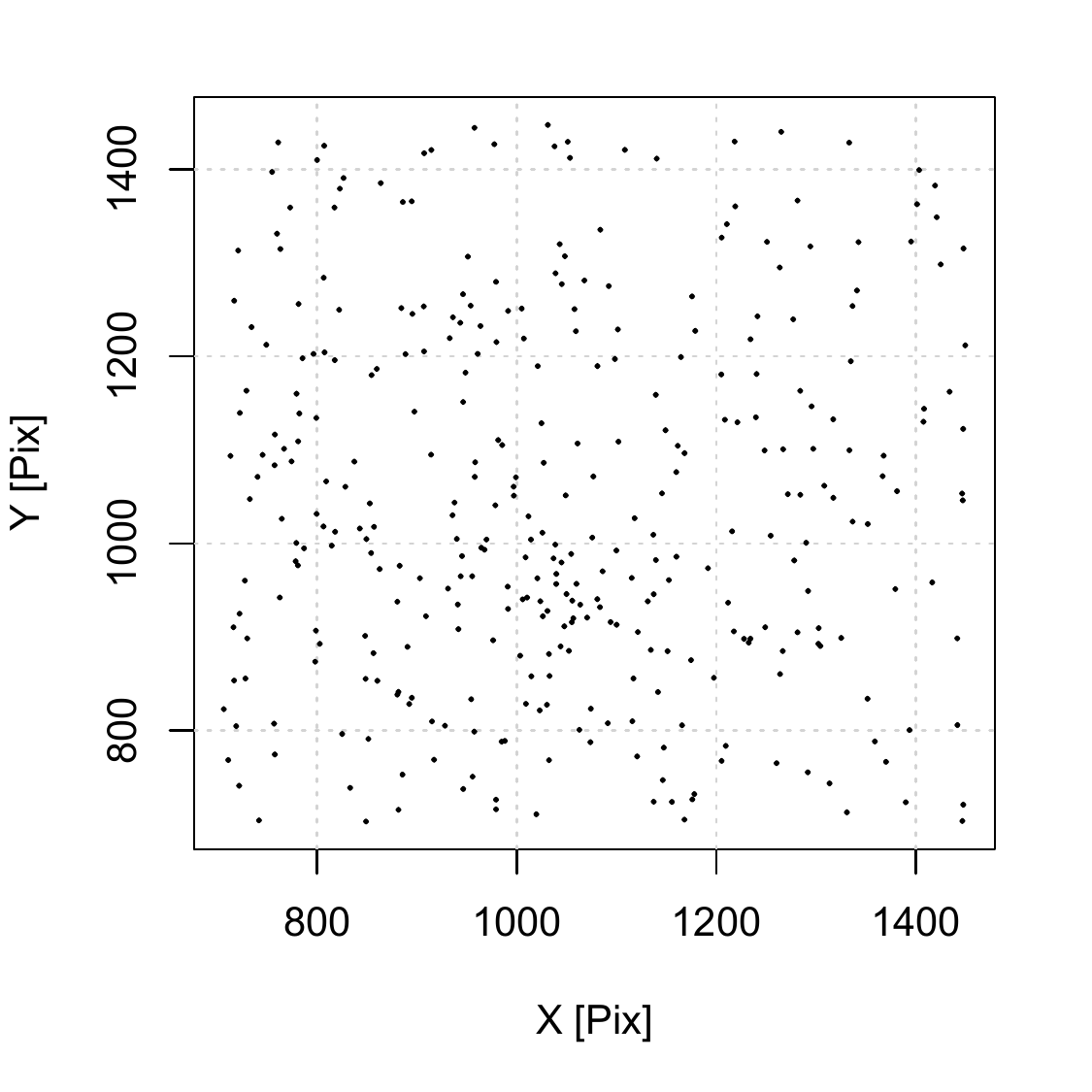}
  }
  \subfloat[]{
    \includegraphics[width=0.5 \linewidth, trim=0cm 0.5cm 0cm 0.5cm, clip=true]{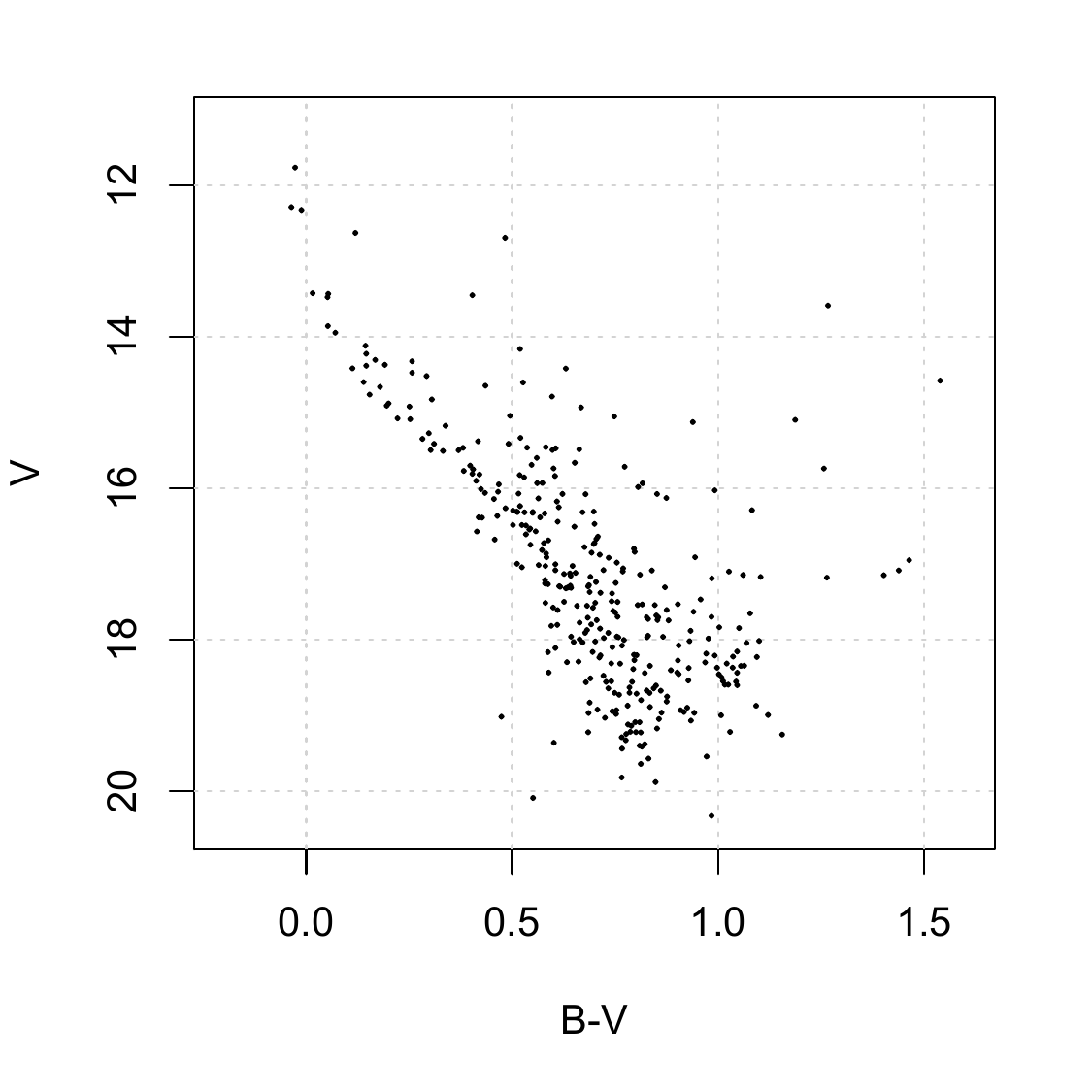}
  }\\
  \subfloat[]{
    \includegraphics[width=0.5 \linewidth, trim=0cm 0.5cm 0cm 0.5cm, clip=true]{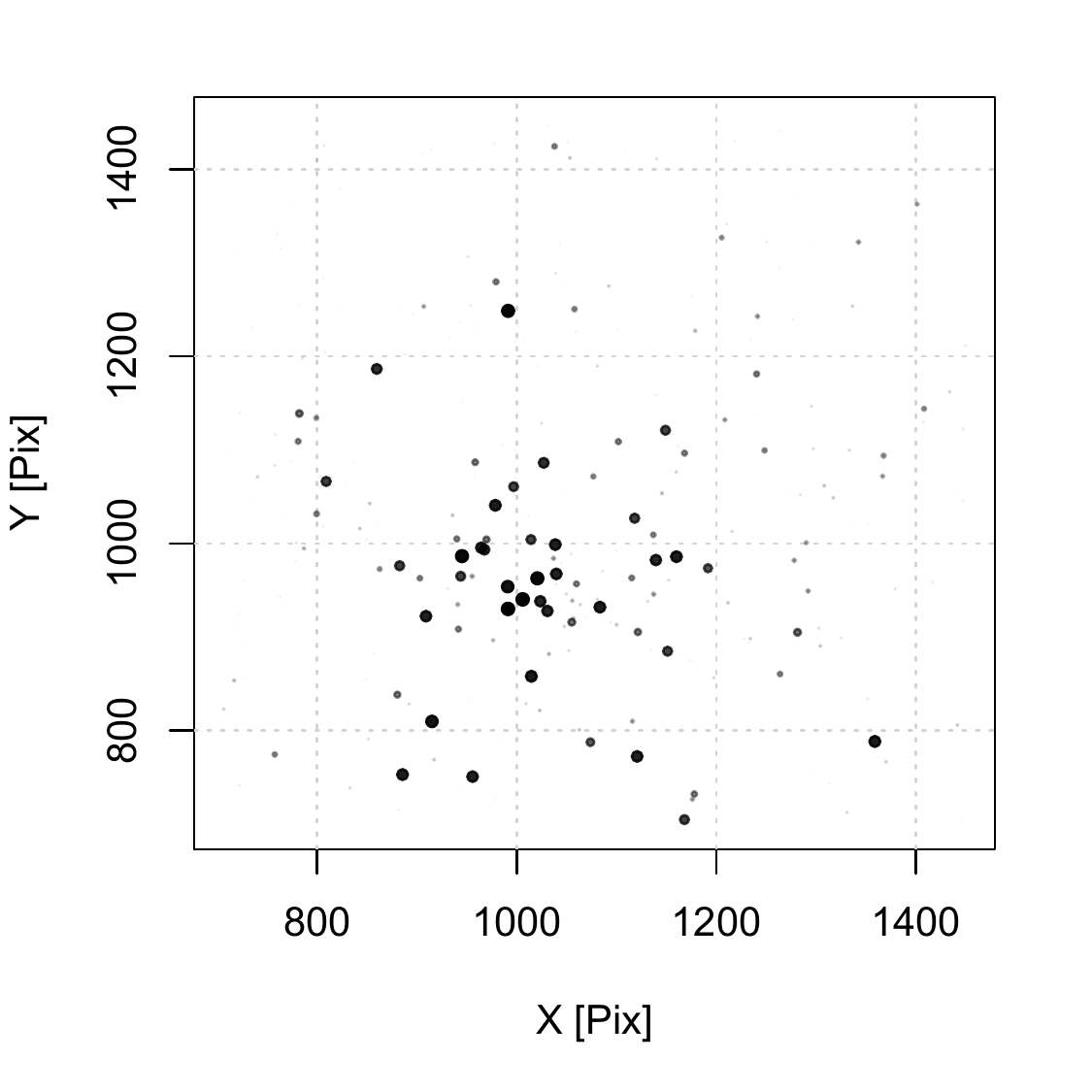}
  }
  \subfloat[]{
    \includegraphics[width=0.5 \linewidth, trim=0cm 0.5cm 0cm 0.5cm, clip=true]{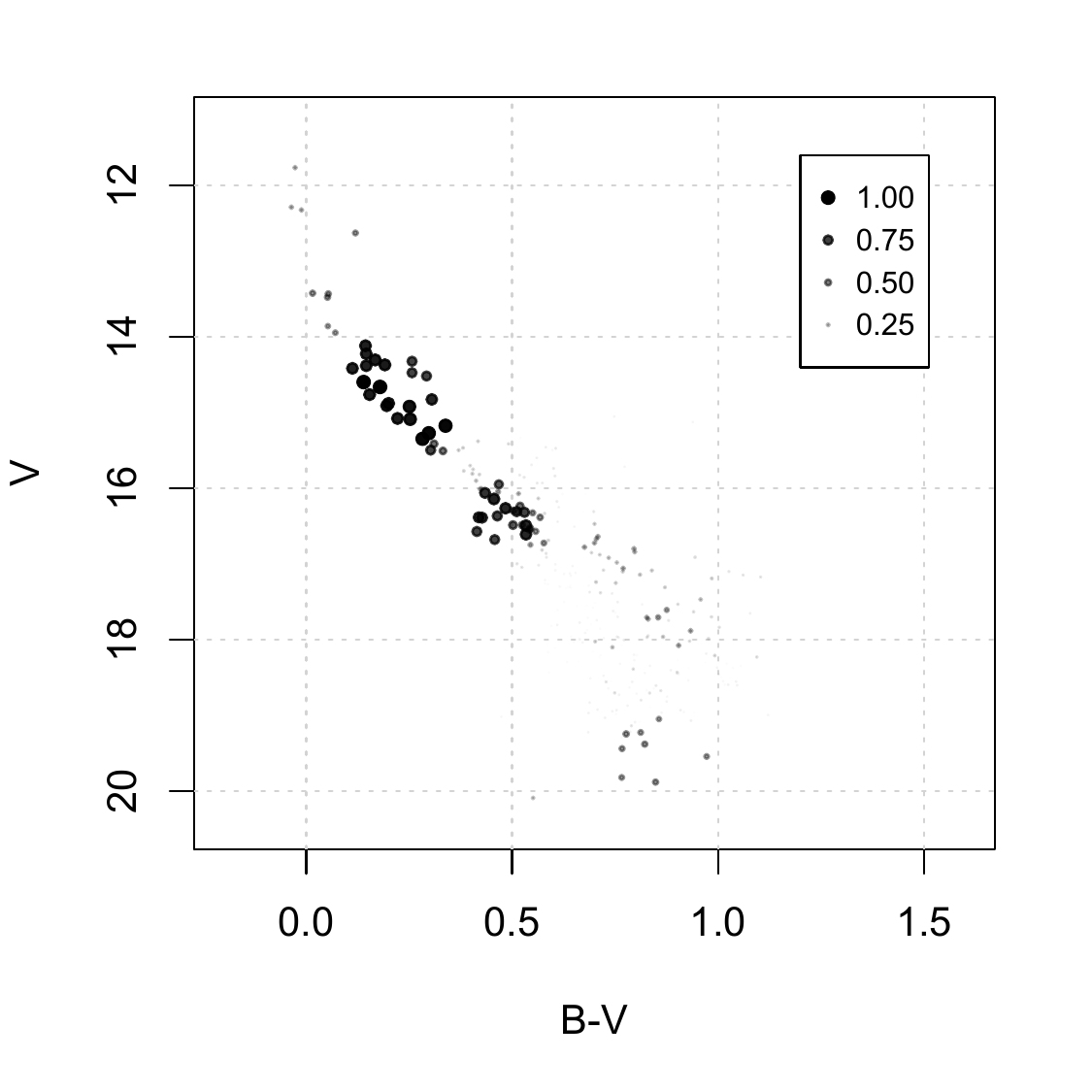}
 }\\
  \subfloat[]{
    \includegraphics[width=0.5 \linewidth, trim=0cm 0.5cm 0cm 0.5cm, clip=true]{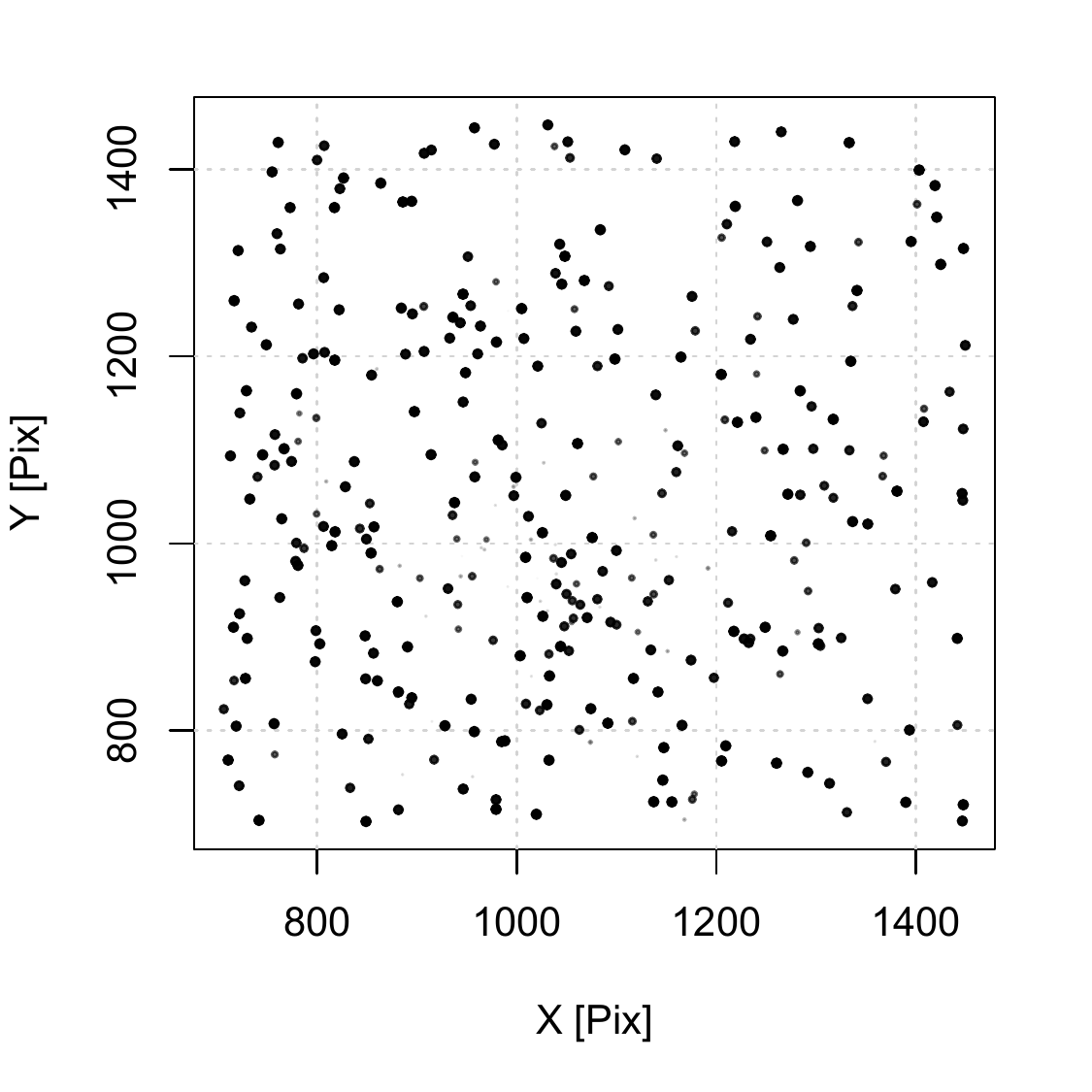}
  }
  \subfloat[]{
    \includegraphics[width=0.5 \linewidth, trim=0cm 0.5cm 0cm 0.5cm, clip=true]{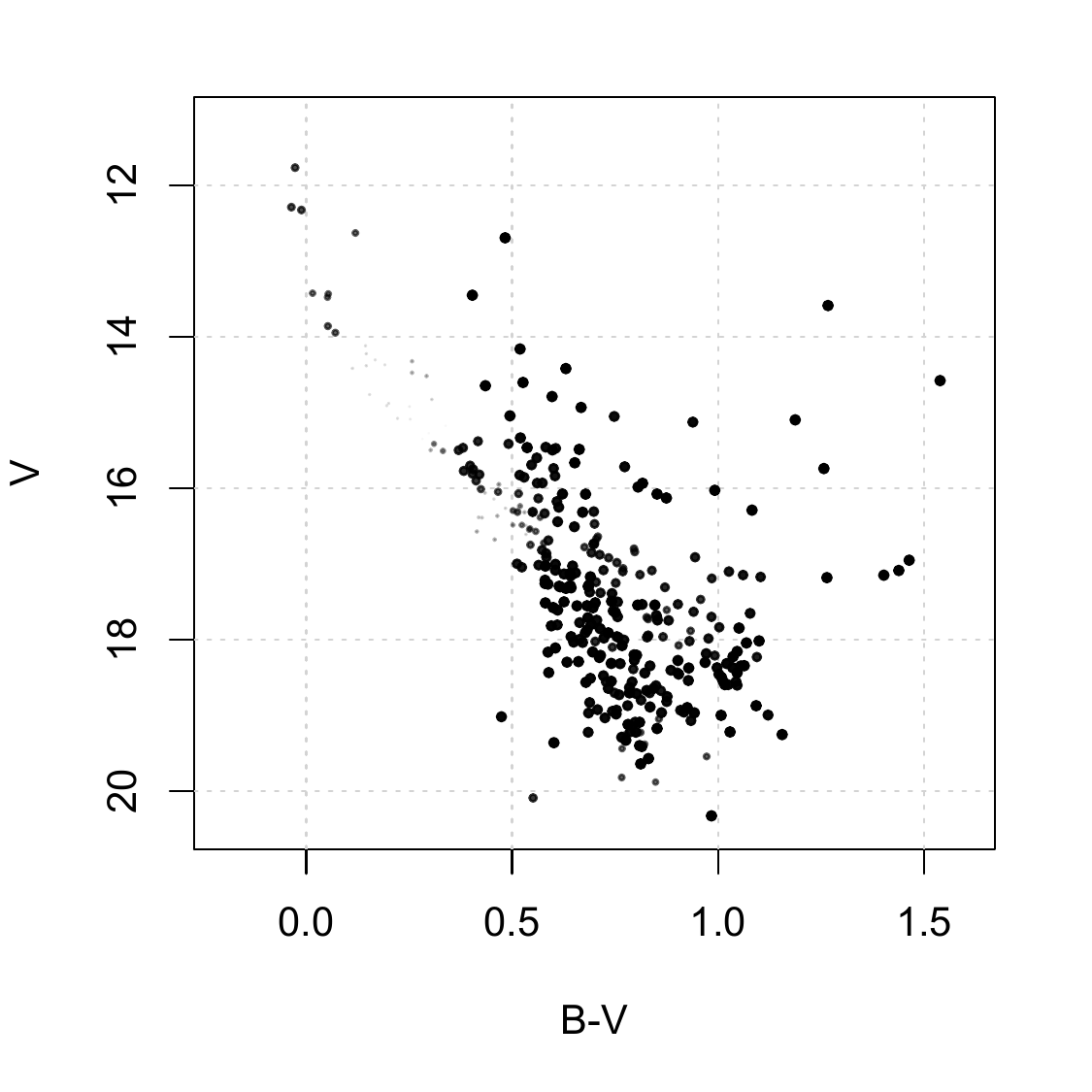}
 }
  \end{center}
      \caption{UPMASK results for the central region of the Haffner~16 field. Panels as in Fig.~\ref{Fig:Haffner16}.}
         \label{Fig:Haffner16_66}
   \end{figure}

   \begin{figure}
   \centering
  \begin{center}
  \subfloat[]{
    \includegraphics[width=0.5\linewidth, trim=0cm 0.5cm 0cm 0.5cm, clip=true]{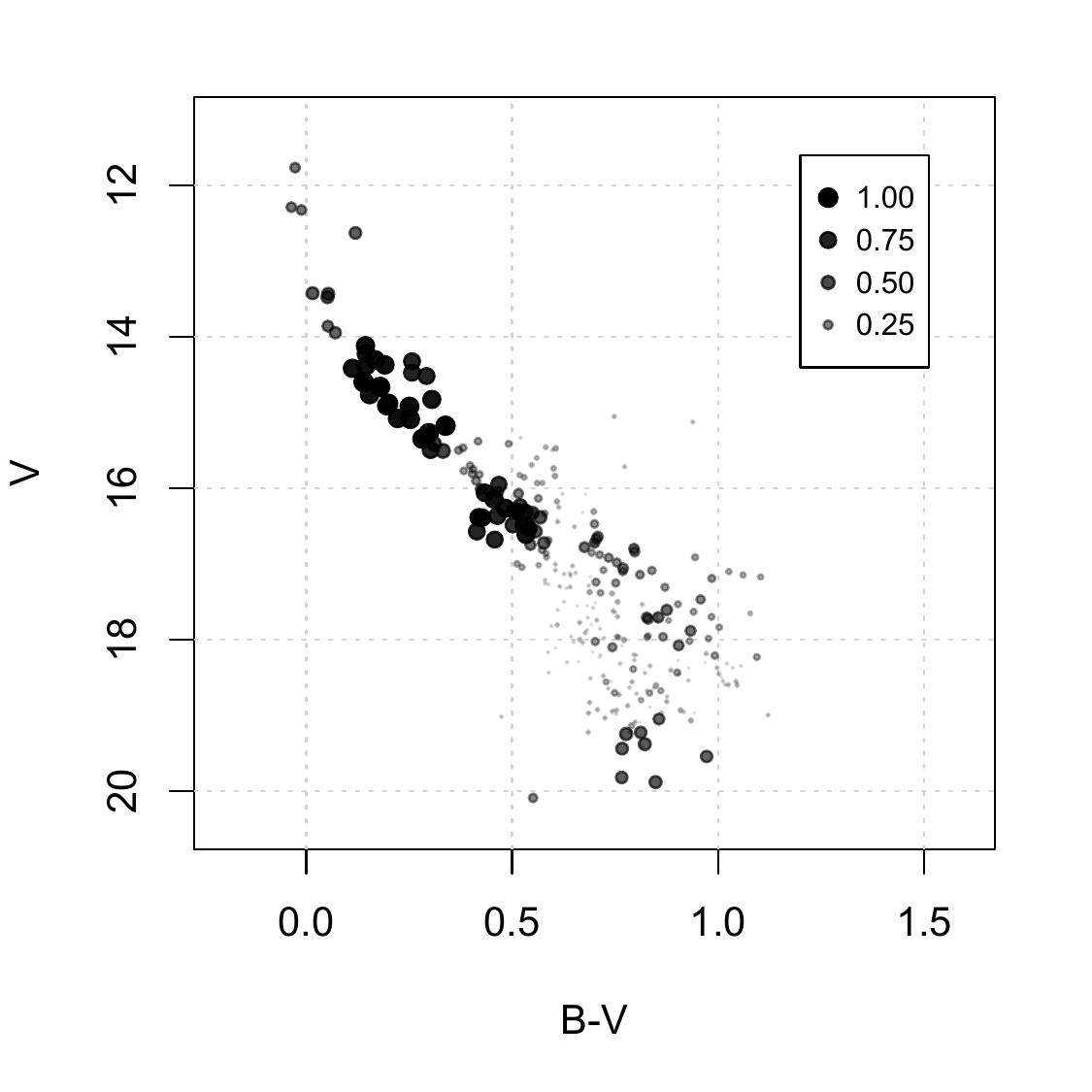}
  }
  \subfloat[]{
    \includegraphics[width=0.5 \linewidth, trim=0cm 0.5cm 0cm 0.5cm, clip=true]{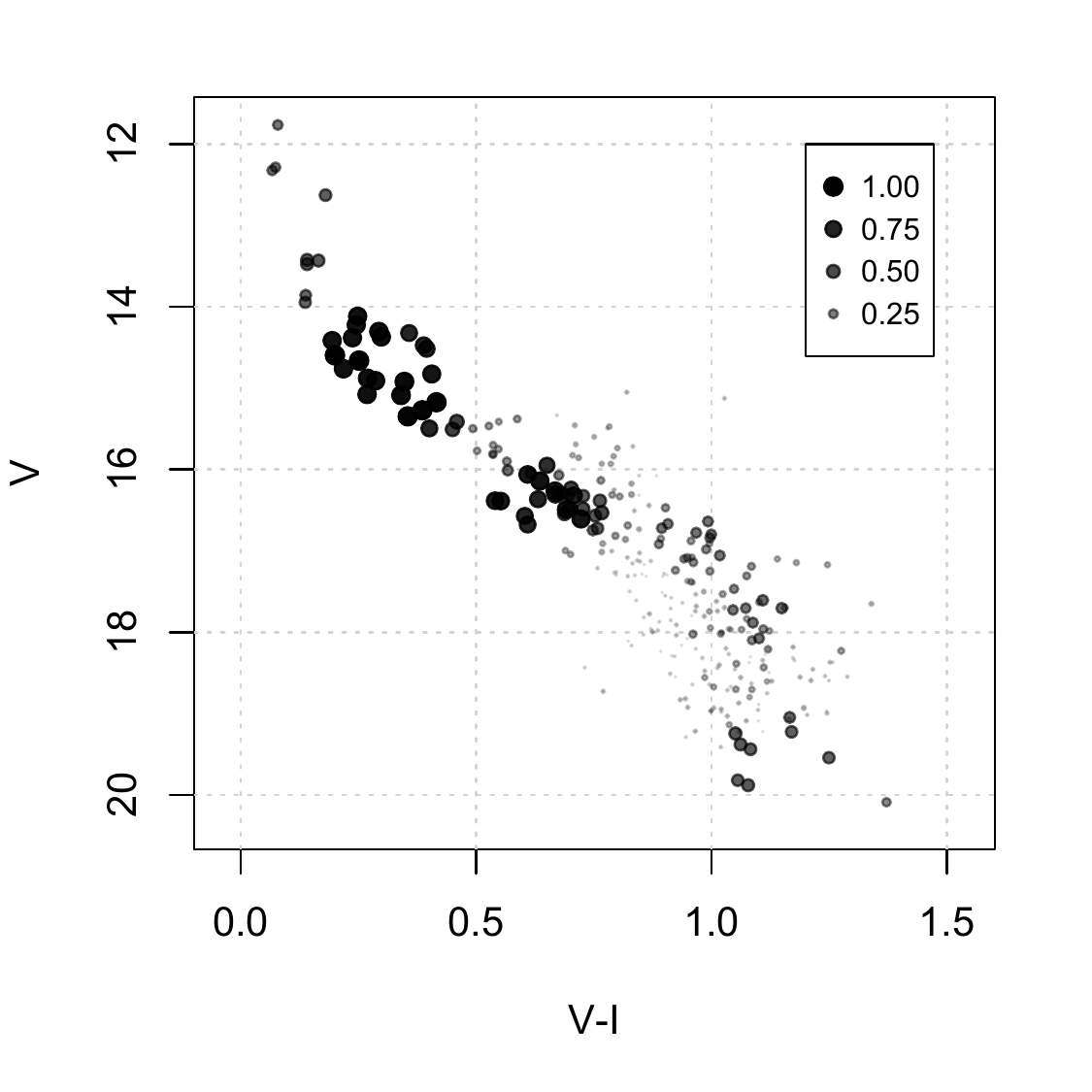}
  }
  \end{center}
      \caption{V vs. (B-V) and V vs. (V-I) CMDs for the central region of the Haffner~16 field. The point size and transparency encode the square-root of the membership probabilities.}
         \label{Fig:Haffner16_66_PMS}
   \end{figure}

As another test, we applied UPMASK only to the most central part of the field, cutting 66\% of the area. The results are presented in Fig. \ref{Fig:Haffner16_66} (c) and (d). Interestingly, the fainter high-probability clump has disappeared and a hint of the cluster sequence departing to the red below $V\sim 17$ can be seen. This red sequence is better distinguished in Fig.~\ref{Fig:Haffner16_66_PMS}. An interesting research possibility offered by applying UPMASK to these data would be to verify whether this redder sequence might be composed of pre-main sequence stars. 

By comparing the results obtained from this reduced field with those from the entire data-set (Fig.~\ref{Fig:Haffner16}), we found that running UPMASK on a fraction of the field influences the membership probabilities. Dependency on the considered area is also known to affect other membership-assessment algorithms \citepads{2010A&A...510A..78S}.

  \begin{figure}
   \centering
  \begin{center}
  \subfloat[]{
    \includegraphics[width=0.5\linewidth, trim=0cm 0.5cm 0cm 0.5cm, clip=true]{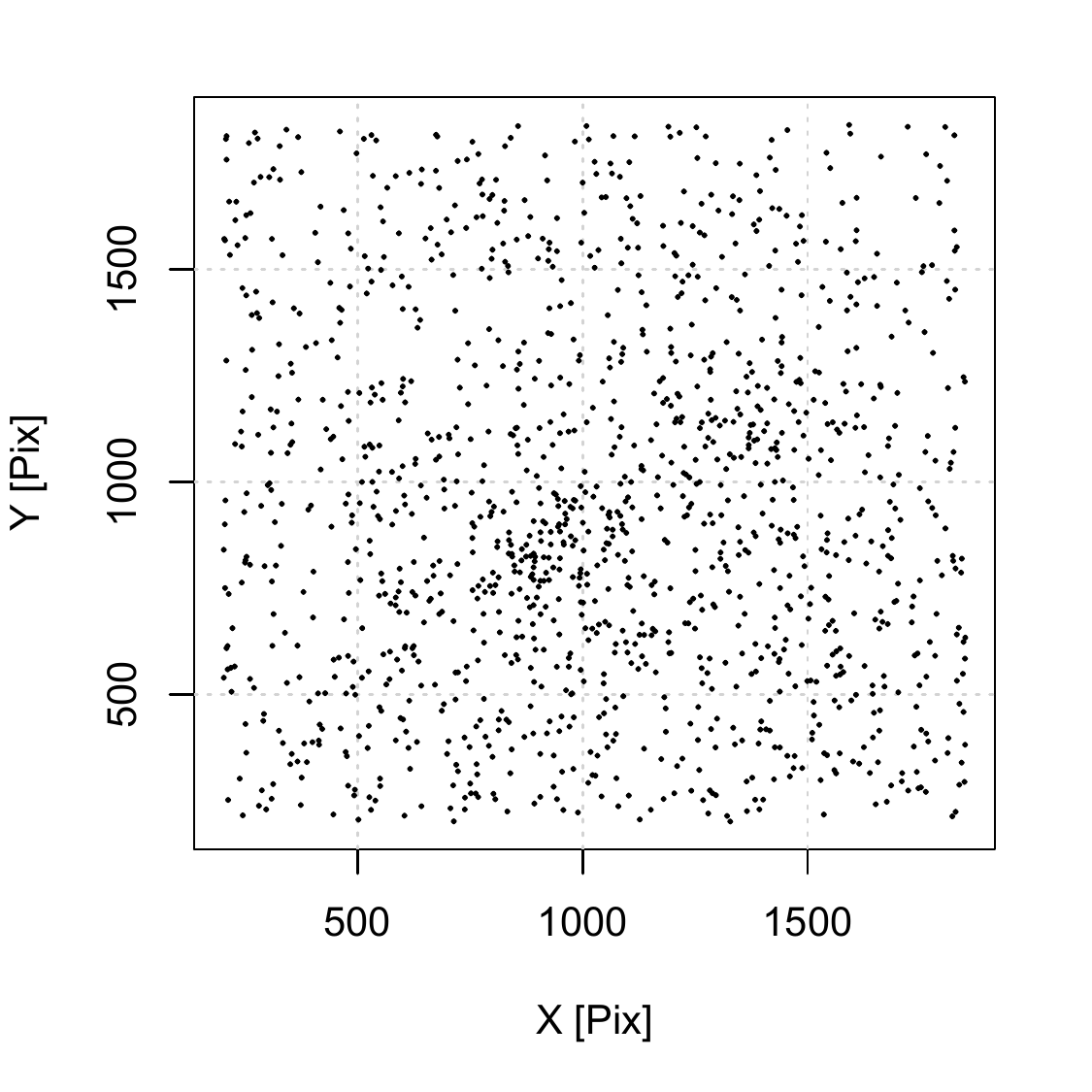}
  }
  \subfloat[]{
    \includegraphics[width=0.5 \linewidth, trim=0cm 0.5cm 0cm 0.5cm, clip=true]{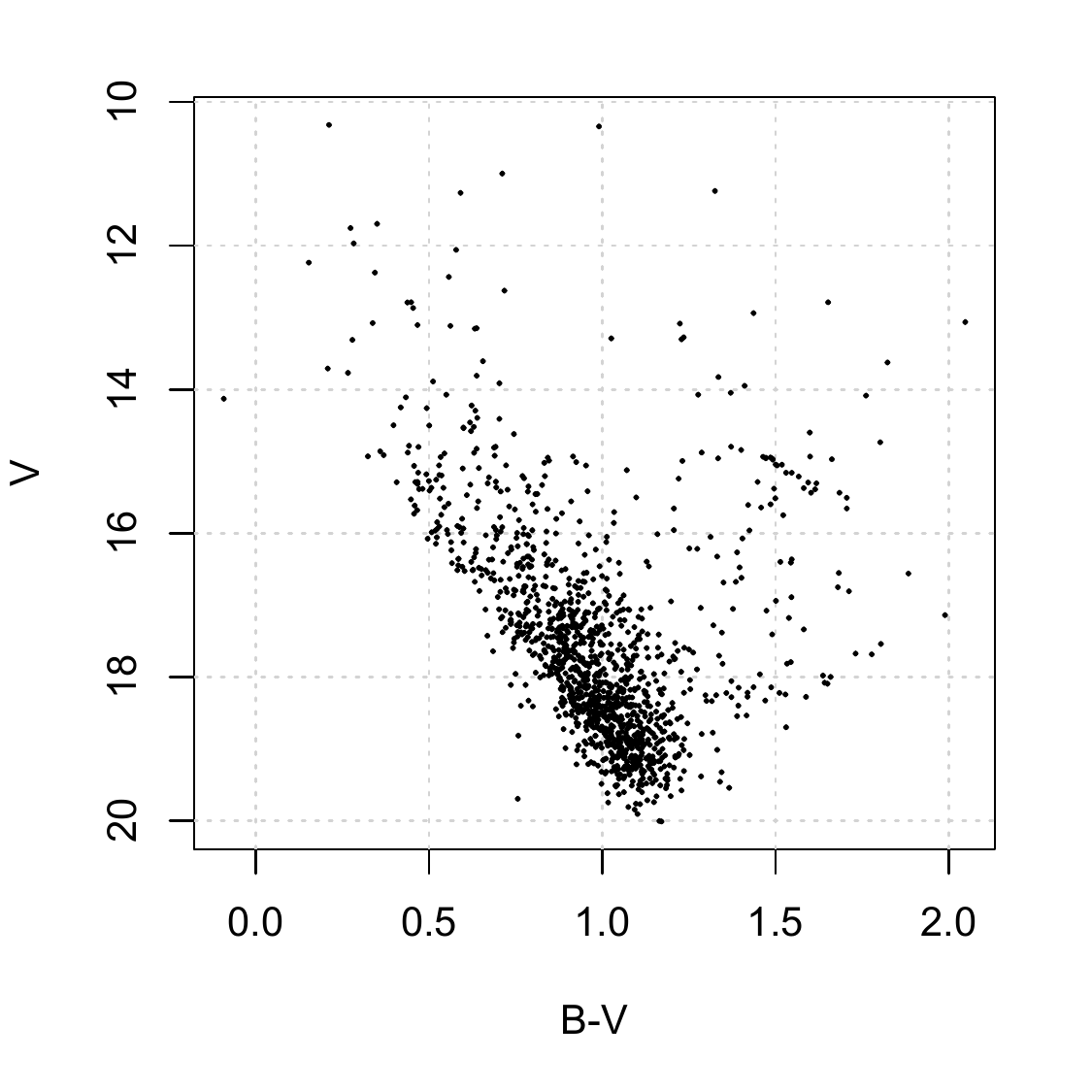}
  }\\
  \subfloat[]{
    \includegraphics[width=0.5 \linewidth, trim=0cm 0.5cm 0cm 0.5cm, clip=true]{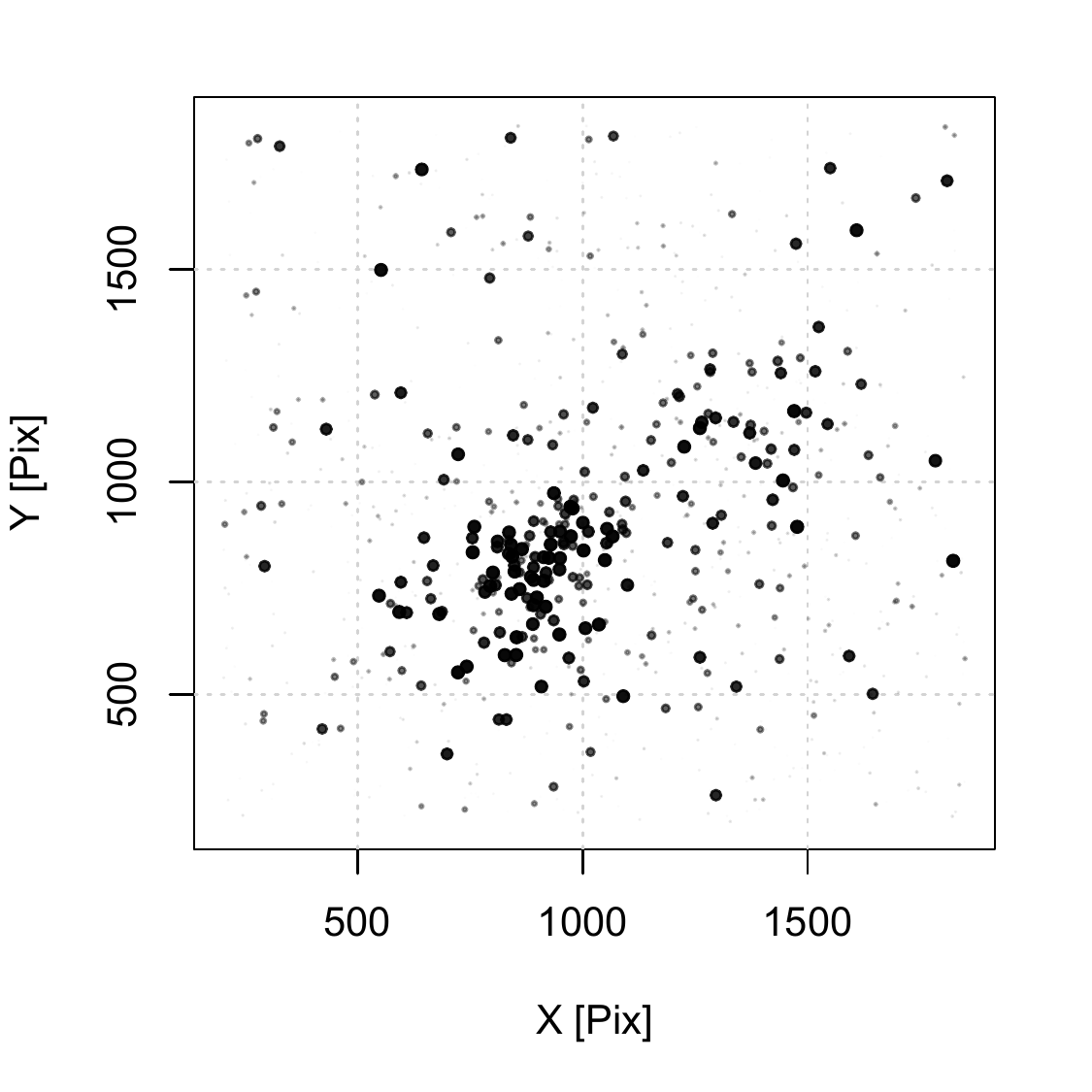}
  }
  \subfloat[]{
    \includegraphics[width=0.5 \linewidth, trim=0cm 0.5cm 0cm 0.5cm, clip=true]{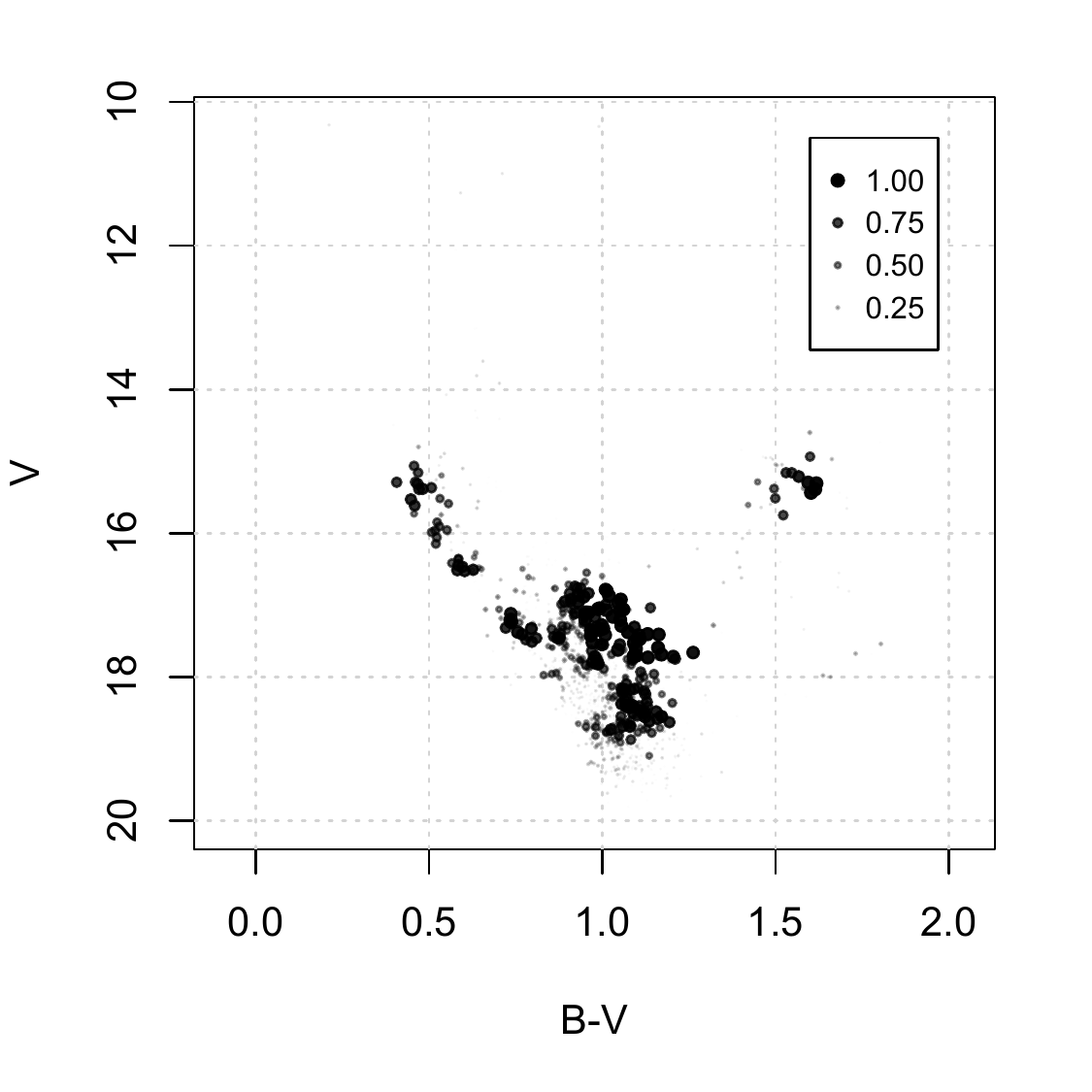}
 }\\
  \subfloat[]{
    \includegraphics[width=0.5 \linewidth, trim=0cm 0.5cm 0cm 0.5cm, clip=true]{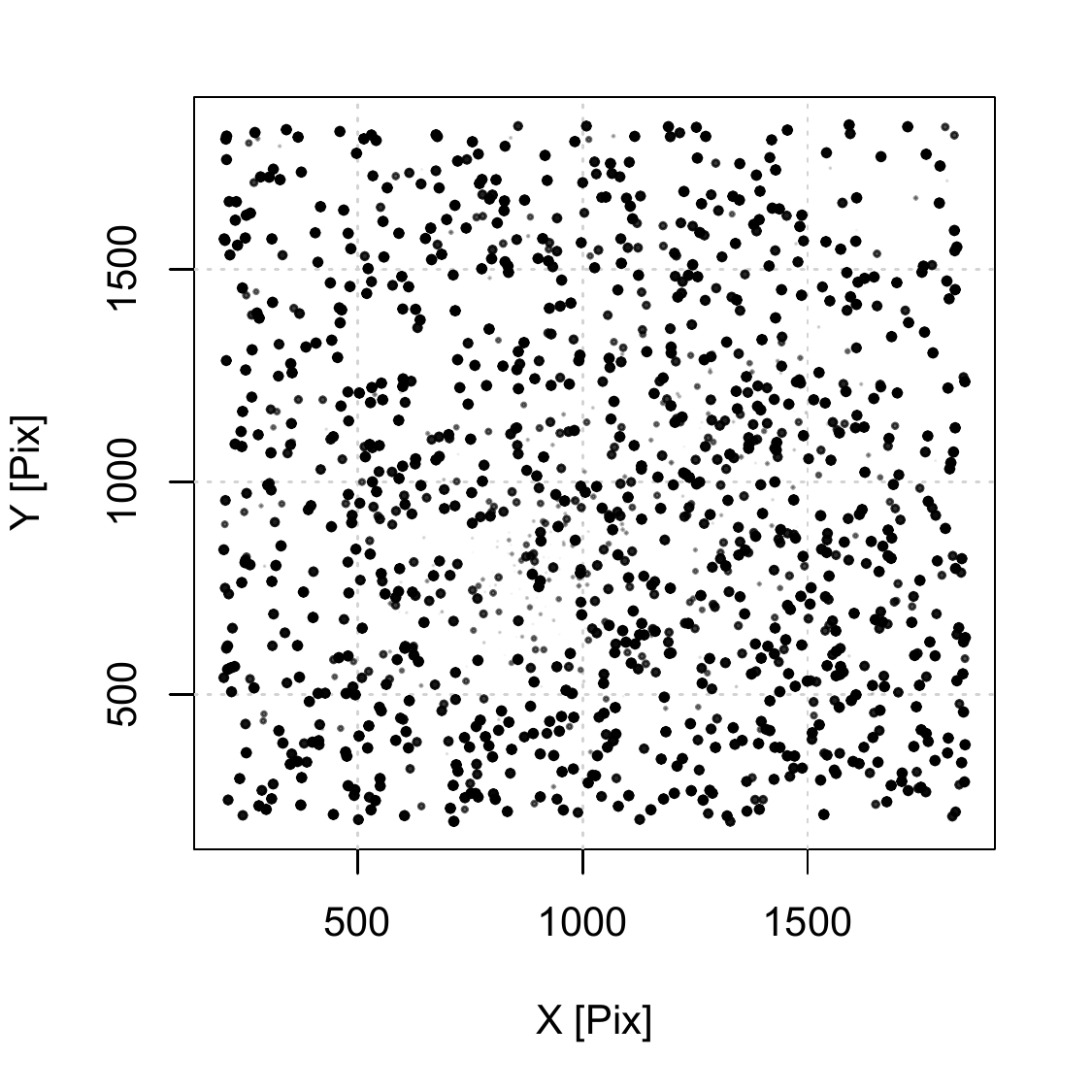}
  }
  \subfloat[]{
    \includegraphics[width=0.5 \linewidth, trim=0cm 0.5cm 0cm 0.5cm, clip=true]{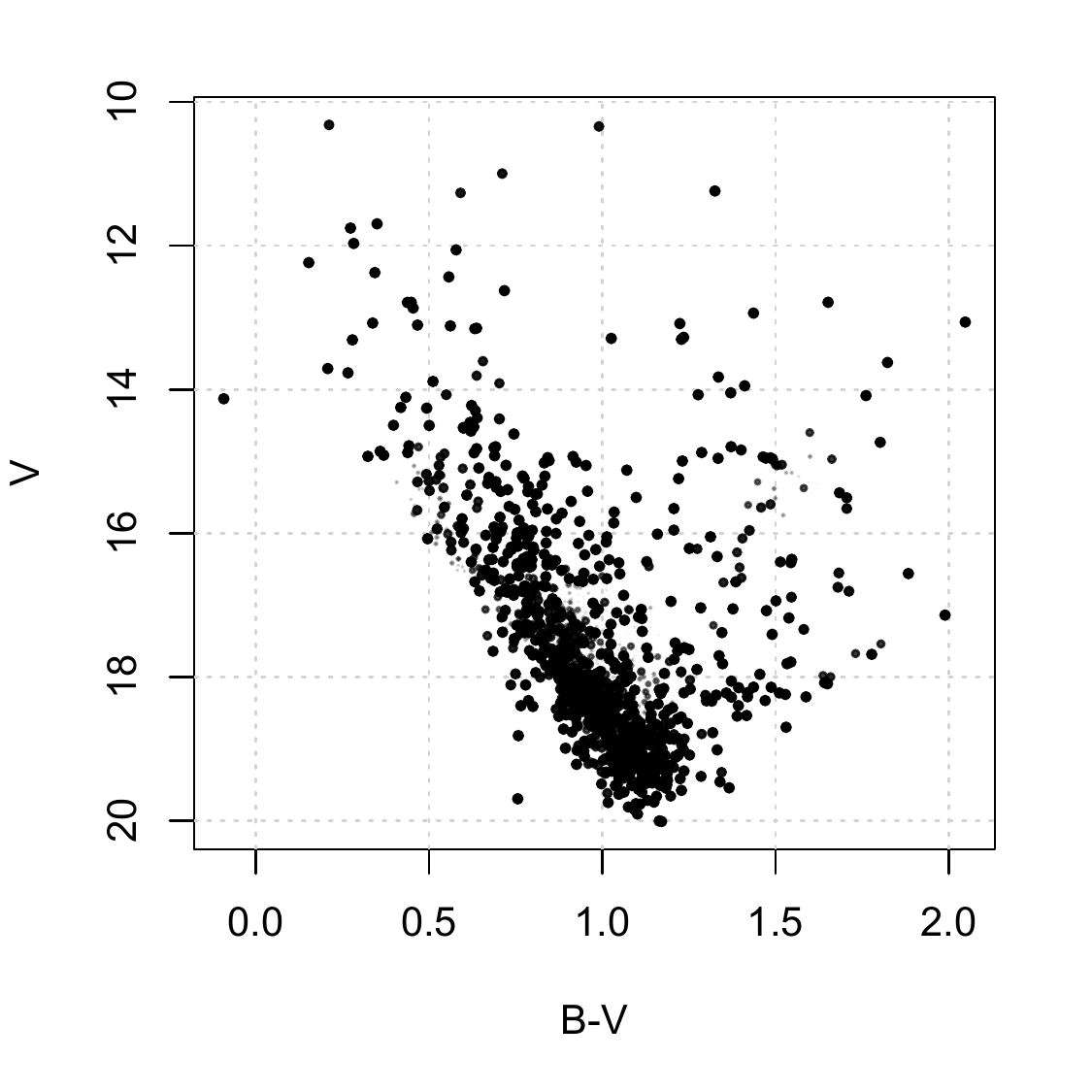}
 }
  \end{center}
      \caption{Same as Fig. \ref{Fig:Haffner16}, but for the field of Haffner 10 and Czernik 29.}
         \label{Fig:Hf10Cz26}
   \end{figure}

\subsection{Haffner 10 and Czernik 29}
The second test using real data was performed on the field of the open clusters Haffner~10 and Czernik~29. These clusters were selected because they are projected close to each other on the plane of the sky and fit in the same CCD frame. As can be seen from Fig.~\ref{Fig:Hf10Cz26} (a), the angular separation of their centres is only $\sim3.8$ arcmin. Another reason was that their sequences are not clearly visible in the CMD of the original data shown in Fig.~\ref{Fig:Hf10Cz26} (b), and thus, these objects pose an additional challenge to fully automatic methods.

Application of UPMASK to the full data-set resulted in the cluster- and field-membership probabilities represented in Figs.~\ref{Fig:Hf10Cz26} (c), (d) and (e), (f), respectively. In panel (c), there are two overlapping regions with a high concentration of objects with high membership probabilities: one in the lower-left (hereafter LL) and another in the upper-right corner (UR).

\begin{figure}
   \centering
  \begin{center}
  \subfloat[]{
    \includegraphics[width=0.5\linewidth, trim=0cm 0.5cm 0cm 0.5cm, clip=true]{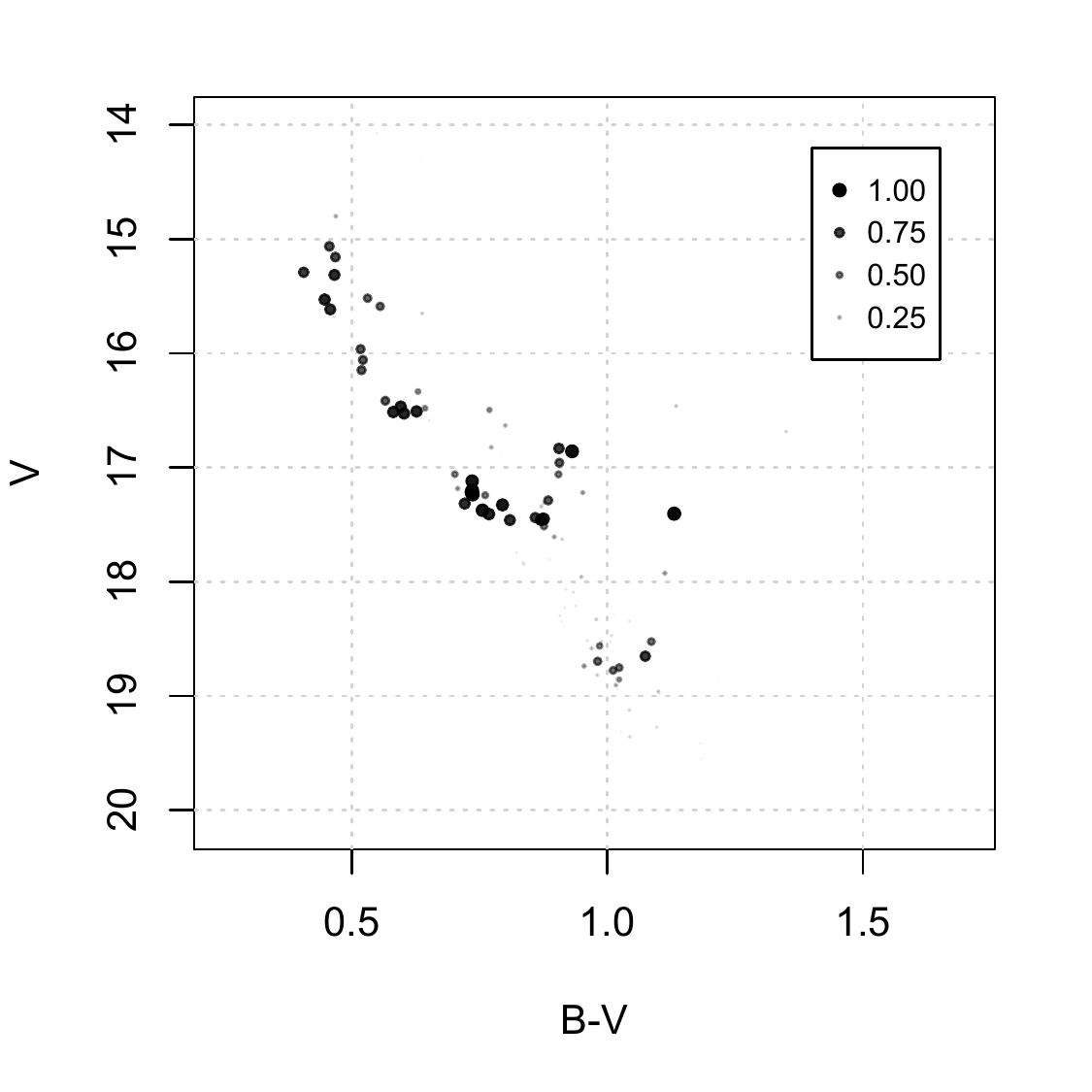}
  }
  \subfloat[]{
    \includegraphics[width=0.5 \linewidth, trim=0cm 0.5cm 0cm 0.5cm, clip=true]{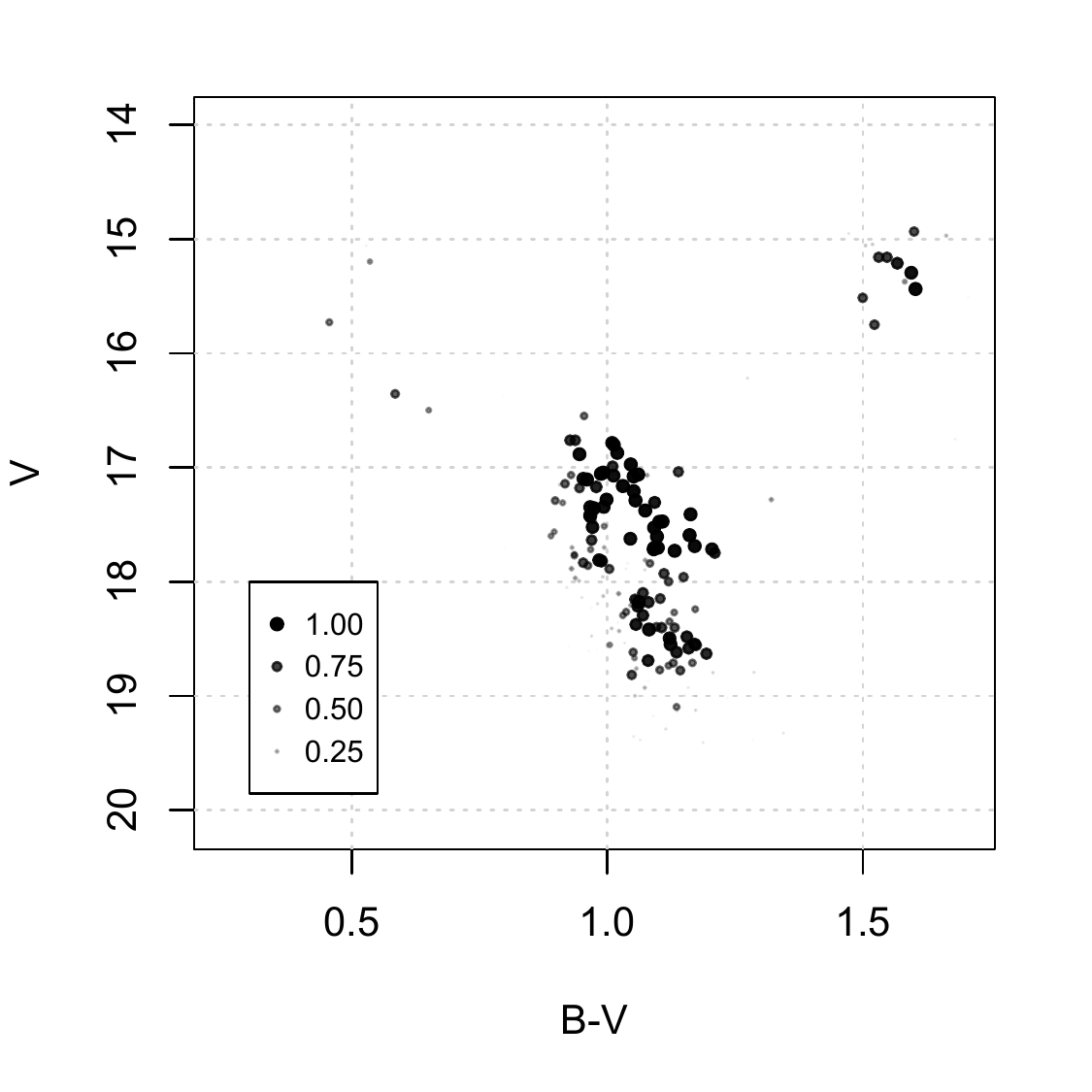}
  }\\
  \end{center}
      \caption{V vs. (B-V) CMD of stars centred on Czernik 29 (left) and Haffner 10 (right). Membership is represented as in Fig.~\ref{Fig:Haffner16}.}
         \label{Fig:Haf10_Cz29_CMDs}
   \end{figure}

The objects defining these two regions occupy different loci in the CMD shown in panel (d): those at the UR region are at $\textrm{(B-V)}<0.9$ mag; while the objects at the LL region are at $\textrm{(B-V)}>0.9$ mag, including those at $\textrm{(B-V)}\sim1.5$ mag. We note that some of the UR objects can also be found at $\textrm{(B-V)}>0.9$ mag. We identify the UR region with Czernik~29 and the LL with Haffner~10. Fig.~\ref{Fig:Haf10_Cz29_CMDs} shows the V vs. (B-V) CMDs of stars centred on these clusters.

Fig.~\ref{Fig:Haf10_Cz29_CMDs} (b) reveals the probable members of Haffner~10 as defining a main sequence with a broad colour spread and a red clump. The spread in the main sequence
could indicate some field contamination among the stars classified as cluster members. Given the faint magnitudes and in turn the larger measurement errors, a certain broadening of the cluster sequence is also expected. 
However, as noted by \citetads{2006MNRAS.365..110P}, the red clump constitutes a tilted branch parallel to the reddening/extinction vector, which these authors argued to be evidence for significant variable reddening. We note that the length of the branch is similar to the spread in the main sequence, which could thus be explained by variable reddening.

The other cluster projected on the same field, Czernik~29, presents a more homogeneous result (Fig.~\ref{Fig:Haf10_Cz29_CMDs}a), although gaps appear in the CMD, for instance, around $\textrm{(B-V)} \sim0.65$ mag, as also occurred with simulated data-sets. This is a consequence of not requiring continuity between the different sets classified as member/non-member stars. However, not requiring continuity is what allows the method to detect two (or more) different clusters projected on the same line-of-sight (as well as the red giant clump of Haffner~10).

With this case study, we have assessed the ability of UPMASK to automatically find and assign membership probabilities to two different overlapping clusters in the same field.

\subsection{Remarks on the application of UPMASK to real data}

The tests presented here pushed UPMASK close to the limits where the simulations indicated that the method would start to perform less well.
Haffner~16 is at $\sim3.7$ kpc \citepads{2006MNRAS.368L..77M}, while Haffner~10 and  Czernik~29 are at $\sim3.7$ kpc and $\sim3.0$ kpc, respectively \citepads{2010A&A...511A..38V}. These distances are at the upper limits of those for which the simulations showed that the method attained high completeness and purity (MRR$_{90}$ and TPR$_{90}$). 
The Haffner~10 and Czernik~29 field added the challenge of including two clusters in the same field-of-view. Moreover, Haffner~10 has been reported to be affected by significant variable extinction \citepads{2006MNRAS.365..110P}, means that its members are not tightly concentrated around a well-defined extinction (one of the hypothesis for distinguishing cluster and field stars).

Notwithstanding, the UPMASK method revealed interesting characteristics of the studied objects in a completely automatic analysis with few assumptions. The results presented in this section indicate that UPMASK might be used with real data up to greater distances limits, providing useful membership probabilities and new insights into the nature of the studied clusters.

\section{Present caveats and possible developments\label{sect:remarks}}

The UPMASK method can take into account missing data during the sampling step in the outer loop described in Sect.~\ref{sect:outiter}. Currently, this is implemented in a very crude way, with a uniform distribution over a specified range of the missing magnitude (which is far from ideal), and therefore this feature was not enabled during the tests presented here. Nonetheless, any distribution can be plugged-in. A better way to address this problem, would be to adopt distributions based on knowledge provided by the available measurements of a given object. A possibility would be to set constraints on the colours that use the missing measurement, instead of allowing for any value. An example of this approach could be to fill the missing values assuming a (reddened) blackbody spectrum constrained by the available photometric data. Treatment of missing data will be the subject of future research.

The UPMASK method, although originally intended for photometric and positional data alone, can also be promptly adapted for use with other types of data, such as proper motions and parallaxes. The parallaxes can be used together with positions in the spatial clustering veto. Ideally, the spatial clustering should be performed on the entire three-dimensional positional space if reliable distance information is available.

The proper motions can be used in an intermediate step between the spatial cluster veto and the photometric membership or as variables additional to the principal components during the $k$-means clustering step.

A possible future development could be to run the method independently on cells of the photometric parameter space and then combine the results a posteriori. The rationale behind this idea is to enhance the clustering in principal component space. As discussed in Sect.~\ref{sect:dimred}, the PCA produces a linear transformation, the same for all the data. However, the problem is non-linear and different spectral types project onto different principal components. Using cells means that when a cell has a high fraction of star cluster members, there will be a dominant spectral type in the cell and thus the principal components will be optimised for that spectral type.

From the technical point of view, as expected from methods based on Monte-Carlo-like heuristics, UPMASK is compute-intensive. Thus, in order to profit from multicore and shared-memory architectures, our R implementation parallelises the outer loop via the multicore package \citepads{multicore}. This allows an almost linear speed-up with the number of cores. As an indication, at the time of the writing, running the UPMASK kernel 120 times on the field of Haffner  (2096 stars with UBVRI photometry) took about $7$ minutes using 12 Intel Xeon E7-8837 cores (2.67GHz).

\section{Summary and conclusions\label{sect:conclusions}}

We have described the method UPMASK, which has been designed to identify star cluster members in highly contaminated fields using a minimal number of assumptions. Unlike field stars, members of star clusters share common properties that make them cluster in the spaces of observables related to those properties.

A PCA step was introduced to minimise overweighting observables and to maximise the separation of the different physical parameters measured by the observables. For  stellar UBVRI photometry, these physical parameters are reddening, distance, temperature, and to lesser degrees metallicity and surface gravity for some spectral types. The PCA is specially important for redundant information brought by different, but correlated, colours.

The method was tested on UBVRI photometry and positions of simulated and real star cluster fields. Results on simulations showed that UPMASK is capable of assigning star cluster members with good true positive rates and member recovery rates for a broad range of open cluster ages, masses, distances from the Sun, and contamination by field stars. Application to real data showed that the method works and produces good results when pushed to the limits indicated by the simulations.

Although the UPMASK implementation and tests presented in this paper exclusively address photometry and positions, the framework is more general and can be easily adapted to other types of data. It might also be applied to clustering studies of objects of a different nature. A possible use could be with extragalactic data or any type of data where the following basic premises apply:

\begin{itemize}
\item The \emph{member objects} are located in the same region of positional (or any physically favoured) space.
\item The \emph{member objects} share other common characteristics. They should be clustered in some arbitrary parameter spaces.
\item The \emph{non-member objects} are not (significantly) clustered in those parameter spaces.
\end{itemize}

Finally, we note that the source code of the implementation presented in this paper will be distributed under an open-source licence. As other R codes, it will be available through the CRAN archive as the UPMASK package. The simulated data used in this paper can be requested from the authors.

 \begin{acknowledgements}
      This work was supported by the Portuguese \emph{Funda\c c\~ao para Ci\^encia e Tecnologia, FCT}, project numbers SFRH/BPD/74697/2010 and PTDC/CTE-SPA/118692/2010.
      This work has made use of the computing facilities of the Laboratory of Astroinformatics (IAG/USP, NAT/Unicsul), whose purchase was made possible by the Brazilian agency FAPESP (grant 2009/54006-4) and the INCT-A.
      This work was written using the collaborative on-line ScribTeX platform.
\end{acknowledgements}

\bibliography{clusters}

\end{document}